\documentclass{article}
\usepackage{graphicx}  % standard LaTeX graphics tool;
\usepackage{amsmath}   % For getting proper math eqns
\usepackage[compress]{cite}
\usepackage{amssymb}   % Used for getting various symbols eg. gtrsim, lesssim etc.
\usepackage{bm} % For bold math (esp of lower greek letters)
\usepackage{dcolumn}% Align table columns on decimal point
\usepackage{color}
\usepackage{mathrsfs}
\usepackage{amsfonts}
\usepackage{varioref}
\usepackage{braket}
\usepackage{comment}

\RequirePackage[colorlinks,citecolor=blue,urlcolor=magenta,linkcolor=blue]{hyperref}
\newcounter{mnotecount}[section]

\renewcommand{\themnotecount}{\thesection.\arabic{mnotecount}}

\newcommand{\mnotex}[1]%{}
{\protect{\stepcounter{mnotecount}}$^{\mbox{\footnotesize
$%\!\!\!\!\!\!\,
\bullet$\themnotecount}}$ \marginpar{
\raggedright\small\em
$\!\!\!\!\!\!\,\bullet$\themnotecount: #1} }
\labelformat{section}{Section #1} 
\labelformat{subsection}{Section #1} 
\labelformat{subsubsection}{Section #1}
\labelformat{subsubsubsection}{Section #1}
\labelformat{equation}{Eq.~(#1)} 
\labelformat{figure}{Fig.~#1} 
\labelformat{subfigure}{Fig.~\thefigure#1} 
\labelformat{table}{Tab.~#1} 
\labelformat{appendix}{Appendix #1}
\title{No-boundary Wave Function, Wheeler-DeWitt Equation and Path Integral Analysis of the Bouncing `Quantum' Cosmology}
\author{Karthik Rajeev\footnote{krthkrajeev@gmail.com}$~^{1}$, Vikramaditya Mondal\footnote{vikram.iisermail@gmail.com}$~^{1}$ and Sumanta Chakraborty\footnote{sumantac.physics@gmail.com}$~^{1}$
\\
\\
{$~^{1}$\small{School of Physical Sciences}}\\
{\small{Indian Association for the Cultivation of Science, Kolkata-700032, India}}
}
\date{ }  %% This command  will supress printing the date. 
\begin{document}

\maketitle
%%%%%%%%%%%%%%%%%%%%%%%%%%%%%%%%%%%%%%%%%%%%%%%%%%%%%%%%%%%%%%%%%%%%%%%%%%%%%%%%%%%%%%%%%%%%%%%%%%%
%%%%%%%%%%%%%%%%%%%%%%%%%%%%%%%%%%%%%%%%%%%%%%%%%%%%%%%%%%%%%%%%%%%%%%%%%%%%%%%%%%%%%%%%%%%%%%%%%%%
%%%%%%%%%%%%%%%%%%%%%%%%%%%%%%%%%%%%%%%%%%%%%%%%%%%%%%%%%%%%%%%%%%%%%%%%%%%%%%%%%%%%%%%%%%%%%%%%%%%
\begin{abstract}
	Bouncing models are alternatives to inflationary cosmology that replace the initial Big-Bang singularity by a `bouncing' phase. A deeper understanding of the initial conditions of the universe, in these scenarios, requires knowledge of quantum aspects of bouncing models. In this work, we propose two classes of bouncing models that can be studied with great analytical ease and hence, provide test-bed for investigating more profound problems in quantum cosmology of bouncing universes. Our model's two key ingredients enable us to do straightforward analytical calculations: (i) a convenient parametrization of the minisuperspace of FRLW spacetimes and (ii) two distinct choices of the effective perfect fluids that source the background geometry of the bouncing universe. We study the quantum cosmology of these models using both the Wheeler-de Witt equations and the path integral approach. In particular, we found a bouncing model analogue of the no-boundary wave function and presented a Lorentzian path integral representation for the same. We also discuss the introduction of real scalar perturbations.    
\end{abstract}
%%%%%%%%%%%%%%%%%%%%%%%%%%%%%%%%%%%%%%%%%%%%%%%%%%%%%%%%%%%%%%%%%%%%%%%%%%%%%%%%%%%%%%%%%%%%%%%%%%%
%%%%%%%%%%%%%%%%%%%%%%%%%%%%%%%%%%%%%%%%%%%%%%%%%%%%%%%%%%%%%%%%%%%%%%%%%%%%%%%%%%%%%%%%%%%%%%%%%%%
%%%%%%%%%%%%%%%%%%%%%%%%%%%%%%%%%%%%%%%%%%%%%%%%%%%%%%%%%%%%%%%%%%%%%%%%%%%%%%%%%%%%%%%%%%%%%%%%%%%	
\newpage

\tableofcontents
%%%%%%%%%%%%%%%%%%%%%%%%%%%%%%%%%%%%%%%%%%%%%%%%%%%%%%%%%%%%%%%%%%%%%%%%%%%%%%%%%%%%%%%%%%%%%%%%%%%
%%%%%%%%%%%%%%%%%%%%%%%%%%%%%%%%%%%%%%%%%%%%%%%%%%%%%%%%%%%%%%%%%%%%%%%%%%%%%%%%%%%%%%%%%%%%%%%%%%%
%%%%%%%%%%%%%%%%%%%%%%%%%%%%%%%%%%%%%%%%%%%%%%%%%%%%%%%%%%%%%%%%%%%%%%%%%%%%%%%%%%%%%%%%%%%%%%%%%%%	
\section{Introduction}	

The inflationary scenario gives rise to significant improvements to the Standard Big-Bang cosmology (SBC)\cite{Guth:1980zm,Sato:1980yn,Linde:1981mu,Albrecht:1982wi,Starobinsky:1980te}. Inflation is credited for solving several issues of SBC that include the horizon and flatness problems. The causal mechanism of structure formation that the theory of inflation offers has made it a proper science of precise and observationally verifiable predictions. Despite these desirable features, inflation is not devoid of serious conceptual challenges. For instance, if we consider the scalar field inflation in the context of Einstein gravity, it can be shown that there is an inevitable singularity before the onset of inflation \cite{Hawking:1969sw,Borde:1993xh}. Another challenge to the inflationary models is the TransPlanckian issue\cite{Martin:2000xs,Brandenberger_2013}.  The term refers to the fact that if the universe expanded slightly over 65 e-folds during inflation, that is a little more than what is required to resolve certain issues of SBC, then one can show that the fluctuations which are within the Hubble radius today were originally at sub-Planckian length scales during the onset of inflation. The TransPlanckian problem challenges the validity of `quantum field theory in classical spacetime' approximation, which is at the heart of analysis of perturbations in inflationary models.  Deficiencies of the inflationary paradigm, such as these, have inspired the investigation of several viable alternative models for the early Universe.  

`Bouncing cosmology' refers to a broad class of cosmological models characterized by a `bounce', i.e., a smooth transition from a contracting to expanding phase. Bouncing models aim to resolve the fundamental problems of SBC without invoking inflation. For instance, by construction, bouncing cosmologies resolve the singularity problem as it replaces the singularity by a bounce.  Bouncing models also avoids the TransPlanckian problem, since, in the bouncing scenario, a given length scale of fluctuation contracts to a non-zero minimum, thereby, assuring that all wavelengths of relevance to cosmology today have initially been far greater than the Planck length during the bounce.
Nevertheless, bouncing models resolve the horizon problem, since, in a typical bouncing scenario, the sub-Hubble wavelengths of the seed fluctuation modes exit the Hubble radius and then re-enter at a later time to emerge as a scale of current cosmological interest, just as in inflation (see, for instance, \cite{Finelli:2001sr,Brandenberger:2012zb,Gasperini:1992em}).  It is worth mentioning that there are challenges in bouncing models as well. For a recent review of progress and problems in bouncing models, the reader may consult \cite{Brandenberger:2016vhg}. 

Our aim, in this work, is to explore the quantum gravitational aspects of bouncing scenarios. There are, however, several models of bouncing cosmologies and, most of them are conceptually disparate \cite{Battefeld:2014uga}. To see this, recall that the inevitable singularity that arises in homogeneous and isotropic cosmologies in Einstein gravity is a consequence of Hawking-Penrose singularity theorems. Hence, the several ways of realizing a bounce can be viewed as essentially corresponding to the several ways of bypassing Hawking-Penrose singularity theorems' assumptions. In order to realize a bouncing scenario, therefore, one might resort to any of the diverse models of unconventional physics which may include, among other things, concepts from modified gravity theories\cite{Brandenberger:2009yt,Brandenberger:2012zb,Bamba:2013fha,Desai:2015haa}, modified matter \cite{ArkaniHamed:2003uy,Cai:2008qw,Raveendran:2018why,Raveendran:2017vfx}, asymptotically safe gravity \cite{Bonanno:2017gji}, string and string-inspired theories \cite{Bamba:2014zoa,Basile:2021amb,Brandenberger:1988aj} and loop quantum cosmology\cite{Haro:2015oqa,Ashtekar:2008ay,WilsonEwing:2012pu,Cai:2014zga}. Consequently, it is not easy to study the entire family of bouncing cosmologies in a single framework. Here, we adopt a simple phenomenological approach, which we can use to study a range of bouncing models of interest to theoretical cosmology.

Our starting point is the minisuperspace of homogeneous and isotropic spacetimes, supplemented by an effective potential $U_{\rm eff}(a)$ that determines the classical dynamics of `$a$', the scale factor. The functional form of $U_{\rm eff}(a)$ is such that the scale factor's classical dynamics gives rise to a bouncing scenario. We introduce this effective potential as a proxy for any of the unconventional physics that can accomplish a bounce. An obvious advantage of this approach is that we do not have to directly subscribe to any particular bouncing model. It turns out that for several models a suitable $U_{\rm eff}(a)$ may be chosen to capture its relevant features effectively. Our strategy then is to apply principles of quantum cosmology to study the quantum aspects of our model.     

The paper's structure is as follows: In \ref{gen_setup_minisuperspace}, we briefly review the minisuperspace model of cosmology and introduce a convenient parametrization of the metric. In \ref{gen_setup_matter}, we introduce two classes of perfect fluid densities that give rise to a wide range of bouncing cosmologies. We chose specific forms for these densities such that they generate effective potentials which, when rewritten in terms of a suitable variable, transform into either a linear or a quadratic form . Quantum aspects of these models are then analyzed in \ref{wavefunction_bounce} using the Wheeler-de Witt equation and a recent approach to quantum cosmology, for instance, prescribed in \cite{Feldbrugge:2017kzv,DiTucci:2019dji,Lehners2019}. Followed by that, we introduce a massive scalar field conformally coupled to the background spacetime in \ref{intro_matter}. We conclude with the summary and discussion of our results in \ref{discussion}. (We shall henceforth work in units with $c=\hbar=1$, unless otherwise specified.)

%%%%%%%%%%%%%%%%%%%%%%%%%%%%%%%%%%%%%%%%%%%%%%%%%%%%%%%%%%%%%%%%%%%%%%%%%%%%%%%%%%%%%%%%%%%%%%%%%%%
%%%%%%%%%%%%%%%%%%%%%%%%%%%%%%%%%%%%%%%%%%%%%%%%%%%%%%%%%%%%%%%%%%%%%%%%%%%%%%%%%%%%%%%%%%%%%%%%%%%
%%%%%%%%%%%%%%%%%%%%%%%%%%%%%%%%%%%%%%%%%%%%%%%%%%%%%%%%%%%%%%%%%%%%%%%%%%%%%%%%%%%%%%%%%%%%%%%%%%%
\section{Minisuperspace model of cosmology: The general setup}\label{gen_setup_minisuperspace}

We consider the minisuperspace of homogeneous and isotopic FRLW spacetimes with flat spatial sections. A convenient parametrization for this class of metrics is given by,
%%%%%%%%%%%%%%%%%%%%%%%%%%%%%%%%%%%%%%%%%%%%%%%%%%%%%%%%%%%%%%%%
\begin{align}\label{def_minisuperspace}
ds^{2}=-\frac{\mathcal{N}^2(t)}{q(t)^{p}}dt^2+q(t)^{b}|d\mathbf{x}|^2~,
\end{align}
%%%%%%%%%%%%%%%%%%%%%%%%%%%%%%%%%%%%%%%%%%%%%%%%%%%%%%%%%%%%%%%%
where $p$ and $b$ are both real numbers and are arbitrary at this point. We shall, however, fix the values of $p$ and $b$ later, based on the necessity to have a quadratic action. As we shall see shortly, the analysis we hope to pursue here greatly simplifies when the action is a quadratic functional. Notice that the scale factor is given by $a(t)=q(t)^{b/2}$. Moreover, the time coordinate $t$, introduced above, is neither the cosmic time $\tau$, nor the conformal time $\eta$. The time coordinate $t$ is related to the cosmic time $\tau$ and the conformal time $\eta$ through the following relations,
%%%%%%%%%%%%%%%%%%%%%%%%%%%%%%%%%%%%%%%%%%%%%%%%%%%%%%%%%%%%%%%%
\begin{align}
d\tau=\frac{\mathcal{N}(t)dt}{q(t)^{p/2}}~; 
\qquad
d\eta=\frac{\mathcal{N}(t)dt}{q(t)^{(p+b)/2}}~.
\end{align}
%%%%%%%%%%%%%%%%%%%%%%%%%%%%%%%%%%%%%%%%%%%%%%%%%%%%%%%%%%%%%%%%
If we assume that the function $\mathcal{N}$ is a constant, then it is evident that as the parameters $p$ and $b$ take values such that $p+b=0$ and $p=0$, then the time coordinate $t$ reduces to $\eta$ and $\tau$, respectively. 

For the metric ansatz presented in \ref{def_minisuperspace}, the Einstein-Hilbert action takes the following form,
%%%%%%%%%%%%%%%%%%%%%%%%%%%%%%%%%%%%%%%%%%%%%%%%%%%%%%%%%%%%%%%%
\begin{align}
\mathcal{S}_{\rm EH}=\int d^{4}x\sqrt{-g}R=V_{3}\left(\frac{3b}{2}\right)\int dt ~ q(t)^{\frac{3b+p-4}{2}}\left[\frac{1}{\mathcal{N}}\left(p+2b-2\right)\dot{q}^{2}+2\frac{d}{dt}\left(\frac{1}{\mathcal{N}}q\dot{q}\right)-\frac{2}{\mathcal{N}}\dot{q}^{2} \right]~,
\end{align}
%%%%%%%%%%%%%%%%%%%%%%%%%%%%%%%%%%%%%%%%%%%%%%%%%%%%%%%%%%%%%%%%
where, the over-dot denotes derivative with respect to $t$ and $V_3$ is the spatial volume\footnote{For a non-compact and flat FRLW universe, the spatial volume $V_3$ is infinite. We can, however, circumvent potential issues arising out of infinite $V_3$ by assuming that the spatial slices are locally flat but, \textit{globally} compact, like a torus \cite{Louko:1988zb,Louko:1989up}. The spacial volume $V_3$, in such cases, is finite.}. However, unless the action functional is quadratic in the dynamical variable $q(t)$, it is difficult to employ the path-integral techniques. Thus, we demand that the parameter $p$ and $b$ are constrained by the following relation: $3b+p=4$. This constraint means that the two independent parameters $p$ and $b$ in \ref{def_minisuperspace} boils down to a single parameter, which we choose to be $b$. After imposing the constraint $3b+p=4$, the Einstein-Hilbert action $S_{\rm EH}$, together with the Gibbons-Hawking-York boundary term for a non-null boundary, acquires the following form:
%%%%%%%%%%%%%%%%%%%%%%%%%%%%%%%%%%%%%%%%%%%%%%%%%%%%%%%%%%%%%%%%
\begin{align}\label{EH_Action}
\mathcal{S}_{\rm EH}&= \frac{1}{2\kappa}\int dx^4\sqrt{-g} R-\frac{1}{\kappa}\int_{\textrm{Boundary}}d^3y \sqrt{h}K
\nonumber
\\
&=\frac{V_{3}}{2\kappa}\int dt~\left(-\frac{3b^{2}}{2\mathcal{N}}\right)\dot{q}^{2}
+3bV_{3}\left(\frac{q\dot{q}}{\mathcal{N}}\right)_{\rm Boundary}-\frac{1}{\kappa}\int_{\textrm{Boundary}}d^3y \sqrt{h}K
\\
\nonumber
&=V_{3}\int\left[-\frac{M }{2\mathcal{N}}\left(\frac{dq}{dt}\right)^2\right]dt~,
\end{align}
%%%%%%%%%%%%%%%%%%%%%%%%%%%%%%%%%%%%%%%%%%%%%%%%%%%%%%%%%%%%%%%%
where, we have defined an ``effective mass" $M\equiv\{(3b^{2})/(2\kappa)\}$ with $\kappa=8\pi G$. Note that, in the second line of the above equation, the extrinsic curvature terms evaluated on the $t=\textrm{constant}$ boundary hypersurfaces exactly cancels the total derivative term, thereby, reducing $\mathcal{S}_{\rm EH}$  to the final expression. The parametrization of FRLW metric we presented in \ref{def_minisuperspace} contains several scenarios that have appeared in earlier works in the literature. These include the choice  $b=1$, which has been used to study de Sitter cosmology in \cite{Halliwell1988} and more recently in \cite{Lehners2019} to describe the Lorentzian path integral approach to quantum cosmology. Additionally, for $b=4/3$, the above parametrization has also appeared in \cite{Rajeev:2019ivq} to rewrite the minisuperspace Lagrangian into a quadratic form. Finally, $b=2$ corresponds to the standard representation of the metric in terms of the conformal time coordinate. 

We have already described the Einstein-Hilbert part of the gravitational action, along with the Gibbons-Hawking-York boundary term, for the parametrization of the line element presented in \ref{def_minisuperspace}. We shall now introduce the matter sector into the picture. The most straightforward inclusion corresponds to the addition of a perfect fluid, with energy density $\rho_{\rm eff}$ and pressure $p_{\rm eff}\propto \rho_{\rm eff}$, for which the complete action describing gravity plus perfect fluid system is given by \cite{hawking_ellis_1973},
%%%%%%%%%%%%%%%%%%%%%%%%%%%%%%%%%%%%%%%%%%%%%%%%%%%%%%%%%%%%%%%%
\begin{align}\label{action_eff}
\mathcal{S}_{\rm perfect}&=\frac{1}{2\kappa}\int d^{4}x\sqrt{-g} R-\frac{1}{\kappa}\int_{\textrm{Boundary}}d^{3}y \sqrt{h}K-\int d^{4}x \sqrt{-g}~\rho_{\rm eff}(q) 
\nonumber
\\
&=V_{3}\int dt\left[-\frac{M}{2\mathcal{N}}\left(\frac{dq}{dt}\right)^2+\mathcal{N} U_{\rm eff}(q)\right]~,
\end{align}
%%%%%%%%%%%%%%%%%%%%%%%%%%%%%%%%%%%%%%%%%%%%%%%%%%%%%%%%%%%%%%%%
where, we have used \ref{EH_Action} and have defined the effective `potential' $U_{\rm eff}$ as\footnote{Although we have restricted ourselves to flat FRLW models, one can easily extend our analysis to closed as well as open FRLW cosmologies. In fact, one could imagine that $U_{\rm eff}(q)$ contains a contribution of the form $-kq^{-2 + 2 b}\propto\sqrt{-g}\frac{k}{a^2}$ (where $k=0,\pm 1$ for flat, closed/open spatial slices), to account for the non-zero spatial curvature.},
%%%%%%%%%%%%%%%%%%%%%%%%%%%%%%%%%%%%%%%%%%%%%%%%%%%%%%%%%%%%%%%%
\begin{align}\label{Ueff}
U_{\rm eff}(q)&\equiv-q^{(3b-2)}\rho_{\rm eff}(q)~.
\end{align}
%%%%%%%%%%%%%%%%%%%%%%%%%%%%%%%%%%%%%%%%%%%%%%%%%%%%%%%%%%%%%%%%
As evident from \ref{action_eff}, the Lagrangian resembles that of a point particle, with generalized coordinate $q$, moving in the effective potential $U_{\rm eff}(q)$.

Another possible addition to our `gravity-perfect fluid system' is a non-minimally coupled scalar field of mass $m$, such that, the total action takes the following form,
%%%%%%%%%%%%%%%%%%%%%%%%%%%%%%%%%%%%%%%%%%%%%%%%%%%%%%%%%%%%%%%%
\begin{align}\label{Full_Action}
\mathcal{S}_{\rm conformal}&=\frac{1}{2\kappa}\int d^{4}x\sqrt{-g}~R-\frac{1}{\kappa}\int_{\textrm{Boundary}}d^{3}y~\sqrt{h}K-\int d^{4}x \sqrt{-g}~\rho_{\rm eff}(q) 
\nonumber
\\
&\hskip 2 cm +\int d^{4}x\left[-\frac{1}{2}g^{\mu \nu}\partial_{\mu}\phi \partial_{\nu}\phi-V(\phi)-\frac{\xi}{2}R\phi^{2} \right]~,
\end{align}  
%%%%%%%%%%%%%%%%%%%%%%%%%%%%%%%%%%%%%%%%%%%%%%%%%%%%%%%%%%%%%%%%
where, $\xi=0$ corresponds to the minimally coupled scalar field and $\xi=(1/6)$ will yield the conformally coupled scalar field. Using the Fourier transform for the scalar field $\phi$ and using the fact that the scalar field is real, the scalar field action takes the following form,
%%%%%%%%%%%%%%%%%%%%%%%%%%%%%%%%%%%%%%%%%%%%%%%%%%%%%%%%%%%%%%%%
\begin{align}
\mathcal{S}_{\rm scalar}&=\int d^{4}x\left[-\frac{1}{2}g^{\mu \nu}\partial_{\mu}\phi \partial_{\nu}\phi-V(\phi)-\frac{\xi}{2}R\phi^{2} \right]
\nonumber
\\
&=\int dt \int \frac{d^{3}\bm{k}}{(2\pi)^{3}}\left[\frac{1}{2\mathcal{N}}q^{2}|\dot{\phi}_{\bm{k}}|^{2}-\frac{\mathcal{N}}{2}q^{2b-2}\left(|\bm{k}|^{2}+m^{2}q^{b}\right)|\phi_{\bm{k}}|^{2}+\frac{3b^{2}\xi}{4\mathcal{N}}\dot{q}^{2}|\phi_{\bm{k}}|^{2}
+\frac{3b\xi}{2}\frac{1}{2\mathcal{N}}\frac{dq^{2}}{dt}\frac{d|\phi_{\bm{k}}|^{2}}{dt} \right]
\nonumber
\\
&\hskip 2 cm -\frac{3b\xi}{2} \int dt\int \frac{d^{3}\bm{k}}{(2\pi)^{3}}\frac{d}{dt}\left(\frac{1}{2\mathcal{N}}|\phi_{\bm{k}}|^{2}\frac{dq^{2}}{dt}\right)
\end{align}  
%%%%%%%%%%%%%%%%%%%%%%%%%%%%%%%%%%%%%%%%%%%%%%%%%%%%%%%%%%%%%%%%
Redefining, the scalar field as, $\phi_{\bm{k}}=\sqrt{V_3}q^{c}\widetilde{\phi}_{\bm k}$, the above action can be expressed as,
%%%%%%%%%%%%%%%%%%%%%%%%%%%%%%%%%%%%%%%%%%%%%%%%%%%%%%%%%%%%%%%%
\begin{align}\label{action_scalar}
\mathcal{S}_{\rm scalar}&=V_3\int dt \int \frac{d^{3}\bm{k}}{(2\pi)^{3}}\Bigg[\frac{1}{2\mathcal{N}}q^{2+2c}|\dot{\widetilde{\phi}}_{\bm{k}}|^{2}+\frac{c+3b\xi}{4\mathcal{N}}q^{2c}\frac{dq^{2}}{dt}\frac{d|\widetilde{\phi}_{\bm{k}}|^{2}}{dt}+\frac{c^{2}+(3b^{2}\xi/2)+6bc\xi}{2\mathcal{N}}q^{2c}\dot{q}^{2}|\widetilde{\phi}_{\bm{k}}|^{2}
\nonumber
\\
&\hskip 2 cm -\frac{\mathcal{N}}{2}q^{2b-2+2c}\left(|\bm{k}|^{2}+m^{2}q^{b}\right)|\widetilde{\phi}_{\bm{k}}|^{2}\Bigg]-\frac{3b\xi V_3}{2} \int dt\int \frac{d^{3}\bm{k}}{(2\pi)^{3}}\frac{d}{dt}\left(\frac{1}{2\mathcal{N}}|\phi_{\bm{k}}|^{2}\frac{dq^{2}}{dt}\right)
\end{align}  
%%%%%%%%%%%%%%%%%%%%%%%%%%%%%%%%%%%%%%%%%%%%%%%%%%%%%%%%%%%%%%%%
Note that, if we wish to remove the second term in the matter action presented above then, the constant $c$ must be chosen such that $c=-3b\xi$. In that case, the coefficient of the third term in the right-hand side of \ref{action_scalar} yields $(b^{2}/2)(-18\xi^{2}+3\xi)$, which vanish iff $\xi=0$ or $\xi=(1/6)$. Therefore, for conformally coupled scalar field, $\mathcal{S}_{\rm scalar}$ simplifies considerably and reduces to the following form,
%%%%%%%%%%%%%%%%%%%%%%%%%%%%%%%%%%%%%%%%%%%%%%%%%%%%%%%%%%%%%%%%
\begin{align}\label{Full_Action_Final}
\mathcal{S}_{\rm conformal}&=V_{3}\int dt\left[-\frac{M}{2\mathcal{N}}\left(\frac{dq}{dt}\right)^2+\mathcal{N} U_{\rm eff}(q)\right]
+V_3\int dt \int\frac{d^3\mathbf{k}}{(2\pi)^3}\frac{d }{dt}\left(\frac{1}{2\mathcal{N}}~\mathcal{S}_{\bm{k}}\right)
\nonumber
\\
&\hskip 2 cm +V_3\int dt \int\frac{d^3\mathbf{k}}{(2\pi)^3}\left[\frac{\mu(q)}{2\mathcal{N}}\left|\frac{d\widetilde{\phi}_{\bm{k}}}{dt}\right|^2-\frac{\mathcal{N}\mu(q)\omega^2_{\bm{k}}(q)}{2}|\widetilde{\phi}_{\bm{k}}|^{2}\right]~,
\end{align}  
%%%%%%%%%%%%%%%%%%%%%%%%%%%%%%%%%%%%%%%%%%%%%%%%%%%%%%%%%%%%%%%%
where, the functions $\mu(q)$, $\omega_{\bm{k}}(q)$ and $\mathcal{S}_{\bm{k}}$ are defined as, 
%%%%%%%%%%%%%%%%%%%%%%%%%%%%%%%%%%%%%%%%%%%%%%%%%%%%%%%%%%%%%%%%
\begin{align}
\mu(q)=q^{2-b}~; \quad 
\omega^{2}_{\bm{k}}(q)=q^{2b-4}\left(k^2+m^2 q^{b}\right)~; \quad
\mathcal{S}_{\bm{k}}=-\frac{b}{4}|\phi_{\bm{k}}^{2}|\frac{dq^2}{dt}~.
\end{align}

%%%%%%%%%%%%%%%%%%%%%%%%%%%%%%%%%%%%%%%%%%%%%%%%%%%%%%%%%%%%%%%%
Thus, the action for the scalar field $\phi_{\bm{k}}$, as presented above, is also quadratic and each of the Fourier modes $\phi_{\bm{k}}$ depicts a harmonic oscillator with a time-dependent mass $\mu(q)$ and a time-dependent frequency $\omega^{2}_{\bm{k}}(q)$. Therefore, the equation of motion obtained by varying $\phi_{\bm k}$ is given by:
\begin{align}
\frac{d}{dT}\left(\mu(q)\frac{d\phi_{\bm k}}{dT}\right)+\omega^2_{\bm{k}}(q)\phi_{\bm k}&=0
\end{align}
where, $dT\equiv\mathcal{N}dt$. Furthermore, the momentum conjugate to $q$ and $\widetilde{\phi}_{\bm{k}}$ are, respectively, given by  $\pi_{q}=-(M/\mathcal{N})(dq/dt)$ and $\pi_{\bm{k}}=(\mu(q)/2)(d\widetilde{\phi}_{\bm{k}}/dt)$. We shall now determine the Wheeler-DeWitt equation from the action functional. To this end, we first determine the Hamiltonian constraint, obtained by varying the action $\mathcal{S}_{\rm conformal}$ with respect to the lapse function $\mathcal{N}$ and then setting $\mathcal{N}=1$. This yields:
%%%%%%%%%%%%%%%%%%%%%%%%%%%%%%%%%%%%%%%%%%%%%%%%%%%%%%%%%%%%%%%%
\begin{align}\label{classical_constraint}
-\frac{MV_{3}}{2}\left(\frac{dq}{dt}\right)^{2}-V_{3}U_{\rm eff}(q)+V_3\int\frac{d^3\bm{k}}{(2\pi)^3}\left[\frac{\mu(q)}{2}\left(\frac{dQ_{\bm k}}{dt}\right)^{2}+\frac{\mu(q)\omega^2_{\bm{k}}(q)}{2}Q_{\bm k}^{2}\right]=0~.
\end{align}
%%%%%%%%%%%%%%%%%%%%%%%%%%%%%%%%%%%%%%%%%%%%%%%%%%%%%%%%%%%%%%%%
where, we have defined:
\begin{align}
Q_{\bm k}\equiv\begin{cases}
\sqrt{2}{\rm Re}[\tilde{\phi}_{\bm k}];\quad k_z\geq 0\\
\sqrt{2}{\rm Im}[\tilde{\phi}_{\bm k}];\quad k_z<0
\end{cases}
\end{align}
Expressing $\dot{q}$ and $\dot{Q}_{\bm{k}}$ in terms of the conjugate momentum $\pi_{q}$ and $\pi_{\bm{k}}$, followed by promoting them to appropriate operators, namely, $\pi_{q}\equiv-i\partial_q$ and $\pi_{\bm{k}}\equiv-i\partial_{Q_{\bm{k}}}$ (for ease of notation and since further confusion is unlikely to arise, we have henceforth renamed $Q_{\bm{k}}$ as $\phi_{\bm{k}}$), we obtain the following Wheeler-DeWitt equation,
%%%%%%%%%%%%%%%%%%%%%%%%%%%%%%%%%%%%%%%%%%%%%%%%%%%%%%%%%%%%%%%%
\begin{align}\label{WdW}
\left\{\frac{1}{2MV_{3}}\frac{\partial^{2}}{\partial q^{2}}-V_{3}U_{\rm eff}(q)+V_3\int\frac{d^{3}\bm{k}}{(2\pi)^3}\left[-\frac{1}{2\mu(q)}\frac{\partial^{2}}{\partial \phi_{\bm{k}}^{2}}+\frac{\mu(q)\omega^{2}_{\bm{k}}(q)}{2}\phi^{2}_{\bm{k}}\right]\right\}\Psi(q,\{\phi_{\bm{k}}\})=0~,
\end{align} 
%%%%%%%%%%%%%%%%%%%%%%%%%%%%%%%%%%%%%%%%%%%%%%%%%%%%%%%%%%%%%%%%
where, $\Psi(q,\{\phi_{\bm{k}}\})$ is the wave function of the universe describing the evolution of the scale factor (through $q(t)$) and also the evolution of the matter field $\phi_{\bm{k}}$ in a coherent manner. 

The first two terms of \ref{WdW}, which account for the purely gravitational part, governs the leading order quantum dynamics of the system considered here. We shall treat the scalar field $\phi_{\bm{k}}$ as a test field at the quantum level. As is evident from \ref{WdW}, the dynamics of the gravity sector is determined by the effective potential $U_{\rm eff}$, which in turn is determined by the effective energy density $\rho_{\rm eff}$ of the perfect fluid system. In the next section, we will consider a family of perfect fluid models that can accomplish the bouncing scenario in the classical regime. Following that, we will discuss the associated quantum scenario. 

%%%%%%%%%%%%%%%%%%%%%%%%%%%%%%%%%%%%%%%%%%%%%%%%%%%%%%%%%%%%%%%%%%%%%%%%%%%%%%%%%%%%%%%%%%%%%%%%%%%
%%%%%%%%%%%%%%%%%%%%%%%%%%%%%%%%%%%%%%%%%%%%%%%%%%%%%%%%%%%%%%%%%%%%%%%%%%%%%%%%%%%%%%%%%%%%%%%%%%%
%%%%%%%%%%%%%%%%%%%%%%%%%%%%%%%%%%%%%%%%%%%%%%%%%%%%%%%%%%%%%%%%%%%%%%%%%%%%%%%%%%%%%%%%%%%%%%%%%%%
\section{Effective matter content of bouncing models: General analysis}\label{gen_setup_matter}

In the previous section, we discussed the general aspects of minisuperspace model of cosmology consisting of a single dynamical variable $q(t)$. The ansatz for the spacetime metric was chosen such that the gravitational part of the action reduces to a quadratic functional of the dynamical variable $q(t)$. We also considered the introduction of a non-minimally coupled scalar field as well as a perfect fluid with energy density $\rho_{\rm eff}$. Therein, we saw that the variable frequencies of the harmonic oscillator modes of the scalar field are functionals of $q$ alone, if and only if, the coupling is either minimal or conformal. In this section, we will discuss the possible forms of the energy density $\rho_{\rm eff}$, which can retain the quadratic nature of the action presented in \ref{Full_Action_Final}. 

We start by considering an effective energy density $\rho_{\rm eff}$ of the following form, 
%%%%%%%%%%%%%%%%%%%%%%%%%%%%%%%%%%%%%%%%%%%%%%%%%%%%%%%%%%%%%%%%
\begin{align}\label{energy_density}
\rho_{\rm eff}(a)=\rho_{0}\left[\frac{c_{1}}{a^{n_{1}}}+\frac{c_{2}}{a^{n_{2}}}+\frac{c_{3}}{a^{n_{3}}} \right]~,
\end{align} 
%%%%%%%%%%%%%%%%%%%%%%%%%%%%%%%%%%%%%%%%%%%%%%%%%%%%%%%%%%%%%%%%
where, $\rho_{0}$ is some energy density scale, $(c_{1},c_{2},c_{3})$ are \textit{real} constants, while $(n_{1},n_{2},n_{3})$ are \textit{positive} numbers, not necessarily integers. Recall that the scale factor is related to $q(t)$ via $a(t)=q(t)^{b/2}$. Hence, the effective energy density $U_{\rm eff}(q)$ that appears in the action presented in \ref{Full_Action_Final}, takes the following form,
%%%%%%%%%%%%%%%%%%%%%%%%%%%%%%%%%%%%%%%%%%%%%%%%%%%%%%%%%%%%%%%%
\begin{align}\label{effective_U_01}
U_{\rm eff}(q)=-q^{3b-2}\rho_{\rm eff}(q^{b/2})
=-\rho_{0}\left[c_{1}q^{3b-2-\frac{b}{2}n_{1}}+c_{2}q^{3b-2-\frac{b}{2}n_{2}}+c_{3}q^{3b-2-\frac{b}{2}n_{3}}\right]~.
\end{align} 
%%%%%%%%%%%%%%%%%%%%%%%%%%%%%%%%%%%%%%%%%%%%%%%%%%%%%%%%%%%%%%%%
In order to facilitate explicit evaluation of the path integral over $q$, we shall now choose the powers $n_{1},n_{2}$ and $n_{3}$ judiciously. Let us start by setting the power of $q$ in the term involving $c_{1}$, in the last line of \ref{effective_U_01}, to be such that $3b-2-(b/2)n_{1}=0$. This yields,
%%%%%%%%%%%%%%%%%%%%%%%%%%%%%%%%%%%%%%%%%%%%%%%%%%%%%%%%%%%%%%%%
\begin{align}\label{choice_b}
b=\frac{4}{6-n_{1}}~;\quad n_{1}=\frac{2(3b-2)}{b}~.
\end{align} 
%%%%%%%%%%%%%%%%%%%%%%%%%%%%%%%%%%%%%%%%%%%%%%%%%%%%%%%%%%%%%%%%
Substituting the expression for $b$, presented above, in \ref{effective_U_01}, the effective potential due to the matter field becomes
%%%%%%%%%%%%%%%%%%%%%%%%%%%%%%%%%%%%%%%%%%%%%%%%%%%%%%%%%%%%%%%%
\begin{align}\label{effective_U_02}
U_{\rm eff}=-\rho_{0}\left[c_{1}+c_{2}q^{\frac{2(n_{1}-n_{2})}{6-n_{1}}}+c_{3}q^{\frac{2(n_{1}-n_{3})}{6-n_{1}}}\right]~.
\end{align} 
%%%%%%%%%%%%%%%%%%%%%%%%%%%%%%%%%%%%%%%%%%%%%%%%%%%%%%%%%%%%%%%%
Setting the power of $q$ in the term involving $c_{2}$ to identity and the power of $q$ in the term involving $c_{3}$ to be quadratic, we obtain the powers $n_{2}$ and $n_{3}$ as a function of $n_{1}$ as,
%%%%%%%%%%%%%%%%%%%%%%%%%%%%%%%%%%%%%%%%%%%%%%%%%%%%%%%%%%%%%%%%
\begin{align}\label{effective_U_final}
U_{\rm eff}=-\rho_{0}\left[c_{1}+c_{2}q+c_{3}q^{2}\right]~; \quad n_{2}=\frac{3}{2}\left(n_{1}-2\right)~;\quad n_{3}=2n_{1}-6~.
\end{align} 
%%%%%%%%%%%%%%%%%%%%%%%%%%%%%%%%%%%%%%%%%%%%%%%%%%%%%%%%%%%%%%%%
On the other hand, we could have also taken $n_{2}$ to be fundamental and have expressed the powers $n_{1}$ and $n_{3}$ in terms of $n_{2}$, which yield, $b=6/(6-n_{2})$ and $n_{1}=(2/3)(n_{2}+3)$ as well as $n_{3}=(2/3)(2n_{2}-3)$. Similarly, choosing $n_{3}$ to be fundamental, one can express $n_{1}$ and $n_{2}$ in terms of it: $b=8/(6-n_{3})$, $n_{1}=(1/2)(n_{3}+6)$ and $n_{2}=(3/2)\{1+(n_{3}/2)\}$. A list of such convenient choices of the parameter $b$ and the corresponding constraints among $(n_1,n_2,n_3)$ is given in \ref{tab_1}. We conclude that, for the choice of parameters, as given in \ref{tab_1}, the action becomes quadratic in the variable $q(t)$ and thus path integral can be explicitly computed.  
\begin{table}
	\centering
\begin{tabular}{ |c|c|c| }
	\hline
$b$& constraint 1& constraint 2\\
	\hline
	$\dfrac{4}{6-n_{1}}$ & $n_{2}=\frac{3}{2}\left(n_{1}-2\right)$    &$n_{3}=2n_{1}-6$\\ \hline
	$\dfrac{6}{6-n_{2}}$&   $n_{1}=\frac{2}{3}(n_{2}+3)$ & $n_{3}=\frac{2}{3}(2n_{2}-3)$   \\ \hline
	$\dfrac{8}{6-n_{3}}$&$n_{1}=\frac{1}{2}(n_{3}+6)$ & $n_{2}=\frac{3}{2}\left(1+\frac{n_{3}}{2}\right)$\\
	\hline
\end{tabular}
\caption{The table shows the choices of the parameter $b$ and the corresponding constraints among the parameters $(n_1,n_2,n_3)$, which may be implemented to reduce the effective potential to the simple quadratic form $U_{\rm eff}(q)=-\rho_{0}(c_2+c_2 q+c_3 q^2)$.}
\label{tab_1}
\end{table}

So far, our analysis has been fairly general. In what follows, we will be focussing on scenarios that give rise to `bounce', by choosing the energy density $\rho_{\rm eff}$ appropriately. A generic bouncing scenario may be thought of as being sourced by two species of matter fields; one of them satisfying the energy condition, while the other violating the same. Moreover, the latter is the factor that enables the bounce. A possible form of $\rho_{\rm eff}(a)$, that can realise a classical bouncing scenario, consists of components of both positive as well as negative energy densities which, in a generic situation, takes the following form:
%%%%%%%%%%%%%%%%%%%%%%%%%%%%%%%%%%%%%%%%%%%%%%%%%%%%%%%%%%%%%%%%
\begin{align}\label{rho_eff}
\rho_{\rm eff}(a)=\sum_{n}\frac{\rho^{+}_n}{a^n}-\sum_{m}\frac{\rho^{-}_{m}}{a^m}~,
\end{align}  
%%%%%%%%%%%%%%%%%%%%%%%%%%%%%%%%%%%%%%%%%%%%%%%%%%%%%%%%%%%%%%%%
where $\rho^{+}_{n}$ and $\rho^{-}_{m}$ as positive real numbers. Note that the negative energy density terms could either be arising from an exotic matter field (as, for instance, in \cite{Raveendran:2017vfx,Raveendran:2018why,Cai:2007qw}) or emerging from the corrections to the Einstein-Hilbert action introduced by a UV-complete quantum gravity theory (as, for instance, in \cite{Brandenberger:2009yt}). We would like to emphasize that we shall not concern ourself with the origin of the effective density $\rho_{\rm eff}(a)$ here. We take the point of view that an appropriate $\rho_{\rm eff}(a)$ effectively captures most of the essential aspects relevant to the bouncing cosmology and, hence, we focus only on its repercussions to quantum cosmology. 

\begin{figure}[t]
	\centering
	\includegraphics[scale=.25]{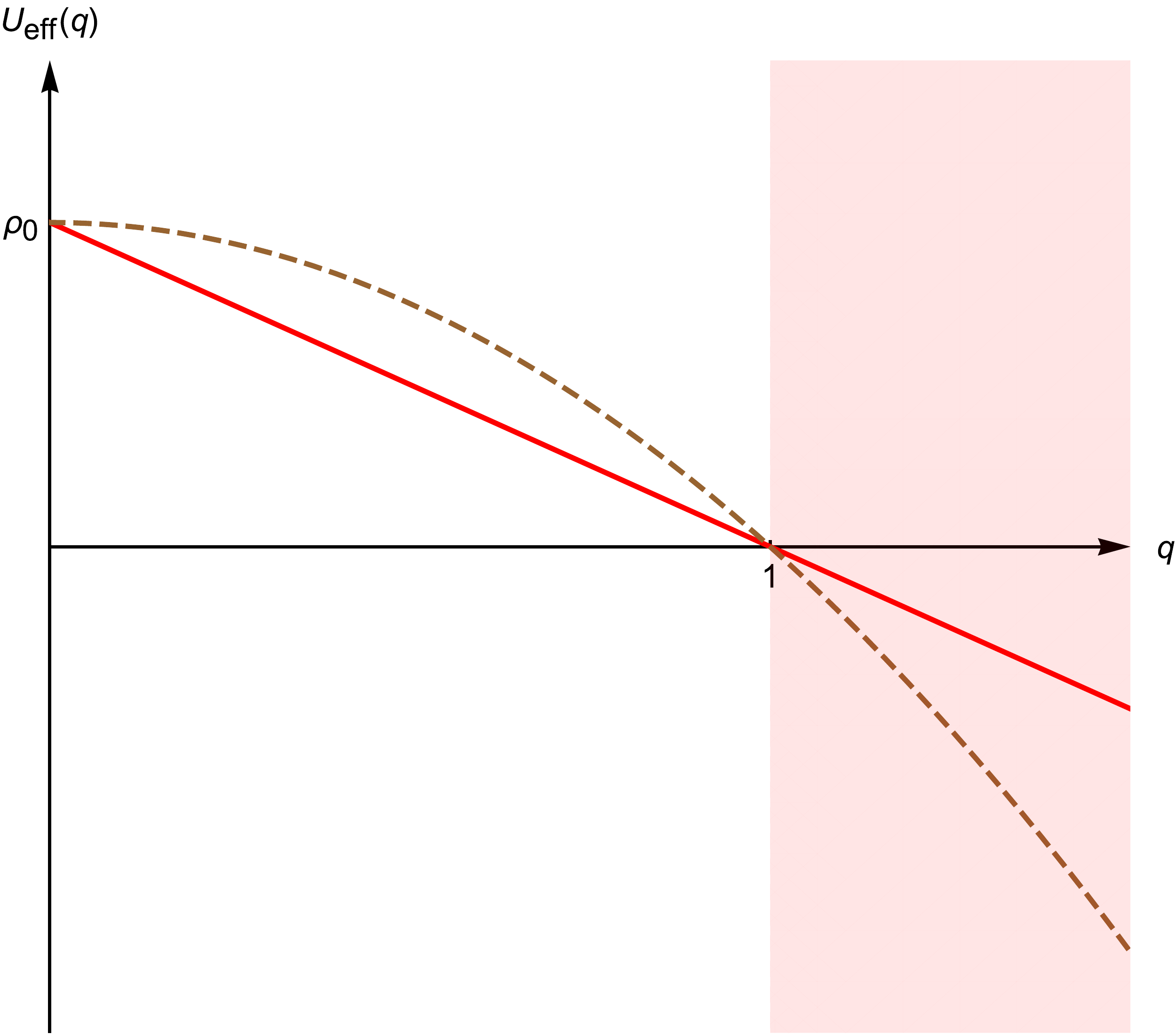}
	\caption{The red graph shows the first kind of effective potential $U^{\rm (I)}_{\rm eff}(q)$ and the brown dashed graph shows the second kind of effective potential $U_{\rm eff}^{\rm (II)}(q)$. The shaded portion, with $q>1$, denotes the classically allowed region, describing a bouncing universe.}\label{Ueff_plot}
\end{figure}

A reasonable scenario corresponds to the case where the dominant contribution to the effective energy density near the bounce is from only two components, namely, (1) a certain type of `normal matter', with density scaling as $a^{-n_+}$ and (2) a certain type of `phantom matter' with density scaling as $a^{-n_-}$. Motivated by this, we shall henceforth consider effective densities of the form,
%%%%%%%%%%%%%%%%%%%%%%%%%%%%%%%%%%%%%%%%%%%%%%%%%%%%%%%%%%%%%%%%
\begin{align}
\rho_{\rm eff}(a)= \frac{\rho^{+}_{n_+}}{a^{n_+}}-\frac{\rho^{-}_{n_-}}{a^{n_-}}~,
\end{align}  
%%%%%%%%%%%%%%%%%%%%%%%%%%%%%%%%%%%%%%%%%%%%%%%%%%%%%%%%%%%%%%%%
with $n_{-}>n_{+}$. As we have already pointed out, for convenient evaluation of path integrals, it is desirable to choose the values of $n_{+}$ and $n_{-}$ such that the action of the gravity-fluid system $S_{\rm perfect}[\mathcal{N},q]$ is a quadratic functional of $q$. In light of our analysis leading to \ref{effective_U_final} and \ref{tab_1}, the powers $n_{+}$ and $n_{-}$ may be chosen such that, we have the following two classes of effective energy densities:
%%%%%%%%%%%%%%%%%%%%%%%%%%%%%%%%%%%%%%%%%%%%%%%%%%%%%%%%%%%%%%%%
\begin{align}
\rho^{\rm (I)}_{\rm eff}(a)&=\rho_0\left(\frac{1}{a^n}-\frac{1}{a^{\frac{(6+2n)}{3}}}\right)~;
\quad b^{\rm (I)}=\frac{6}{6-n}~,
\label{rho_I}
\\
\rho^{\rm (II)}_{\rm eff}(a)&=\rho_0\left(\frac{1}{a^n}-\frac{1}{a^{\frac{(6+n)}{2}}}\right)~;
\quad b^{\rm (II)}=\frac{8}{6-n}=\frac{4}{3}b^{\rm (I)}~,
\label{rho_II}
\end{align} 
%%%%%%%%%%%%%%%%%%%%%%%%%%%%%%%%%%%%%%%%%%%%%%%%%%%%%%%%%%%%%%%%
where, $n$ and $\rho_0$ are real constants and, we have scaled the scale factor such that the bounce happens at $a=1$. The necessary condition for both the above effective energy densities to describe a bounce is $n<6$. It is worth mentioning that these two classes of energy densities together cover a wide range of physically relevant bouncing scenarios. For example, substituting $n=3$ in the expression for $\rho^{\rm (I)}_{\rm eff}(a)$, we obtain an effective energy density that describes a `matter bounce' scenario (denoted by subscript `mb') that may be realized, for instance, in the context of Horava-Lifshitz cosmology \cite{Brandenberger:2009yt} or by invoking a non-canonical, ghost field \cite{Raveendran:2017vfx}:
%%%%%%%%%%%%%%%%%%%%%%%%%%%%%%%%%%%%%%%%%%%%%%%%%%%%%%%%%%%%%%%%
\begin{align}\label{rho_mb}
\rho^{\rm (mb)}_{\rm eff}(a)&=\rho_0\left(\frac{1}{a^3}-\frac{1}{a^{4}}\right)~.
\end{align}  
%%%%%%%%%%%%%%%%%%%%%%%%%%%%%%%%%%%%%%%%%%%%%%%%%%%%%%%%%%%%%%%%
Further, it is easy to see that the substitution of $n=0$ in the first class of energy density, we obtain $\rho_{\rm eff}(a)=\rho_{0}(1-a^{2})$, that is relevant for studying de Sitter spacetime in closed slicing\cite{Hartle:1983ai,WADA1986729,Vilenkin:1987kf} . Referring to \ref{tab_1} and \ref{energy_density}, we see that two cases of energy densities $\rho^{\rm (I)}_{\rm eff}(a)$ and $\rho^{\rm (II)}_{\rm eff}(a)$ correspond to $c_{2}=1=-c_{1}$ with $c_{3}=0$ and $c_{3}=1=-c_{1}$ with $c_{2}=0$, respectively. Consequently, the corresponding effective potentials arising out of $\rho^{\rm (I)}_{\rm eff}(a)$ and $\rho^{\rm (II)}_{\rm eff}(a)$, respectively, reduce to the following simple forms:
%%%%%%%%%%%%%%%%%%%%%%%%%%%%%%%%%%%%%%%%%%%%%%%%%%%%%%%%%%%%%%%%
\begin{align}\label{U_eff_I}
U^{\rm (I)}_{\rm eff}(q)&=\rho_{0}\left(1-q\right)~; 
\\
\label{U_eff_II}
U^{\rm (II)}_{\rm eff}(q)&=\rho_{0}\left(1-q^{2}\right)~,
\end{align} 
%%%%%%%%%%%%%%%%%%%%%%%%%%%%%%%%%%%%%%%%%%%%%%%%%%%%%%%%%%%%%%%%
Hence, we shall henceforth refer to the bouncing models described by the effective densities $\rho^{\rm (I)}_{\rm eff}(a)$ and $\rho^{\rm (II)}_{\rm eff}(a)$ as `the linear models' and `the quadratic models', respectively.  
Keeping our later purposes in mind, it will be useful to introduce another constant $h_{n}$, which in terms of $\rho_0$ takes the following form,
%%%%%%%%%%%%%%%%%%%%%%%%%%%%%%%%%%%%%%%%%%%%%%%%%%%%%%%%%%%%%%%%
\begin{align}\label{def_h}
h_{n}^{2}&\equiv\frac{1}{108} \kappa  (n-6)^2 \rho_0
\end{align}
%%%%%%%%%%%%%%%%%%%%%%%%%%%%%%%%%%%%%%%%%%%%%%%%%%%%%%%%%%%%%%%%
The constant $h_{n}$ may be viewed as the Hubble parameter associated with the constant energy density $\rho_{0}$ and will play a significant role in the subsequent discussion. To summarize, we have essentially brought down the analysis of a wide class of bouncing cosmologies to that of a particle in one dimensional quadratic or linear potentials! For completeness, the effective potentials $U^{\rm (I)}_{\rm eff}(q)$ and $U^{\rm (II)}_{\rm eff}(q)$ have been plotted in \ref{Ueff_plot}. The point corresponding to $q=1$ stands for the bounce and the region of parameter space with $q>1$ depicts the classically allowed region.  
%%%%%%%%%%%%%%%%%%%%%%%%%%%%%%%%%%%%%%%%%%%%%%
%%%%%%%%%%%%%%%%%%%%%%%%%%%%%%%%%%%%%%%%%%%%%%
%%%%%%%%%%%%%%%%%%%%%%%%%%%%%%%%%%%%%%%%%%%%%%
%%%%%%%%%%%%%%%%%%%%%%%%%%%%%%%%%%%%%%%%%%%%%%
%%%%%%%%%%%%%%%%%%%%%%%%%%%%%%%%%%%%%%%%%%%%%%
%%%%%%%%%%%%%%%%%%%%%%%%%%%%%%%%%%%%%%%%%%%%%%
%%%%%%%%%%%%%%%%%%%%%%%%%%%%%%%%%%%%%%%%%%%%%%
%%%%%%%%%%%%%%%%%%%%%%%%%%%%%%%%%%%%%%%%%%%%%%

We shall now look at the classical spacetimes sourced by the energy densities $\rho^{\rm (I)}_{\rm eff}(a)$ and $\rho^{\rm (II)}_{\rm eff}(a)$. The classical background spacetime can be found by solving the corresponding Friedman equations. If we choose our time coordinate such that the bounce happens at $t=0$, then the background scale factors corresponding to the linear and the quadratic models take the following forms, respectively:
%%%%%%%%%%%%%%%%%%%%%%%%%%%%%%%%%%%%%%%%%%%%%%%%%%%%%%%%%%%%%%%%
\begin{align}\label{a_I}
a_{\rm (I)}(t)&=\left(1+h_{n}^{2}t^{2}\right)^{\frac{3}{6-n}}~,
\\
\label{a_II}
a_{\rm (II)}(t)&=\left[\cosh\left(\frac{3}{2}h_{n}t\right)\right]^{\frac{4}{6-n}}~.
\end{align}
%%%%%%%%%%%%%%%%%%%%%%%%%%%%%%%%%%%%%%%%%%%%%%%%%%%%%%%%%%%%%%%%
We caution the reader, once again, that the time coordinate $t$ is \textit{not} the cosmic time. Consequently, the apparent simple forms of the scale factors in \ref{a_I} and \ref{a_II} is rather deceptive, that is to say, the scale factors $a_{\rm (I)}(t)$ and $a_{\rm (II)}(t)$ could, in general, be complicated functions of the cosmic time $\tau$.  As an illustration, in the case of the `matter-bounce' scenario described by the effective density in \ref{rho_mb}, it turns out that the time coordinate $t$ coincides with the conformal time $\eta$ and hence the background scale factor is given by:
%%%%%%%%%%%%%%%%%%%%%%%%%%%%%%%%%%%%%%%%%%%%%%%%%%%%%%%%%%%%%%%%
\begin{align}
a_{\rm (mb)}(t)=\left(1+h^{2}_{3}t^{2}\right)=\left(1+h^{2}_{3}\eta^{2}\right)~.
\end{align} 
%%%%%%%%%%%%%%%%%%%%%%%%%%%%%%%%%%%%%%%%%%%%%%%%%%%%%%%%%%%%%%%%
However, the same scale factor, rewritten as a function of the cosmic time $\tau$, acquires a more complicated form:
%%%%%%%%%%%%%%%%%%%%%%%%%%%%%%%%%%%%%%%%%%%%%%%%%%%%%%%%%%%%%%%%
\begin{align}
a_{\rm (mb)}(\tau)=\left(\sqrt{\frac{9h_{3}^{2}\tau^{2}}{4}+1}-\frac{3h_{3}\tau}{2}\right)^{2/3}+\left(\sqrt{\frac{9h_{3}^{2}\tau^{2}}{4}+1}+\frac{3h_{3}\tau}{2}\right)^{2/3}-1~.
\end{align}
%%%%%%%%%%%%%%%%%%%%%%%%%%%%%%%%%%%%%%%%%%%%%%%%%%%%%%%%%%%%%%%%
Generically, the functional dependence of the scale factor on the cosmic time and the conformal time is significantly complicated.

In summary, we proposed two classes of bounce-enabling effective energy densities in this section, namely, $\rho^{\rm (I)}_{\rm eff}(a)$ and $\rho^{\rm (II)}_{\rm eff}(a)$. These energy densities have the desirable characteristic that one can reduce the dynamics of the background spacetime, in the presence of these densities, to that of a particle in one-dimensional quadratic/linear potential. In what follows, we will provide a quantum treatment of the bouncing universe, by first solving the Wheeler-DeWitt equation and then using the path-integral technique. As we shall shortly see, the specific forms of the densities $\rho^{\rm (I)}_{\rm eff}(a)$ and $\rho^{\rm (II)}_{\rm eff}(a)$ facilitate easy evaluation of the path integrals leading to the ground state wave function of the universe. Hence, we hope that these classes of bouncing universes furnish test-bed to study deeper aspects of bouncing cosmologies. 

%%%%%%%%%%%%%%%%%%%%%%%%%%%%%%%%%%%%%%%%%%%%%%%%%%%%%%%%%%%%%%%%%%%%%%%%%%%%%%%%%%%%%%%%%%%%%%%%%%%
%%%%%%%%%%%%%%%%%%%%%%%%%%%%%%%%%%%%%%%%%%%%%%%%%%%%%%%%%%%%%%%%%%%%%%%%%%%%%%%%%%%%%%%%%%%%%%%%%%%
%%%%%%%%%%%%%%%%%%%%%%%%%%%%%%%%%%%%%%%%%%%%%%%%%%%%%%%%%%%%%%%%%%%%%%%%%%%%%%%%%%%%%%%%%%%%%%%%%%%
\section{The groundstate wave function of the bouncing universe}\label{wavefunction_bounce}

In this section, we study the quantum aspects of the `linear' and the `quadratic' models of bounce introduced in \ref{gen_setup_matter}. To this end, we attempt to evaluate the relevant `groundstate wave function of the universe' and its properties. We shall take two different approaches for this purpose --- (a) using the Wheeler-DeWitt equation and (2) the path integral approach. We shall also make comparisons of the results of these two approaches at appropriate junctures.

%%%%%%%%%%%%%%%%%%%%%%%%%%%%%%%%%%%%%%%%%%%%%%%%%%
%%%%%%%%%%%%%%%%%%%%%%%%%%%%%%%%%%%%%%%%%%%%%%%%%%
%%%%%%%%%%%%%%%%%%%%%%%%%%%%%%%%%%%%%%%%%%%%%%%%%%
%%%%%%%%%%%%%%%%%%%%%%%%%%%%%%%%%%%%%%%%%%%%%%%%%%
\subsection{The Wheeler-de Witt equation and its solution}\label{wavefunction_bounce_WdW}

We have already seen that the gravity sector can be conveniently described by the dynamical variable $q(t)$. The corresponding Wheeler-de Witt equation for the gravity sector reduces to the following one-dimensional time-independent Schr\"{o}dinger equations for, respectively, the linear and the quadratic models:
%%%%%%%%%%%%%%%%%%%%%%%%%%%%%%%%%%%%%%%%%%%%%%%%%%%%%%%%%%%%%%%%
\begin{align}
\left[-\frac{\partial^{2}}{\partial q^{2}}+\alpha_{n}^{2}\left(1-q\right)\right]\Psi^{\rm (I)}(q)&=0~,
\label{WdW_I}
\\
\left[-\frac{\partial^{2}}{\partial q^{2}}+\frac{16}{9}\alpha_{n}^{2}\left(1-q^{2}\right)\right]\Psi^{\rm (II)}(q)&=0~,
\label{WdW_II}
\end{align} 
%%%%%%%%%%%%%%%%%%%%%%%%%%%%%%%%%%%%%%%%%%%%%%%%%%%%%%%%%%%%%%%%
where, the factor $(16/9)$ comes from the relation between $b^{\rm (I)}$ and $b^{\rm (II)}$, as presented in \ref{rho_II}. The constant $\alpha_{n}$, on the other hand, can be expressed as $\beta_{n}V_{3}$ where
%%%%%%%%%%%%%%%%%%%%%%%%%%%%%%%%%%%%%%%%%%%%%%%%%%%%%%%%%%%%%%%%
\begin{align}\label{def_beta_n}
\beta_{n}^{2}\equiv 2M^{\rm (I)}\rho_{0}=\frac{9h_{n}^{2}}{\kappa^{2}\left(\frac{n}{6}-1\right)^{4}}
\equiv \frac{h_{n}^{2}}{\ell_{\rm p}^{4}\left(\frac{n}{6}-1\right)^{4}}~.
\end{align}
%%%%%%%%%%%%%%%%%%%%%%%%%%%%%%%%%%%%%%%%%%%%%%%%%%%%%%%%%%%%%%%%
Here, we have introduced a new length scale $\ell_{\rm p}=\sqrt{\kappa/3}$, which corresponds to the Planck length associated with the quantum nature of the gravitational interaction. 

Let us first consider the WdW equation associated with the linear model. The general solution to \ref{WdW_I} can be written down in terms of the Airy functions involving two arbitrary constants. Except for an overall normalization constant, the other constant can be fixed by imposing appropriate boundary conditions. Following \cite{PhysRev.160.1113}, or recently \cite{Damour:2019iyi}, one can impose the boundary condition $\Psi(a=0)=0$, which corresponds to the choice that the wave function must vanish at the singular point $a=0$. Under this boundary condition, the WdW wave function is given by:
%%%%%%%%%%%%%%%%%%%%%%%%%%%%%%%%%%%%%%%%%%%%%%%%%%%%%%%%%%%%%%%%
\begin{align}
\Psi^{\rm (I)}_{0}(q)=\mathcal{N}^{\rm (I)}_{0}\Bigg\{\textrm{Bi}\left(\alpha_{n}^{2/3}\right)\textrm{Ai}\left[\alpha_{n}^{2/3}\left(1-q\right)\right]-\textrm{Ai}\left(\alpha_{n}^{2/3}\right)\textrm{Bi}\left[\alpha_{n}^{2/3}(1-q)\right]\Bigg\}~,
\end{align} 
%%%%%%%%%%%%%%%%%%%%%%%%%%%%%%%%%%%%%%%%%%%%%%%%%%%%%%%%%%%%%%%%
where, $\mathcal{N}^{\rm (I)}_{0}$ is a normalisation constant. For $q>1$, i.e., in the classical regime, one is interested in the limit $(\alpha_{n}/\ell_{\rm p}^{2})\gg 1$. In which case, we can use the following expansions of the Airy functions,
%%%%%%%%%%%%%%%%%%%%%%%%%%%%%%%%%%%%%%%%%%%%%%%%%%%%%%%%%%%%%%%%
\begin{align}\label{expansion_airy}
\textrm{Ai}(x)&\sim\frac{\sqrt{\pi}}{\left(-x\right)^{1/4}}\cos \left[-\frac{2}{3}\left(-x\right)^{3/2}+\frac{\pi}{4} \right]~; \quad (x<0,|x|\gg 1)\\\nonumber
\textrm{Ai}(x)&\sim\frac{1}{\sqrt{\pi}x^{1/4}}\exp\left(-\frac{2}{3}x^{3/2}\right)~;\qquad (x\gg 1)
\end{align} 
%%%%%%%%%%%%%%%%%%%%%%%%%%%%%%%%%%%%%%%%%%%%%%%%%%%%%%%%%%%%%%%%
along with the fact that the expansion of $\textrm{Bi}(x)$ for negative, but large $x$ is identical to that of $\textrm{Ai}(x)$, while for large positive $x$, $\textrm{Bi}(x)$ is exponentially growing, the exact opposite of the behaviour of $\textrm{Ai}(x)$, presented in \ref{expansion_airy}. Thus for $q>1$ and for large $(\alpha_{n}/\ell_{\rm p}^{2})$ limit, the asymptotic expansion of the wave function $\Psi^{\rm (I)}(q)$ takes the form:
%%%%%%%%%%%%%%%%%%%%%%%%%%%%%%%%%%%%%%%%%%%%%%%%%%%%%%%%%%%%%%%%
\begin{align}\label{Psi_0_I}
\Psi^{\rm (I)}_{0}(q)\approx\frac{\mathcal{N}^{\rm (I)}_{0}}{\sqrt[4]{\alpha_{n}^{2/3}\left(q-1\right)}} \exp\left(\frac{2}{3}\alpha_{n}\right)\cos \left[\frac{2}{3}\alpha_{n}\left(q-1\right)^{3/2}-\frac{\pi}{4}\right]+\mathcal{O}\left\{\exp\left(-\frac{2}{3}\alpha_{n}\right)\right\}~.
\end{align}
%%%%%%%%%%%%%%%%%%%%%%%%%%%%%%%%%%%%%%%%%%%%%%%%%%%%%%%%%%%%%%%%
Notice that there is a noticeable similarity between the above solution of the Wheeler-DeWitt equation and that of the ground-state wave function of Hartle and Hawking (HH), associated with the `no-boundary' proposal \cite{Hawking:1969sw}, except for an exponentially suppressed term $\sim \exp\left(-\frac{2}{3}\alpha_{n}\right)$, which can be neglected in the semi-classical limit.

It is also possible to impose the Hartle-Hawking boundary condition on the solution of the WdW equation. This will result into a particular solution $\Psi^{\rm (I)}$ that may be regarded as the natural analogue of the Hartle-Hawking wave function for bouncing scenario. To this end, we demand that, in the $\alpha\gg 1$ limit, $\Psi^{\rm (I)}$ is an exponentially growing function in the range $q<1$ (see, for instance, \cite{Vilenkin:1987kf}). The corresponding solution is given by:
%%%%%%%%%%%%%%%%%%%%%%%%%%%%%%%%%%%%%%%%%%%%%%%%%%%%%%%%%%%%%%%%
\begin{align}
\Psi^{\rm (I)}(q)=\frac{{\rm N}^{\rm (I)}}{\textrm{Ai}\left(\alpha_{n}^{2/3}\right)}\textrm{Ai}\left[\alpha_{n}^{2/3}\left(1-q\right)\right]~,
\end{align}
%%%%%%%%%%%%%%%%%%%%%%%%%%%%%%%%%%%%%%%%%%%%%%%%%%%%%%%%%%%%%%%%
where ${\rm N}^{\rm (I)}$ is a normalisation constant. Given the above asymptotic expansion of the Airy function, it is evident that the above particular solution to the WdW equation allows both ingoing and outgoing modes in the classical regime, as fit for the Hartle-Hawking wave function. Thus, in the region $q>1$, the asymptotic expansion of the wave function $\Psi^{\rm (I)}(q)$ for $\alpha_{n}\gg 1$ limit takes the form:
%%%%%%%%%%%%%%%%%%%%%%%%%%%%%%%%%%%%%%%%%%%%%%%%%%%%%%%%%%%%%%%%
\begin{align}\label{Psi_I}
\Psi^{\rm (I)}(q)\approx\frac{{\rm N}^{\rm (I)}}{\sqrt[4]{\alpha_{n}^{2/3}\left(q-1\right)}} \exp\left(\frac{2}{3}\alpha_{n}\right)\cos \left[\frac{2}{3}\alpha_{n}\left(q-1\right)^{3/2}-\frac{\pi}{4}\right]~.
\end{align}
%%%%%%%%%%%%%%%%%%%%%%%%%%%%%%%%%%%%%%%%%%%%%%%%%%%%%%%%%%%%%%%%
While, in the classically forbidden region $(q<1)$ as well, the wave function is exponentially suppressed. Thus this Hartle-Hawking wave function differs from $\Psi^{\rm (I)}_{0}(q)$ by an exponentially small correction. In particular, in the context of de Sitter cosmology, the value of the no-boundary wave function at $a=0$ is an exponentially small, nevertheless, non-zero quantity. Therefore, the condition that the wave function \textit{exactly} vanishes at $a=0$ may be regarded as a convenient approximation for the exact no-boundary wave function, with the errors being negligible in the semi-classical limit. In fact, for the numerical evaluation of the `ground-state wave function', Hartle-Hawking also assumes the condition $\Psi(a\rightarrow0)=0$ in their original work \cite{Hartle:1983ai}. In a similar manner, in the present context, we may regard $\Psi^{\rm (I)}_{0}(q)$ as a convenient approximation of $\Psi^{\rm (I)}(q)$. The precise connection of the above wave function $\Psi^{\rm (I)}(q)$ with the no-boundary proposal will be made clear when we analyze the above problem using the path integral formalism.

We shall now look at the WdW equation for the quadratic model. The general solution to \ref{WdW_II} can be written down in terms of the parabolic cylinder functions $D_{\nu}(x)$. However, for our purpose it is convenient to work with another set of linearly independent solutions \cite[p.314-p.317]{olver2010nist}, namely, $W(a,-x)$ and $W(a,-x)$. These functions are constructed from linear combinations of two linearly independent sets of standard parabolic cylinder functions. Again, the general solution of the WdW equation will involve two arbitrary constants and can be written as a linear combination of $W(a,-x)$ and $W(a,-x)$. In order to derive particular solutions to the WdW equations, as in the case of linear potential, we may impose the condition $\Psi(a=0)=0$. However, the resulting particular solution will be different from the Hartle-Hawking prescription by exponentially small quantities. Thus we directly look for the particular solution that may be regarded as the analogue of the no-boundary wave function, which is given by:
%%%%%%%%%%%%%%%%%%%%%%%%%%%%%%%%%%%%%%%%%%%%%%%%%%%%%%%%%%%%%%%%
\begin{align}
\Psi^{\rm (II)}(q)={\rm N}^{\rm (II)}~W\left(\frac{2}{3}\alpha_{n},-\frac{2\sqrt{2}}{\sqrt{3}}\sqrt{\alpha_{n}}q\right)~,
\end{align}
%%%%%%%%%%%%%%%%%%%%%%%%%%%%%%%%%%%%%%%%%%%%%%%%%%%%%%%%%%%%%%%%
where, ${\rm N}^{\rm (II)}$ is a normalisation constant. For $q>1$, i.e., in the classically allowed region, the asymptotic expansion of the above solution in the $\alpha_{n}\gg 1$ limit is given by \cite{olver2010nist},
%%%%%%%%%%%%%%%%%%%%%%%%%%%%%%%%%%%%%%%%%%%%%%%%%%%%%%%%%%%%%%%%
\begin{align}\label{psi_II}
\Psi^{\rm (II)}(q)\approx\frac{{\rm N}^{\rm (II)}}{\sqrt[4]{\frac{4}{3}\alpha_{n}\left(q^{2}-1\right)}} \exp\left(\frac{\pi}{3}\alpha_n\right)
\cos \left[\frac{4}{3}\alpha_{n}\xi(q)-\frac{\pi}{4}\right]~,
\end{align}
%%%%%%%%%%%%%%%%%%%%%%%%%%%%%%%%%%%%%%%%%%%%%%%%%%%%%%%%%%%%%%%%
where,
%%%%%%%%%%%%%%%%%%%%%%%%%%%%%%%%%%%%%%%%%%%%%%%%%%%%%%%%%%%%%%%%
\begin{align}\label{def_xi_q}
\xi(q)\equiv\frac{1}{2}q\sqrt{q^2-1}-\frac{1}{2}\log (q+\sqrt{q^2-1})~.
\end{align}
%%%%%%%%%%%%%%%%%%%%%%%%%%%%%%%%%%%%%%%%%%%%%%%%%%%%%%%%%%%%%%%%
As evident this particular solution involves both ingoing and outgoing modes in the classically allowed region, as fit for a Hartle-Hawking wave function.

It is worth emphasizing that even for a general quadratic potential of the form $U_{\rm eff}(q)=-\rho_{0}(c_1+c_2 q+c_3q^2)$, with the coefficients $\{c_1,c_2,c_3\}$ such that classically a bouncing scenario is feasible, the solution to the Wheeler-DeWitt equation would have a structure identical to that presented in \ref{psi_II}. Thus, the results obtained above is general enough to encompass aspects of all classes of bouncing models whose effective potentials can be reduced to a polynomial in $q$, of degree at most 2. Once again, we see that the asymptotic expression for the wave function has a close resemblance to that of the Hartle-Hawking wave function, compatible with the no-boundary proposal. As we shall shortly see in the next subsection, our path integral analysis will reveal that this resemblance is not accidental.

%%%%%%%%%%%%%%%%%%%%%%%%%%%%%%%%%%%%%%%%%%%%%%%%%%
%%%%%%%%%%%%%%%%%%%%%%%%%%%%%%%%%%%%%%%%%%%%%%%%%%
%%%%%%%%%%%%%%%%%%%%%%%%%%%%%%%%%%%%%%%%%%%%%%%%%%
%%%%%%%%%%%%%%%%%%%%%%%%%%%%%%%%%%%%%%%%%%%%%%%%%%
\subsection{Path integral approach}\label{wavefunction_bounce_PI}

Having described the `wave function of the universe' for the linear and quadratic models using the Wheeler-de Witt equation, let us now try to study the quantum aspects of these models using the path integral formalism. Prior to that, let us briefly review some fundamentals of the path integral approach to the minisuperspace model. 

Recall that the action $\mathcal{S}_{\rm perfect}[q,\mathcal{N}]$ describing the gravity sector is given by \ref{action_eff}. Following \cite{Halliwell1988}, it is convenient to use a gauge in which $\dot{\mathcal{N}}=0$ and hence the `physical time' $T$, defined through the following relation, $T=\int \mathcal{N}dt$, becomes $T=\mathcal{N}t$. For simplicity, but without losing generality, we will restrict ourselves within the time interval $0\leq t\leq 1$, which translates into $0\leq T \leq \mathcal{N}$. We can then construct a path integral kernel from the gravitational action $\mathcal{S}_{\rm perfect}$, with the boundary condition that $q(T=\mathcal{N})=q_{1}$ and $q(T=0)=q_{0}$, as follows:
%%%%%%%%%%%%%%%%%%%%%%%%%%%%%%%%%%%%%%%%%%%%%%%%%%%%%%%%%%%%%%%%
\begin{align}\label{kernal_def}
K(q_{1},\mathcal{N};q_{0},0)\equiv \int_{q(T=0)=q_{0}}^{q(T=\mathcal{N})=q_{1}} \mathcal{D}[q]~\exp\left\{iV_{3}\int_{0}^{\mathcal{N}} dT'\left[-\frac{M}{2}\left(\frac{dq}{dT'}\right)^{2}+U_{\rm eff}(q)\right]\right\}~,
\end{align}
%%%%%%%%%%%%%%%%%%%%%%%%%%%%%%%%%%%%%%%%%%%%%%%%%%%%%%%%%%%%%%%%
One can check that the Hamiltonian constraint indeed generates translation in the `physical time' coordinate $T$, such that the Kernel $K(q_{1},\mathcal{N};q_{0},0)$ satisfies the following differential equation: $\mathcal{H}_{\rm perfect}K(q_{1},\mathcal{N};q_{0},0)=i\partial_{\mathcal{N}}K(q_{1},\mathcal{N};q_{0},0)$. Here, $\mathcal{H}_{\rm perfect}$ is the Hamiltonian associated with the bouncing model with perfect fluid source, given by,
%%%%%%%%%%%%%%%%%%%%%%%%%%%%%%%%%%%%%%%%%%%%%%%%%%%%%%%%%%%%%%%%
\begin{align}\label{H_perfect}
\mathcal{H}_{\rm perfect}=\frac{1}{2MV_3}\partial^2_{q_1}-V_3U_{\rm eff}(q_1)
\end{align}
%%%%%%%%%%%%%%%%%%%%%%%%%%%%%%%%%%%%%%%%%%%%%%%%%%%%%%%%%%%%%%%%
Let us now consider a wave function $\Psi(q_{1})$, constructed out of a square integrable function $\psi(q_0)$ and the kernel $K(q_{1},\mathcal{N};q_{0},0)$, in the following manner:
%%%%%%%%%%%%%%%%%%%%%%%%%%%%%%%%%%%%%%%%%%%%%%%%%%%%%%%%%%%%%%%%
\begin{align}
\Psi(q_1)=\int_{\mathcal{C}}\left[\int_{-\infty}^{\infty} K(q_{1},\mathcal{N};q_{0},0)\psi(q_{0})dq_{0}\right]d\mathcal{N}
\end{align}
%%%%%%%%%%%%%%%%%%%%%%%%%%%%%%%%%%%%%%%%%%%%%%%%%%%%%%%%%%%%%%%%
where, the integration over $\mathcal{N}$ is performed along a contour $\mathcal{C}$ in complex $\mathcal{N}$-plane, which is yet to be fixed. Using the evolution equation of the Kernel along the lapse function $\mathcal{N}$, which is generated by the Hamiltonian constraint, as depicted above, we obtain,
%%%%%%%%%%%%%%%%%%%%%%%%%%%%%%%%%%%%%%%%%%%%%%%%%%%%%%%%%%%%%%%%
\begin{align}
\mathcal{H}_{\rm perfect}\Psi(q_1)&=\int_{\mathcal{C}}\left[\int_{-\infty}^{\infty} i\partial_{\mathcal{N}}K(q_{1},\mathcal{N};q_{0},0)\psi(q_{0})dq_{0}\right]d\mathcal{N}
\end{align}
%%%%%%%%%%%%%%%%%%%%%%%%%%%%%%%%%%%%%%%%%%%%%%%%%%%%%%%%%%%%%%%%
where, $\mathcal{H}_{\rm perfect}$ is the Hamiltonian defined in \ref{H_perfect}. Therefore, if we further demand that $\Psi(q_1)$ should satisfy the Wheeler-DeWitt equation, i.e., $\mathcal{H}_{\rm perfect}\Psi(q_1)=0$, then we can perform the integration over the lapse function $\mathcal{N}$ and hence arrive at the following condition,
%%%%%%%%%%%%%%%%%%%%%%%%%%%%%%%%%%%%%%%%%%%%%%%%%%%%%%%%%%%%%%%%
\begin{align}\label{kernel_condition}
\int_{-\infty}^{\infty} K(q_{1},\mathcal{N}_f;q_{0},0)\psi(q_0)dq_0-\int_{-\infty}^{\infty} K(q_{1},\mathcal{N}_i;q_{0},0)\psi(q_0)dq_0=0~,
\end{align}
%%%%%%%%%%%%%%%%%%%%%%%%%%%%%%%%%%%%%%%%%%%%%%%%%%%%%%%%%%%%%%%%
where, $\mathcal{N}_i$ and $\mathcal{N}_f$ are the endpoints of the contour $\mathcal{C}$. One way of achieving this condition is to choose the endpoints such that, the kernel identically vanishes at these points, as, for instance, is the case for the contours used in \cite{Halliwell1988,DiazDorronsoro:2017hti,Lehners2019}. Another possibility would be to choose the contour of $\mathcal{N}$ integration to be a closed one \cite{DiazDorronsoro:2018wro}. Even though both of these possibilities mathematically are viable, for the purpose of our current discussion, we shall see that the former choice is more natural for the structure of the path integral kernel in both linear and quadratic models. The seminal work of Hartle and Hartle \cite{Hartle:1983ai}, on the other hand, may be viewed as an approach based on a contour along the imaginary axis of the complex $\mathcal{N}$-plane and, therefore, is usually referred to as the Euclidean path integral approach. Recently\cite{Feldbrugge:2017fcc,Feldbrugge:2017mbc}, however, evidences have emerged of serious problems with the approach based on the imaginary line contour. One might argue that the most natural choice of the contour is the one along the real line. Moreover, recent investigations based on the real line contour, dubbed the `Lorentzian quantum cosmology', seems to be devoid of the issues encountered in the Euclidean approach\cite{Feldbrugge:2017kzv,Lehners2019} and, hence, shall be the basis of our analysis in this section.  

%%%%%%%%%%%%%%%%%%%%%%%%%%%%%%%%%%
%%%%%%%%%%%%%%%%%%%%%%%%%%%%%%%%%%
%%%%%%%%%%%%%%%%%%%%%%%%%%%%%%%%%%
%%%%%%%%%%%%%%%%%%%%%%%%%%%%%%%%%%
\subsubsection{The linear model}

The path integral kernel for the linear model is essentially the kernel of a one dimensional particle, subject to a constant force and hence can be evaluated explicitly. The explicit form of the kernel being,
%%%%%%%%%%%%%%%%%%%%%%%%%%%%%%%%%%%%%%%%%%%%%%%%%%%%%%%%%%%%%%%%
\begin{align}
K^{\rm (I)}(q_{1},\mathcal{N};q_{0},0)=\sqrt{\frac{iM}{2\pi \mathcal{N}}}~\exp\left[i \mathcal{S}^{\rm (I)}_{\rm cl}[q_{1},q_{0};\mathcal{N}]\right]
\end{align}
%%%%%%%%%%%%%%%%%%%%%%%%%%%%%%%%%%%%%%%%%%%%%%%%%%%%%%%%%%%%%%%%
where, $\mathcal{S}^{\rm (I)}_{\rm cl}$ is just the action $\mathcal{S}_{\rm perfect}[q]$ evaluated at the classical solution satisfying the boundary conditions $q(T=0)=q_{0}$ and $q(T=\mathcal{N})=q_{1}>1$. The explicit form of $\mathcal{S}^{\rm (I)}_{\rm cl}$ in terms of $q_{0}$, $q_{1}$ and the `lapse' $\mathcal{N}$ is given by,
%%%%%%%%%%%%%%%%%%%%%%%%%%%%%%%%%%%%%%%%%%%%%%%%%%%%%%%%%%%%%%%%
\begin{align}
\frac{\mathcal{S}^{\rm (I)}_{\rm cl}}{V_{3}}=\frac{1}{2\ell^{2}_{\rm p}\left(1-\frac{n}{6}\right)^2}\left[-\frac{(q_1-q_0)^2}{2\mathcal{N}}
-h_{n}^{2}\left(q_{1}+q_{0}\right)\mathcal{N}+\frac{h_{n}^{4}}{6}\mathcal{N}^{3}+2h_{n}^{2}\mathcal{N}\right]~.
\end{align}
%%%%%%%%%%%%%%%%%%%%%%%%%%%%%%%%%%%%%%%%%%%%%%%%%%%%%%%%%%%%%%%%
In the context of de Sitter spacetime, the original no-boundary prescription of HH for defining a wave of the universe corresponds to choosing an initial wave function $\psi(q_0)\propto\delta(q_0)$. Another possibility would be to choose $\psi(q_0)\propto\delta(q_0-\bar{q})$, followed by taking $\bar{q}\rightarrow 0$ limit \cite{DiazDorronsoro:2017hti}.  However, recent, more careful mathematical considerations have shown that such approaches lead to physical and mathematical issues\cite{Feldbrugge:2017fcc,Feldbrugge:2017mbc,Feldbrugge:2018gin}. It is worth mentioning that principles of loop quantum cosmology has been employed recently to attempt to `rescue' the no-boundary proposal \cite{Bojowald:2018gdt,Bojowald:2020kob}. On the other hand, a promising approach that retains the no-boundary wave function, strictly in the context of canonical quantum cosmology, has been proposed in \cite{DiTucci:2019dji,Lehners2019}. Therein, the authors choose an initial state $\psi(q_0)$ that corresponds to a well defined Euclidean momentum, in which case, the corresponding perturbations can be shown to be Gaussian distributed. 
Motivated by this, we demand that the initial wave function $\psi(q_{0})$ corresponds to that of a momentum eigenstate with momentum $p=-Mv$, where, $v\equiv (dq/dT)_{T=0}=\mathcal{N}^{-1}\dot{q}(0)$. It is then instructive to define the momentum space kernel as the Fourier transform of the position space kernel, yielding,
%%%%%%%%%%%%%%%%%%%%%%%%%%%%%%%%%%%%%%%%%%%%%%%%%%%%%%%%%%%%%%%%
\begin{align}\label{int_q0_linear}
\mathcal{K}^{\rm (I)}(q_{1},\mathcal{N};-Mv,0)&\equiv\int K^{\rm (I)}(q_{1},\mathcal{N};q_{0},0)~e^{-i MvV_{3}q_{0}}~dq_0
\nonumber
\\
&=e^{i\widetilde{\mathcal{S}}^{\rm (I)}_{\rm cl}[q_{1},-Mv;\mathcal{N}]}~,
\end{align}
%%%%%%%%%%%%%%%%%%%%%%%%%%%%%%%%%%%%%%%%%%%%%%%%%%%%%%%%%%%%%%%%
where, $\widetilde{\mathcal{S}}^{\rm (I)}_{\rm cl}$ is the action $\mathcal{S}_{\rm perfect}[q]-\mathcal{S}_{\rm Boundary}[q_{0}]$, evaluated at the classical solution $\widetilde{q}_{\rm cl}^{\rm (I)}(t)$ which satisfies the boundary conditions $\mathcal{N}^{-1}\dot{q}(0)=v$ and $q(T=\mathcal{N})=q_{1}$. The boundary term $\mathcal{S}_{\rm Boundary}[q_{0}]$ has the form: $\mathcal{N}^{-1}MV_{3}q(0)\dot{q}(0)$ and, the dot, as usual, denotes a derivative with respect to $t$. 

%%%%%%%%%%%%%%%%%%%%%%%%%%%%%%%%%%%%%%%%%%%%%%
%%%%%%%%%%%%%%%%%%%%%%%%%%%%%%%%%%%%%%%%%%%%%%
%%%%%%%%%%%%%%%%%%%%%%%%%%%%%%%%%%%%%%%%%%%%%%
%%%%%%%%%%%%%%%%%%%%%%%%%%%%%%%%%%%%%%%%%%%%%%

%%%%%%%%%%%%%%%%%%%%%%%%%%%%%%%%%%%%%%%%%%%%%%
%%%%%%%%%%%%%%%%%%%%%%%%%%%%%%%%%%%%%%%%%%%%%%
%%%%%%%%%%%%%%%%%%%%%%%%%%%%%%%%%%%%%%%%%%%%%%
%%%%%%%%%%%%%%%%%%%%%%%%%%%%%%%%%%%%%%%%%%%%%%

%%%%%%%%%%%%%%%%%%%%%%%%%%%%%%%%%%%%%%%%%%%%%%
%%%%%%%%%%%%%%%%%%%%%%%%%%%%%%%%%%%%%%%%%%%%%%
%%%%%%%%%%%%%%%%%%%%%%%%%%%%%%%%%%%%%%%%%%%%%%
%%%%%%%%%%%%%%%%%%%%%%%%%%%%%%%%%%%%%%%%%%%%%%
\begin{figure}[t]
	\centering
	\includegraphics[scale=.5]{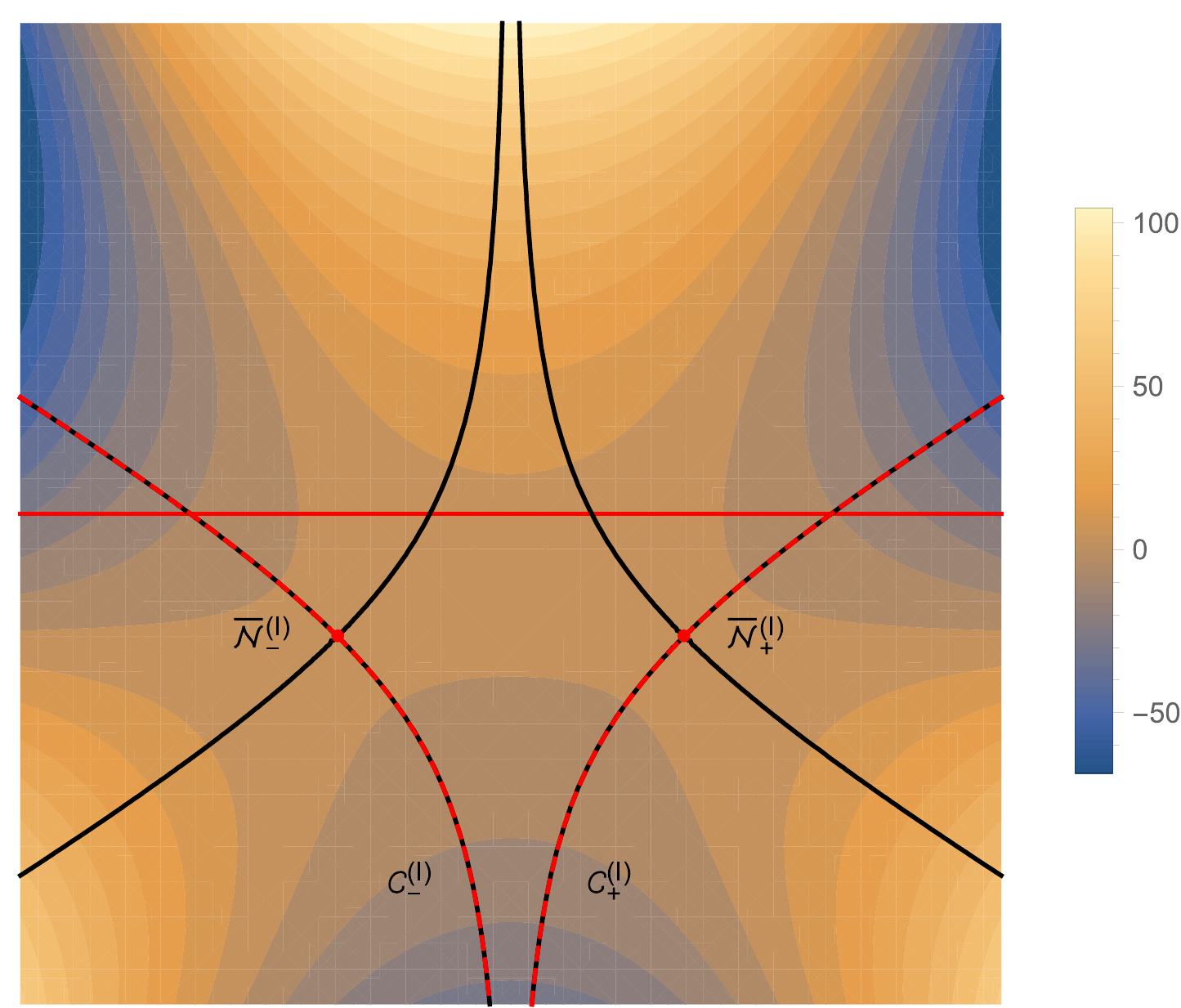}
	\caption{The steepest descent/ascent curves in the complex $\mathcal{N}$-plane, that pass through the saddle points $\bar{\mathcal{N}}^{\rm (I)}_{\pm}$, for the linear model are represented by the black curves. The contour-plots for the function $\textrm{Re}[i2\ell_{\rm p}^{2}\{(n/6)-1\}^{2}V_{3}^{-1}\tilde{\mathcal{S}}^{\rm (I)}_{\rm cl}]$ are also presented, using the colour coding scheme given by the right inset, to better visualise the descent/ascent directions (the plots are for the parameter values $h_{n}=1$ and $q_{1}=3$). The dark red, horizontal line is the real line contour over which the integration of the lapse function needs to be performed to evaluate the no-boundary wave function. The real line contour can be smoothly deformed into the union of dashed red contours $\mathcal{C}^{\rm (I)}_{+}$ and $\mathcal{C}^{\rm (I)}_{-}$, thereby, enabling us to perform saddle point approximation. See text for more discussion.}
	\label{linear_contour}
\end{figure}
%%%%%%%%%%%%%%%%%%%%%%%%%%%%%%%%%%%%%%%%%%%%%%
%%%%%%%%%%%%%%%%%%%%%%%%%%%%%%%%%%%%%%%%%%%%%%
%%%%%%%%%%%%%%%%%%%%%%%%%%%%%%%%%%%%%%%%%%%%%%
%%%%%%%%%%%%%%%%%%%%%%%%%%%%%%%%%%%%%%%%%%%%%%

The solution to the equation of motion of $q(t)$ with the above boundary conditions takes the following explicit form:
%%%%%%%%%%%%%%%%%%%%%%%%%%%%%%%%%%%%%%%%%%%%%%%%%%%%%%%%%%%%%%%%
\begin{align}
\widetilde{q}^{\rm (I)}_{\rm cl}(t)=h_{n}^{2}\mathcal{N}^{2}t^{2}-\mathcal{N}\left(h_{n}^{2}\mathcal{N}+v\right)+\mathcal{N} t v+q_{1}~.
\end{align}
%%%%%%%%%%%%%%%%%%%%%%%%%%%%%%%%%%%%%%%%%%%%%%%%%%%%%%%%%%%%%%%%
Motivated by the recent results of \cite{Lehners2019} in the context of de Sitter spacetime, where the no-boundary wave function was realised by choosing the initial momentum to be purely imaginary, we demand that the $v=2ih_{n}$. With this choice of the initial velocity $v$ and the classical solution $\widetilde{q}^{\rm (I)}_{\rm cl}(t)$, the explicit form of the classical action $\widetilde{\mathcal{S}}^{\rm (I)}_{\rm cl}$ takes the following form,
%%%%%%%%%%%%%%%%%%%%%%%%%%%%%%%%%%%%%%%%%%%%%%%%%%%%%%%%%%%%%%%%
\begin{align}\label{Stilde_I}
\frac{\widetilde{\mathcal{S}}^{\rm (I)}_{\rm cl}}{V_{3}}=\frac{1}{2\ell^{2}_{\rm p}\left(1-\frac{n}{6}\right)^2}\left[\frac{2h_{n}^{4}}{3} \left(\mathcal{N}+\frac{i}{h_{n}}\right)^{3}-2h_{n}^{2}\left(\mathcal{N}+\frac{i}{h_{n}}\right)(q_1-1)-\frac{4ih_{n}}{3}\right]~.
\end{align}
%%%%%%%%%%%%%%%%%%%%%%%%%%%%%%%%%%%%%%%%%%%%%%%%%%%%%%%%%%%%%%%%
In order to determine the wave function from the path integral kernel, we need to integrate over the lapse function $\mathcal{N}$. The integral should be over a contour for which the kernel vanishes sufficiently fast as one approaches the endpoints, i.e., the real part of $i\widetilde{\mathcal{S}}_{\rm cl}^{\rm (I)}$ should take a large and negative value as we approach the endpoints of the contour. Taking $\{\mathcal{N}+(i/h_{n})\}$ as $re^{i\theta}$, we obtain from \ref{Stilde_I}, that $\textrm{Re}(i\widetilde{S}_{\rm cl}^{\rm (I)})$ takes large negative values, such that $\mathcal{K}^{\rm (I)}(q_{1},\mathcal{N};-Mv,0)$ vanishes, for large values of $r$, if $(\sin^{3}\theta-3\sin \theta \cos^{2}\theta)<0$. Thus the momentum space kernel identically vanishes in the asymptotic regions in the complex $\mathcal{N}$-plane, provided we restrict ourselves to the following regions: $\textrm{Arg}[\left\{\mathcal{N}+i/(h_{n})\right\}]\in(0,\pi/3)\cup(2\pi/3,\pi)\cup(4\pi/3,5\pi/3)$ (These regions have been depicted in \ref{linear_contour} in dark shades of blue). The saddle points of $\widetilde{\mathcal{S}}^{\rm (I)}_{\rm cl}$, which we shall require shortly, turns out to be,
%%%%%%%%%%%%%%%%%%%%%%%%%%%%%%%%%%%%%%%%%%%%%%%%%%%%%%%%%%%%%%%%
\begin{align}\label{HH_Saddle_I}
\bar{\mathcal{N}}^{\rm (I)}_{\pm}=-\frac{i}{h_{n}}\pm\frac{\sqrt{q_{1}-1}}{h_{n}}~,
\end{align}
%%%%%%%%%%%%%%%%%%%%%%%%%%%%%%%%%%%%%%%%%%%%%%%%%%%%%%%%%%%%%%%%
The classical solutions $\widetilde{q}^{\rm (I)}_{\rm cl}(t)$ with the above choice for the lapse function and with $v=2ih_{n}$ is given by:
%%%%%%%%%%%%%%%%%%%%%%%%%%%%%%%%%%%%%%%%%%%%%%%%%%%%%%%%%%%%%%%%
\begin{align}
\widetilde{q}^{\rm (I)}_{\textrm{cl},\pm}(t)=t^{2}\left[\left(q_{1}-2\right)\mp 2i\sqrt{q_{1}-1}\right]+t \left(2\pm 2i\sqrt{q_{1}-1}\right)~.
\end{align}
%%%%%%%%%%%%%%%%%%%%%%%%%%%%%%%%%%%%%%%%%%%%%%%%%%%%%%%%%%%%%%%%
Therefore, we see that the saddle points describe geometries in which the scale factor $a(t)=q(t)^{3/(6-n)}$ vanishes at $t=0$ (since $\widetilde{q}^{\rm (I)}_{\textrm{cl},\pm}(t=0)=0$) and equals $q_1^{3/(6-n)}$ at $t=1$ (since $\widetilde{q}^{\rm (I)}_{\textrm{cl},\pm}(t=1)=q_{1}$). 

The Lorentzian contour, namely the real line, is the most natural choice for the $\mathcal{N}$ integration that evaluates the wavefuntion. Fortunately, from \ref{linear_contour}, we see that the $\mathcal{N}$-integral along the real line is convergent. In order to find the asymptotic expression for the wave function, however, we will employ the saddle approximation. The steepest descent/ascent curves associated with the action $\widetilde{\mathcal{S}}^{\rm (I)}_{\rm cl}$ are shown in \ref{linear_contour} along with the saddle points $\bar{\mathcal{N}}^{\rm (I)}_{\textrm{cl},\pm}$. The regions where the real part of $i\widetilde{\mathcal{S}}^{\rm (I)}_{\rm cl}$ becomes large and negative are also presented. The relevant saddle points that contribute to the evaluation of the integral, according to Picard-Lefschetz theory \cite{Feldbrugge:2017kzv,Lehners2019}, are the ones that can be approached by flowing down (i.e., in view of \ref{linear_contour}, towards the direction of darker shades of blue) the original integration contour. This dictates that the appropriate contour, to which the real line should be deformed to, corresponds to $\mathcal{C}^{\rm (I)}_{+}+\mathcal{C}^{\rm (I)}_{-}$. With the relevant saddle points determined, we can now evaluate the saddle point approximation of the no-boundary wave function for the bouncing model, yielding,
%%%%%%%%%%%%%%%%%%%%%%%%%%%%%%%%%%%%%%%%%%%%%%%%%%%%%%%%%%%%%%%%
\begin{align}\label{HH_wavefunction_I}
\Psi^{\rm (I)}(q_{1})&\sim \left(\frac{e^{i\frac{\pi}{4}}}{\sqrt{|\partial^2_{\mathcal{N}}\widetilde{\mathcal{S}}^{\rm (I)}_{\rm cl}}|}e^{i\widetilde{\mathcal{S}}^{\rm (I)}_{\rm cl}}\right)_{\bar{\mathcal{N}}_{\textrm{cl},+}^{\rm (I)}}
+\left(\frac{e^{-i\frac{\pi}{4}}}{\sqrt{|\partial^2_{\mathcal{N}}\widetilde{\mathcal{S}}^{\rm (I)}_{\rm cl}|}}e^{i\widetilde{\mathcal{S}}^{\rm (I)}_{\rm cl}}\right)_{\bar{\mathcal{N}}^{\rm (I)}_{\textrm{cl},-}}
\propto \exp\left(\frac{2\alpha_{n}}{3}\right) \frac{ \cos \left[\frac{2}{3}\left(q-1\right)^{3/2}\alpha_{n}-\frac{\pi}{4}\right]}{\sqrt[4]{(\alpha_{n})^{2}(q-1)}}~,
\end{align} 
%%%%%%%%%%%%%%%%%%%%%%%%%%%%%%%%%%%%%%%%%%%%%%%%%%%%%%%%%%%%%%%%
which is clearly consistent with \ref{Psi_I}. 

\begin{figure}[t]
	\centering
	\includegraphics[scale=.25]{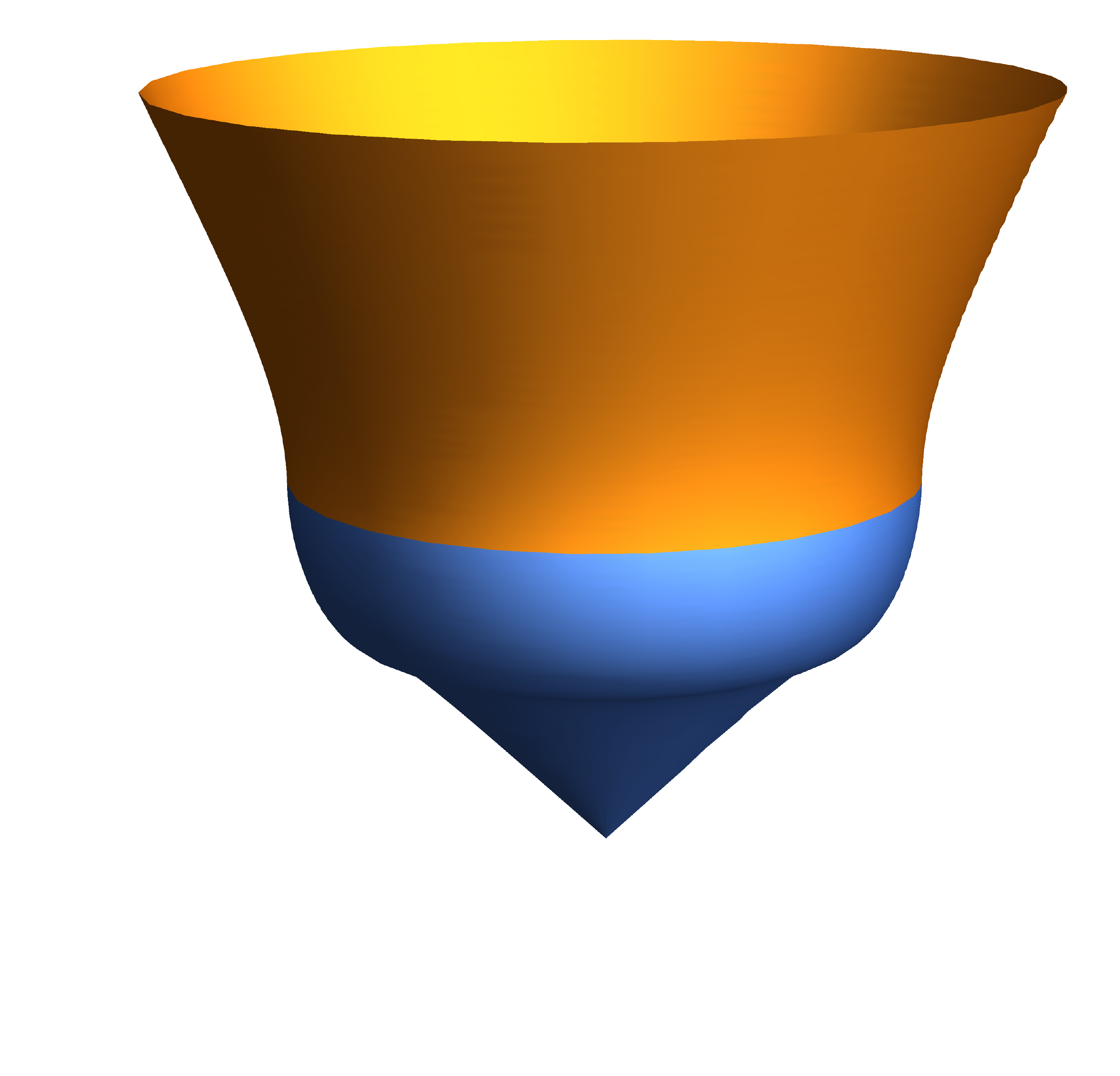}
	\caption{The saddle point geometry that would be relevant to Euclidean path integral approach to matter bounce. The surface denotes the $x-\eta$ plane, with periodic identification assumed along the $x$ direction. The orange-coloured portion denotes the Lorentzian part of the geometry, while blue denotes the Euclidean portion. There is a true singularity at the centre of the Euclidean portion, unlike the Hawking-Hartle geometry which is smooth everywhere. }
	\label{Saddlepoint_geometry}
\end{figure}

Having described the derivation of the no-boundary wave function from the path integral analysis with `mixed' boundary condition, let us discuss its interpretation. We have seen that the saddle point solution $\widetilde{q}^{\rm (I)}_{\textrm{cl},\pm}(t)$ satisfies $\widetilde{q}^{\rm (I)}_{\textrm{cl},\pm}(0)=0$ and $\widetilde{q}^{\rm (I)}_{\textrm{cl},\pm}(1)=q_1$. Therefore, the corresponding geometry may be imagined as describing an evolution from a point-sized to a finite-sized universe. This is reminiscent of the famous Hartle-Hawking saddle point geometry that appears in the context of Euclidean path integral approach to de Sitter cosmology\cite{Hartle:1983ai}. Therein, the corresponding wave function is usually interpreted as describing `tunnelling from nothing'. One might, therefore, be tempted to bestow an analogous interpretation on the wave function $\Psi^{\rm (I)}(q_{1})$, with the Hawking-Hartle geometry replaced by the appropriate generalization of it in the context of bounce (For representation purpose, in \ref{Saddlepoint_geometry}, we have presented the saddle point geometry corresponding to a `matter-bounce' scenario). However, we argue that such an interpretation is questionable. In fact, the wave function $\Psi^{\rm(I)}(q_1)$ has a form similar to the original Hartle-Hawking wave function precisely because the latter also has both the expanding and contracting branches of the de Sitter spacetime. In the latter context, it has already been found in \cite{Lehners2019} that the conventional interpretation of the no-boundary wave function, as describing `tunnelling from nothing', is problematic. In order to appreciate how this translates to in the bouncing scenario, note that even though the dominant contribution to the path integral leading to $\Psi^{\rm (I)}(q_1)$ is from the Hartle-Hawking-like saddle point geometry, corresponding to a spacetime that emerges from zero size, the off-shell geometries can emerge from any size. This fact is evident from \ref{int_q0_linear}, where we have clearly performed a summation over \textit{all} values of the `initial size' $q_0$ to define the wave
function. In the spirit of \cite{Lehners2019}, and in the light of \ref{int_q0_linear} it is more reasonable to interpret $\Psi^{\rm (I)}(q_1)$ as describing a transition from the state of a specific Euclidean momentum. However, it is worth mentioning that the bouncing analogue of the `tunneling wave function' of Vilenkin \cite{Vilenkin:1987kf}, which offers a more appropriate description of `tunneling from nothing', can also be constructed in an analogous manner .

%%%%%%%%%%%%%%%%%%%%%%%%%%%%%%%%%%%%%%%%%%%%%%
%%%%%%%%%%%%%%%%%%%%%%%%%%%%%%%%%%%%%%%%%%%%%%
%%%%%%%%%%%%%%%%%%%%%%%%%%%%%%%%%%%%%%%%%%%%%%
%%%%%%%%%%%%%%%%%%%%%%%%%%%%%%%%%%%%%%%%%%%%%%
%%%%%%%%%%%%%%%%%%%%%%%%%%%%%%%%%%%%%%%%%%%%%%
%%%%%%%%%%%%%%%%%%%%%%%%%%%%%%%%%%%%%%%%%%%%%%
%%%%%%%%%%%%%%%%%%%%%%%%%%%%%%%%%%%%%%%%%%%%%%
%%%%%%%%%%%%%%%%%%%%%%%%%%%%%%%%%%%%%%%%%%%%%%

%%%%%%%%%%%%%%%%%%%%%%%%%%%%%%%%%%
%%%%%%%%%%%%%%%%%%%%%%%%%%%%%%%%%%
%%%%%%%%%%%%%%%%%%%%%%%%%%%%%%%%%%
%%%%%%%%%%%%%%%%%%%%%%%%%%%%%%%%%%
\subsubsection{The quadratic model}

\begin{figure}[t]
	\centering
	\includegraphics[scale=.5]{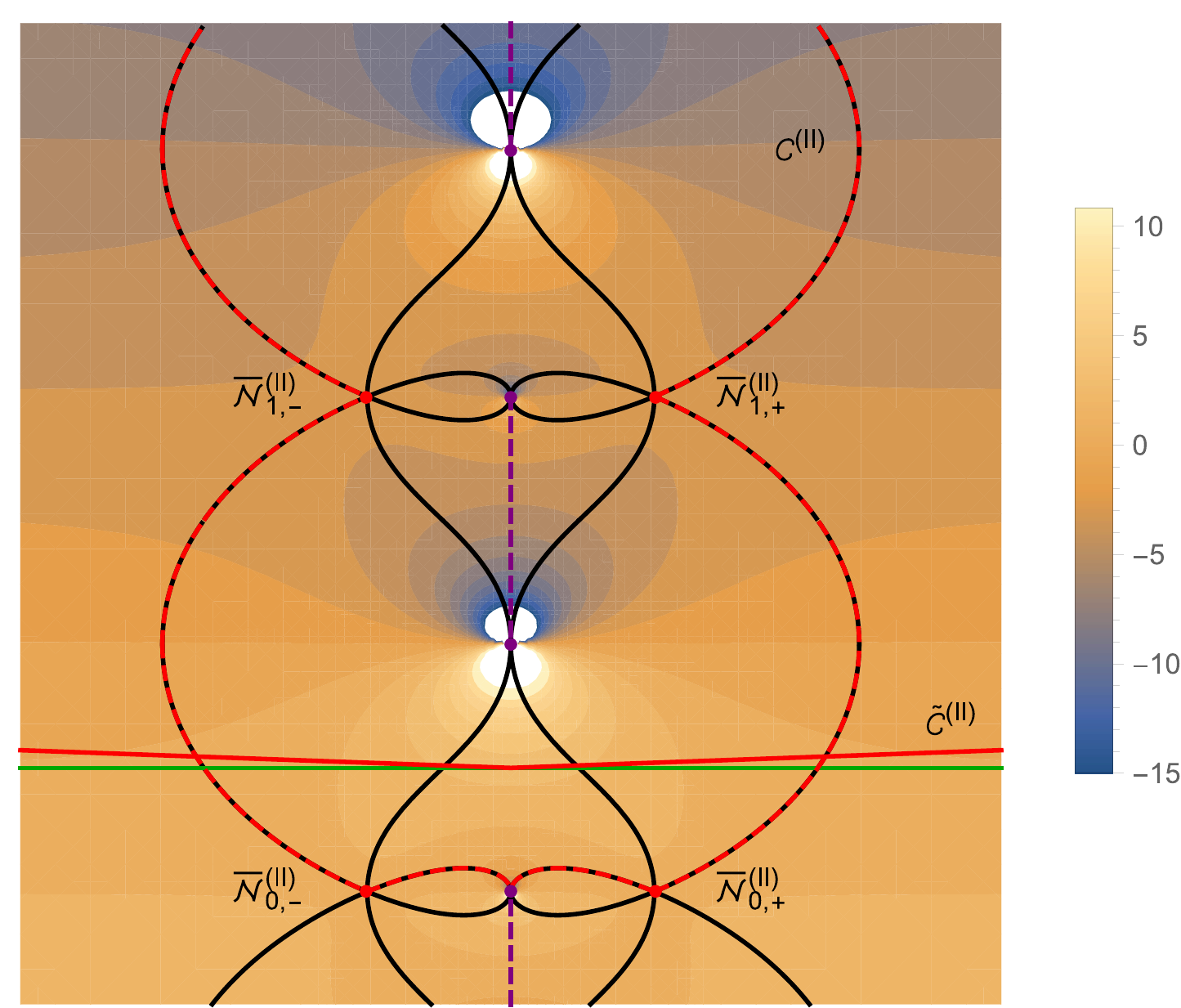}
	\caption{The steepest descent/ascent curves in the complex $\mathcal{N}$-plane for the quadratic model are represented by the black curves. The contour-plots for $\textrm{Re}\left[i(9/8)\ell_{\rm p}^{2}(n/6-1)^{2}V_{3}^{-1}\tilde{\mathcal{S}}^{\rm (II)}_{\rm cl}\right]$ is also presented, with the colour coding scheme given in the right inset, to better visualise the descent/ascent directions (the parameter values for the above plot correspond to: $h_{n}=1$ and $q_{1}=3$). The dashed purple lines are the branch-cuts of the pre-exponential factor $\sqrt{\textrm{sech}(3h_{n}\mathcal{N}/2)}$. The horizontal, green line is the real line contour. However, convergence of the $\mathcal{N}$-integral demands that we choose a slightly modified contour, namely, the red continuous curve $\widetilde{\mathcal{C}}^{\rm (II)}$, joining $-\infty e^{-i0^{+}}$ to $\infty e^{i0^{+}}$. This continuous red contour can be smoothly deformed into the dashed red curve $\mathcal{C}^{\rm (II)}$, that pass through the saddle points, thereby enabling us to evaluate the saddle point approximation. See text for more discussion.}
	\label{quad_contour}
\end{figure}

Having described the linear model in the previous section, we will now present the scenario in which the effective potential appearing in the Wheeler-DeWitt equation for the gravity and the perfect fluid system is quadratic in the dynamical variable $q$. In this model, the action is that of a one-dimensional particle moving in a quadratic potential, for which the path integral kernel can be evaluated explicitly and takes the following form:
%%%%%%%%%%%%%%%%%%%%%%%%%%%%%%%%%%%%%%%%%%%%%%%%%%%%%%%%%%%%%%%%
\begin{align}
K^{\rm (II)}(q_{1},\mathcal{N};q_{0},0)=\sqrt{\frac{3MiV_{3}}{2\sinh\left(\frac{3}{2}h_{n}\mathcal{N}\right)}}~e^{i \mathcal{S}^{\rm (II)}_{\rm cl}[q_{1},q_{0};\mathcal{N}]}
\end{align}
%%%%%%%%%%%%%%%%%%%%%%%%%%%%%%%%%%%%%%%%%%%%%%%%%%%%%%%%%%%%%%%%
where, $\mathcal{S}^{\rm (II)}_{\rm cl}$ is the action $\mathcal{S}_{\rm perfect}[q]$ (see \ref{action_eff}) evaluated at the classical solution satisfying the Dirichlet boundary conditions: $q(t=0)=q_{0}$ and $q(t=1)=q_{1}>1$. The explicit form of the action is given by the following expression,
%%%%%%%%%%%%%%%%%%%%%%%%%%%%%%%%%%%%%%%%%%%%%%%%%%%%%%%%%%%%%%%%
\begin{align}
\frac{\mathcal{S}_{\rm cl}^{\rm (II)}}{V_3}=\frac{8}{9\ell_{\rm p}^{2}\left(\frac{n}{6}-1\right)^2}\left\{-\frac{3h_{n}}{4\sinh\left(\frac{3}{2}h_{n}\mathcal{N}\right)}\left\{\cosh\left(\frac{3}{2}h_{n}\mathcal{N}\right)\left(q_{1}^{2}+q_{0}^{2}\right)-2q_{0}q_{1}\right\}+\frac{9}{8}h_{n}^{2}\mathcal{N}\right\}~.
\end{align}
%%%%%%%%%%%%%%%%%%%%%%%%%%%%%%%%%%%%%%%%%%%%%%%%%%%%%%%%%%%%%%%%
As in the case of linear model, we now demand that the initial wave function $\Psi(q_0)$ corresponds to a state of definite momentum $p=-M v$. This implies that the relevant kernel is the momentum space kernel, with the following expression,
%%%%%%%%%%%%%%%%%%%%%%%%%%%%%%%%%%%%%%%%%%%%%%%%%%%%%%%%%%%%%%%%
\begin{align}\label{int_q0_quadratic}
\mathcal{K}^{\rm (II)}(q_1,\mathcal{N};-M v,0)&\equiv\int K^{\rm (II)}(q_1,\mathcal{N};q_0,0)e^{-i M v V_3q_0}dq_{0}
\nonumber
\\
&=\sqrt{\textrm{sech}\left(\frac{3}{2}h_n\mathcal{N}\right)}~e^{i\widetilde{\mathcal{S}}^{\rm (II)}_{\rm cl}[q_1,-Mv;\mathcal{N}]}~,
\end{align}
%%%%%%%%%%%%%%%%%%%%%%%%%%%%%%%%%%%%%%%%%%%%%%%%%%%%%%%%%%%%%%%%
where, as in the linear case, $\widetilde{\mathcal{S}}^{\rm (II)}_{\rm cl}$ is the action $\mathcal{S}_{\rm perfect}[q]-\mathcal{S}_{\rm Boundary}[q]$, evaluated at the classical solution $\widetilde{q}^{\rm (II)}_{\rm cl}(t)$, which satisfies the boundary condition $\mathcal{N}^{-1}\dot{q}(0)=v$ and $q(1)=q_1$, with the boundary term $\mathcal{S}_{\rm Boundary}[q]$ having the form: $\mathcal{N}^{-1}MV_3q(0)\dot{q}(0)$ and, the dot denotes a derivative with respect to $t$. First of all, the explicit form of the solution $\widetilde{q}^{\rm (II)}_{\rm cl}(t)$, given the above boundary conditions, is given by:
%%%%%%%%%%%%%%%%%%%%%%%%%%%%%%%%%%%%%%%%%%%%%%%%%%%%%%%%%%%%%%%%
\begin{align}
\widetilde{q}^{\rm (II)}_{\rm cl}(t)=\frac{\text{sech}\left(\frac{3h_{n}\mathcal{N}}{2}\right)}{3h_{n}} \left[3h_{n}q_{1}\cosh \left(\frac{3 }{2}h_{n}\mathcal{N}t\right)+2 v \sinh \left(\frac{3}{2} h_{n}\mathcal{N}(t-1)\right)\right]~.
\end{align}
%%%%%%%%%%%%%%%%%%%%%%%%%%%%%%%%%%%%%%%%%%%%%%%%%%%%%%%%%%%%%%%%
Following \cite{Lehners2019} and the analysis in the case of the linear potential, for the quadratic model we have to choose $v=(3ih_{n}/2)$ in order to get the no-boundary wave function for the bouncing model under consideration. With this choice of the velocity $v$, the action on the classical trajectory $\widetilde{q}^{\rm (II)}_{\rm cl}(t)$ becomes:
%%%%%%%%%%%%%%%%%%%%%%%%%%%%%%%%%%%%%%%%%%%%%%%%%%%%%%%%%%%%%%%%
\begin{align}
\frac{\widetilde{\mathcal{S}}^{\rm (II)}_{\rm cl}}{V_3}=\frac{8}{9\ell_{\rm p}^{2}\left(\frac{n}{6}-1\right)^2}\left\{\frac{9h_{n}^{2}\mathcal{N}}{8}-\frac{3}{4}h_{n}\textrm{sech}\left(\frac{3h_{n}\mathcal{N}}{2}\right) \left[\left(q_{1}^{2}+1\right) \sinh \left(\frac{3h_{n}\mathcal{N}}{2}\right)+2iq_{1}\right]\right\}~.
\end{align}
%%%%%%%%%%%%%%%%%%%%%%%%%%%%%%%%%%%%%%%%%%%%%%%%%%%%%%%%%%%%%%%%
Again, we need to find out the asymptotic regions in the complex $\mathcal{N}$-plane, where the endpoints of the contours associated with the $\mathcal{N}$ integration must lie, so that the endpoint contribution to the kernel identically vanishes. This requires $\textrm{Re}[i\mathcal{S}_{\rm cl}^{\rm (II)}]<0$, which for large values of $|\mathcal{N}|$ demands: $\textrm{Im}[\mathcal{N}]>0$. The immediate consequence of this condition is that, unlike in the case of the linear model, the $\mathcal{N}$ integral along real line contour and hence, the strictly Lorentzian path integral, is not convergent. Convergence of the $\mathcal{N}$ integration demands that we choose a slightly modified contour, namely, the continuous curve joining $-\infty e^{-i0^{+}}$ to $\infty e^{i0^{+}}$, represented by the red curve $\tilde{\mathcal{C}}^{\rm (II)}$ in \ref{quad_contour}.

In order to obtain the saddle points, first of all we need to compute the derivative of the classical action with respect to the lapse function $\mathcal{N}$, which yields,
%%%%%%%%%%%%%%%%%%%%%%%%%%%%%%%%%%%%%%%%%%%%%%%%%%%%%%%%%%%%%%%%
\begin{align}\label{derivative_II}
\frac{1}{V_3}\frac{\partial \widetilde{\mathcal{S}}^{\rm (II)}_{\rm cl}}{\partial \mathcal{N}}
&=\frac{9}{8}h_{n}^{2}+\frac{9}{8}h_{n}^{2}\textrm{sech}\left(\frac{3h_{n}\mathcal{N}}{2}\right)\left[-\left(q_{1}^{2}+1\right)\textrm{sech}\left(\frac{3h_{n}\mathcal{N}}{2}\right)+2iq_{1}\textrm{tanh}\left(\frac{3h_{n}\mathcal{N}}{2}\right) \right]
\nonumber
\\
&=\frac{9}{8}h_{n}^{2}\textrm{sech}^{2}\left(\frac{3h_{n}\mathcal{N}}{2}\right)\left[iq_{1}+\textrm{sinh}\left(\frac{3h_{n}\mathcal{N}}{2}\right)\right]^{2}~.
\end{align}
%%%%%%%%%%%%%%%%%%%%%%%%%%%%%%%%%%%%%%%%%%%%%%%%%%%%%%%%%%%%%%%%
Setting the above expression to zero, we obtain, $\textrm{sinh}(3h_{n}\mathcal{N}/2)=-iq_{1}$, solving which we obtain an infinite number of saddle points for the action $\widetilde{\mathcal{S}}^{\rm (II)}_{\rm cl}$ and they are given by:
%%%%%%%%%%%%%%%%%%%%%%%%%%%%%%%%%%%%%%%%%%%%%%%%%%%%%%%%%%%%%%%%
\begin{align}
\bar{\mathcal{N}}^{\rm (II)}_{j,\pm}=\frac{(4j-1)\pi i}{3h_{n}}\pm\frac{2}{3h_n}\log\left(q_{1}+\sqrt{q_{1}^{2}-1}\right)~;\qquad j\in \mathbb{Z}~.
\end{align} 
%%%%%%%%%%%%%%%%%%%%%%%%%%%%%%%%%%%%%%%%%%%%%%%%%%%%%%%%%%%%%%%%
At these saddle points, substitution of the respective values of the lapse function $\mathcal{N}$ yields the following expression for the classical solution $\widetilde{q}^{\rm (II)}_{\rm cl}(t)$,
%%%%%%%%%%%%%%%%%%%%%%%%%%%%%%%%%%%%%%%%%%%%%%%%%%%%%%%%%%%%%%%%
\begin{align}
\widetilde{q}^{\rm (II)}_{\textrm{cl}(j,\pm)}(t)=\sin \left[\left((4j-1)\pi\mp 2i \log(q_{1}+\sqrt{q_{1}^{2}-1})\right)\frac{t}{2}\right]~;\qquad j\in \mathbb{Z}~.
\end{align}
%%%%%%%%%%%%%%%%%%%%%%%%%%%%%%%%%%%%%%%%%%%%%%%%%%%%%%%%%%%%%%%%
It can be easily verified that the above solutions indeed satisfy the necessary boundary conditions, namely, $\widetilde{q}^{\rm (II)}_{\textrm{cl}(j,\pm)}(0)=0$ and $\widetilde{q}^{\rm (II)}_{\textrm{cl}(j,\pm)}(1)=q_{1}$ and this holds for all $j\in\mathbb{Z}$.

%%%%%%%%%%%%%%%%%%%%%%%%%%%%%%%%%%%%%%%%%%%%%%
%%%%%%%%%%%%%%%%%%%%%%%%%%%%%%%%%%%%%%%%%%%%%%
%%%%%%%%%%%%%%%%%%%%%%%%%%%%%%%%%%%%%%%%%%%%%%
%%%%%%%%%%%%%%%%%%%%%%%%%%%%%%%%%%%%%%%%%%%%%%
%%%%%%%%%%%%%%%%%%%%%%%%%%%%%%%%%%%%%%%%%%%%%%
%%%%%%%%%%%%%%%%%%%%%%%%%%%%%%%%%%%%%%%%%%%%%%
%%%%%%%%%%%%%%%%%%%%%%%%%%%%%%%%%%%%%%%%%%%%%%
%%%%%%%%%%%%%%%%%%%%%%%%%%%%%%%%%%%%%%%%%%%%%%

The above analysis determines the saddle points of the classical action, $\widetilde{\mathcal{S}}^{\rm (II)}_{\rm cl}$ in the complex $\mathcal{N}$ plane and the associated classical trajectory. In order to perform saddle point approximation of the no-boundary wave function, we need to first identify the relevant contour, which passes through the saddle points and have vanishing end-point contributions to the kernel. The steepest descent/ascent curves for $i\widetilde{\mathcal{S}}^{\rm (II)}_{\rm cl}(\mathcal{N})$ is depicted in \ref{quad_contour}. Besides, the contours associated with constant values of the $\textrm{Re}[i\widetilde{\mathcal{S}}^{\rm (II)}_{\rm cl}]$ have also been depicted, to better visualize the descent/ascent directions. As prescribed by the Picard-Lefschetz theory, we can deform the original integration contour $\tilde{\mathcal{C}}^{\rm (II)}$ into the red dashed contour $\mathcal{C}^{\rm (II)}$ (see \ref{quad_contour}), without touching the branch-cuts of the pre-exponential factor $\textrm{sech}(3h_{n}\mathcal{N}/2)$ appearing in \ref{derivative_II}. 
Having identified the appropriate contour, we can now evaluate the $\mathcal{N}$-integral using the saddle point approximation in the limit $(V_{3}h_{n}/\ell_{\rm p}^{2})\gg 1$. The dominant contribution to the integral comes from the saddle points $\bar{\mathcal{N}}^{\rm (II)}_{0,+}$ and $\bar{\mathcal{N}}^{\rm (II)}_{0,-}$, since the contributions from $\mathcal{N}^{\rm (II)}_{j,\pm}$ are exponentially suppressed for $j>1$. Therefore, the saddle point approximation of the no-boundary wave function for the quadratic class of bouncing models, is given by:
%%%%%%%%%%%%%%%%%%%%%%%%%%%%%%%%%%%%%%%%%%%%%%%%%%%%%%%%%%%%%%%%
\begin{align}\label{HH_wavefunction_II}
\Psi^{\rm (II)}(q_1)&\sim \left(\frac{e^{i\frac{\pi}{4}}e^{i\widetilde{\mathcal{S}}^{\rm (II)}_{\rm cl}}}{\sqrt{|\cosh\left(\frac{3}{2}h_n\mathcal{N}\right)|}}\right)_{\mathcal{N}_{0,+}}
+\left(\frac{e^{-i\frac{\pi}{4}}e^{i\tilde{\mathcal{S}}^{\rm (II)}_{\rm cl}}}{\sqrt{|\cosh\left(\frac{3}{2}h_n\mathcal{N}\right)|}}\right)_{\mathcal{N}_{0,-}}
\propto \quad e^{\frac{\pi }{3}\alpha_{n}}\frac{ \cos \left[\frac{4}{3}\alpha_{n}\xi(q)-\frac{\pi}{4}\right]}{\sqrt[4]{(\alpha_{n})^{2}(q^2-1)}}~,
\end{align}
%%%%%%%%%%%%%%%%%%%%%%%%%%%%%%%%%%%%%%%%%%%%%%%%%%%%%%%%%%%%%%%%
where,
%%%%%%%%%%%%%%%%%%%%%%%%%%%%%%%%%%%%%%%%%%%%%%%%%%%%%%%%%%%%%%%%
\begin{align}
\xi(q)\equiv\frac{1}{2}q\sqrt{q^2-1}-\frac{1}{2}\log (q+\sqrt{q^2-1})~.
\end{align}
%%%%%%%%%%%%%%%%%%%%%%%%%%%%%%%%%%%%%%%%%%%%%%%%%%%%%%%%%%%%%%%%
Once again, we see that the result is consistent with that obtained from the Wheeler de-Witt equation.

Having derived the no-boundary wave function from the path-integral prescription with an appropriate choice of the integration contours, let us try to provide a physical interpretation to the same. Just as we have obtained in the case of the linear model, the saddle point solution $\widetilde{q}^{\rm (II)}_{\textrm{cl},\pm}(t)$ indeed satisfies the relevant boundary conditions, given by: $\widetilde{q}^{\rm (II)}_{\textrm{cl},\pm}(0)=0$ and $\widetilde{q}^{\rm (II)}_{\textrm{cl},\pm}(1)=q_1$. The corresponding geometry, therefore, acquires the interpretation of an evolution from a point-sized to a finite-sized universe. Once again, one is reminded off the quintessential saddle point geometry of the Euclidean path integral approach to de Sitter cosmology\cite{Hartle:1983ai} and the corresponding wave function of the universe that is usually interpreted as describing tunnelling out of `nothing'. However, with the hindsight of our arguments in the case of the linear model, we maintain that such an interpretation is questionable. This is appreciated once we observe that even though the dominant contribution to the path integral leading to $\Psi^{\rm (II)}(q_1)$ is from the Hartle-Hawking like saddle point geometry, corresponding to a spacetime that emerges from zero size, the off-shell geometries can emerge from any size. This fact is evident from \ref{int_q0_quadratic}, where we have clearly performed a summation over \textit{all} values of the `initial size' $q_0$ to define the wave function. A more reasonable interpretation for $\Psi^{\rm (II)}(q_1)$, in the spirit of \cite{Lehners2019} and in the light of \ref{int_q0_quadratic}, seems to be that it describes a transition from the state of a specific Euclidean momentum. 

%%%%%%%%%%%%%%%%%%%%%%%%%%%%%%%%%%%%%%%%%%%%%%%%%%
%%%%%%%%%%%%%%%%%%%%%%%%%%%%%%%%%%%%%%%%%%%%%%%%%%
%%%%%%%%%%%%%%%%%%%%%%%%%%%%%%%%%%%%%%%%%%%%%%%%%%
%%%%%%%%%%%%%%%%%%%%%%%%%%%%%%%%%%%%%%%%%%%%%%%%%%
\subsection{Shear instability and its quantum analogue}\label{shear_instability}

In \ref{gen_setup_minisuperspace}, we have discussed the metric ansatz, which is spatially homogeneous and isotropic. As a consequence, the geodesics comoving with the Hubble flow are free from any shear. In addition, the effect of any classical shear perturbations to the background FRW spacetime is to add an effective fluid with its density scaling as $a^{-6}$. Therefore, in the standard cosmology, the effects of shear can be safely ignored for sufficiently large values of the scale factor. On the contrary, for bouncing scenarios, the effective energy density arising from the classical shear will inevitably become dominant at sufficiently small values of the scale factor and can potentially affect the bouncing models \cite{Cai_2013}. We shall shortly discuss how this effect translates to in the quantum picture. 

For a brief review of the classical picture along the lines discussed in this work, in particular following the metric ansatz presented in \ref{gen_setup_minisuperspace}, we start with the following parametrization of the Bianchi I universe, 
%%%%%%%%%%%%%%%%%%%%%%%%%%%%%%%%%%%%%%%%%%%%%%%%%%%%%%%%%%%%%%%%
\begin{align}\label{def_Bianchi_I}
ds^{2}=-\frac{\mathcal{N}(t)^{2}}{q(t)^{(4-3b)}}dt^{2}
&+q(t)^{b}\Bigg[\exp\left(\sqrt{\frac{2\kappa}{3}}\left\{\theta_1(t)-\sqrt{3}\theta_2(t)\right\}\right)dx^{2}
\nonumber
\\
&\hskip 1 cm +\exp\left(\sqrt{\frac{2\kappa}{3}} \left\{\theta_1(t)+\sqrt{3}\theta_2(t)\right\}\right)dy^{2}
+\exp\left(-2 \sqrt{\frac{2\kappa}{3}}\theta_1(t)\right)dz^{2}\Bigg]~.
\end{align}
%%%%%%%%%%%%%%%%%%%%%%%%%%%%%%%%%%%%%%%%%%%%%%%%%%%%%%%%%%%%%%%%
It follows that $q(t)^{b/2}$ is the geometric mean of the `scale factors' along the three independent spatial directions. Substitution of the above metric ansatz in the Einstein-Hilbert action with a perfect fluid source, i.e., in the action given by \ref{action_eff} yields:
%%%%%%%%%%%%%%%%%%%%%%%%%%%%%%%%%%%%%%%%%%%%%%%%%%%%%%%%%%%%%%%%
\begin{align}
\mathcal{S}^{\rm (shear)}_{\rm perfect}&=V_3\int dt\left[-\frac{M}{2\mathcal{N}}\left(\frac{dq}{dt}\right)^2+\mathcal{N} U_{\rm eff}(q)\right]dt+V_3\int dt \left[\frac{q^2}{2\mathcal{N}}\left(\frac{d\theta_1}{dt}\right)^2+\frac{q^2}{2\mathcal{N}}\left(\frac{d\theta_2}{dt}\right)^2\right]~.
\end{align}
%%%%%%%%%%%%%%%%%%%%%%%%%%%%%%%%%%%%%%%%%%%%%%%%%%%%%%%%%%%%%%%%
Variation of the above action with respect to the lapse function $\mathcal{N}$, and then setting the lapse function $\mathcal{N}$ to unity, we obtain the classical Hamiltonian constraint to read \footnote{If the spatial slices are closed, in addition to the terms in \ref{Constraint_shear}, we get an extra term given by $q^{2n/(6-n)}U(\theta_1,\theta_2)$ and $q^{(2n+4)/(6-n)}U(\theta_1,\theta_2)$, respectively, for the linear and the quadratic model, with $U(\theta_1,\theta_2)$ describing an effective potential for $\theta_{1,2}$. For sufficiently small values of the scale factor and $\theta_{1,2}$, one can reduce the constraint equation to \ref{Constraint_shear}, if we assume that we can neglect the extra term in comparison to $U_{\rm eff}$. This is possible when $2<n<6$, for both the models.}:
%%%%%%%%%%%%%%%%%%%%%%%%%%%%%%%%%%%%%%%%%%%%%%%%%%%%%%%%%%%%%%%%
\begin{align}\label{Constraint_shear}
-\frac{M }{2}\left(\frac{dq}{dt}\right)^2-U_{\rm eff}(q)+\frac{q^2}{2}\left[\left(\frac{d\theta_1}{dt}\right)^2+\left(\frac{d\theta_2}{dt}\right)^2\right]=0~.
\end{align} 
%%%%%%%%%%%%%%%%%%%%%%%%%%%%%%%%%%%%%%%%%%%%%%%%%%%%%%%%%%%%%%%%
Further, the variation of the action functional with respect to $\theta_{1,2}$ yields the classical equation of motion for $\theta_{1}$ and $\theta_{2}$ as: $\dot{\theta}_{1,2}\propto (1/q^{2})$, where `dot' denotes derivative with respect to $t$. Using this result in the constraint equation, we get:
%%%%%%%%%%%%%%%%%%%%%%%%%%%%%%%%%%%%%%%%%%%%%%%%%%%%%%%%%%%%%%%%
\begin{align}\label{Hamiltonian_shear}
-\frac{M}{2}\left(\frac{dq}{dt}\right)^2- U_{\rm eff}(q)+\frac{\rho_{\theta}}{q^2}=0
\end{align}
%%%%%%%%%%%%%%%%%%%%%%%%%%%%%%%%%%%%%%%%%%%%%%%%%%%%%%%%%%%%%%%%
where, $\rho_{\theta}$ is a constant, related to the proportionality factor between $\dot{\theta}$ and $(1/q^{2})$. Using the expression of the effective energy density $U_{\rm eff}$, as given in \ref{Ueff}, we see that the presence of the $(\rho_{\theta}/q^{2})$ term in \ref{Hamiltonian_shear} can be interpreted as arising from an effective energy density $\rho_{\theta}$, which scales as $q^{-3b}=a^{-6}$. If the value of $\rho_{\theta}$ is sufficiently large, such that $(dq/dt)$ never vanishes, the bouncing scenario becomes classically forbidden. In particular, for the linear model, setting $(dq/dt)=0$, we obtain, $\rho_{0}q^{3}-\rho_{0}q^{2}+\rho_{\theta}=0$. Introducing, a new variable $x$ through the following relation: $q=x+(1/3)$, the above algebraic equation translates into, $x^{3}-(1/3)x+\{-(2/27)+(\rho_{\theta}/\rho_{0})\}=0$. The reduced algebraic equation would have positive real solution consistent with classical bounce, provided $\{-(2/27)+(\rho_{\theta}/\rho_{0})\}^{2}<4(1/3)^{6}$. Thereby, we yield the following condition for a classical bounce to occur:
%%%%%%%%%%%%%%%%%%%%%%%%%%%%%%%%%%%%%%%%%%%%%%%%%%%%%%%%%%%%%%%%
\begin{align}\label{bounce_cond_I}
\frac{\rho_{\theta}}{\rho_0}<\frac{4}{27}~.
\end{align}  
%%%%%%%%%%%%%%%%%%%%%%%%%%%%%%%%%%%%%%%%%%%%%%%%%%%%%%%%%%%%%%%%
On the other hand, for the quadratic model, the relevant algebraic equation becomes, $\rho_{0}q^{4}-\rho_{0}q^{2}+\rho_{\theta}=0$ and hence real and positive solutions for $q$ will exist, provided the following inequality holds,
%%%%%%%%%%%%%%%%%%%%%%%%%%%%%%%%%%%%%%%%%%%%%%%%%%%%%%%%%%%%%%%%
\begin{align}\label{bounce_cond_II}
\frac{\rho_{\theta}}{\rho_0}<\frac{1}{4}~.
\end{align}
%%%%%%%%%%%%%%%%%%%%%%%%%%%%%%%%%%%%%%%%%%%%%%%%%%%%%%%%%%%%%%%%
This finishes our discussion about the effect of shear on the classical FRW spacetime with a bouncing origin, where the contribution from the shear must satisfy the conditions presented in \ref{bounce_cond_I} and \ref{bounce_cond_II} respectively. 

\begin{figure}[t]
	\centering
	\includegraphics[scale=0.25]{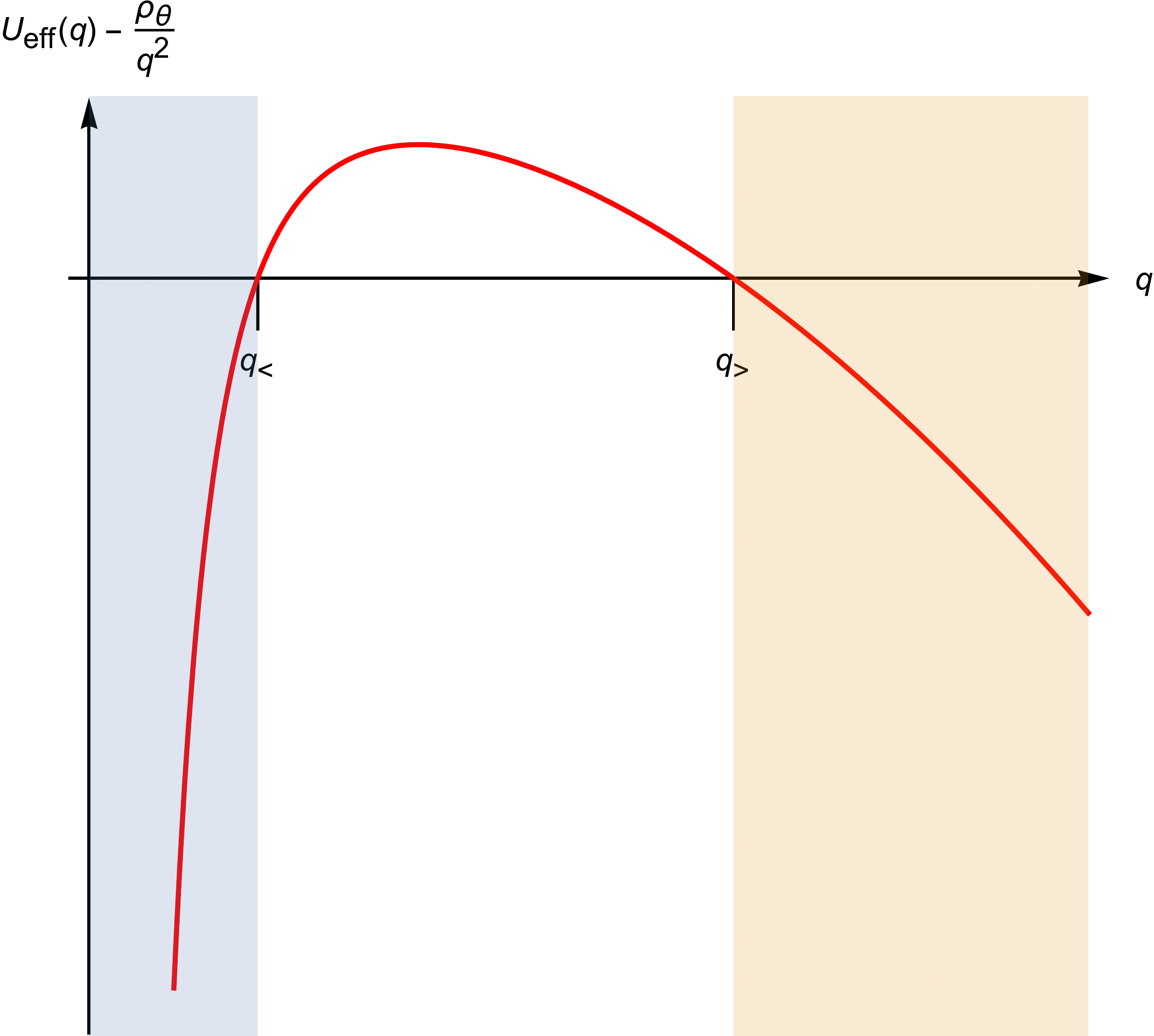}
	\caption{The typical form of the total effective potential $U_{\rm total}=U_{\rm eff}-(\rho_{\theta}/q^{2})$ has been demonstrated, when a classical bouncing scenario is allowed. The two classical turning points are marked $q_{<}$ and $q_{>}$, respectively and the shaded portions with $q>q_{>}$ and $q<q_{<}$ denote the classically allowed regions. The (blue) region to the left of the plot ($0<q<q_{<}$) describes a big crunch scenario, wherein the classical domain the universe started from a zero size, reaches a maximum $q$ value, and then reduces back to zero. While the (red) region to the right ($q_{>}<q<\infty$) describes the bouncing scenario, where the universe reaches a minimum value of $q$ and then re-expands. Classically, a solution of the relevant Einstein's equations describes either of these two scenarios. However, quantum mechanically, tunnelling from one region to the other is allowed.}
	\label{U_T_graph}
\end{figure}

We shall now look at the quantum analysis of the problem and the starting point of the same is the relevant Wheeler-de Witt equation, which in the present context takes the following form,
%%%%%%%%%%%%%%%%%%%%%%%%%%%%%%%%%%%%%%%%%%%%%%%%%%%%%%%%%%%%%%%d
\begin{align}\label{WdW2}
\left[-\frac{1}{2 M V_3}\left(\frac{\partial^2}{\partial q^{2}}\right)+V_3 U_{\rm eff}(q)+\frac{1}{2q^2V_3}\nabla^2\right]\Psi(q,\theta_1,\theta_2)=0~,
\end{align} 
%%%%%%%%%%%%%%%%%%%%%%%%%%%%%%%%%%%%%%%%%%%%%%%%%%%%%%%%%%%%%%%%
where, $\nabla^{2}\equiv(\partial/\partial \theta_{1})^{2}+(\partial/\partial \theta_{2})^{2}$. Given the above differential equation, it is evident that the $(\theta_{1},\theta_{2})$ sector and the main gravitational degree of freedom $q$, do not interact with each other. Hence, it is advisable to look for separable solutions of the Wheeler-DeWitt wave function $\Psi(q,\theta_{1},\theta_{2})$ of the following form,
%%%%%%%%%%%%%%%%%%%%%%%%%%%%%%%%%%%%%%%%%%%%%%%%%%%%%%%%%%%%%%%%
\begin{align}
\Psi(q,\theta_{1},\theta_{2})=\psi_{|\bf{k}|}(q)e^{iV_{3}( k_1\theta_1+k_2\theta_2)}~,
\end{align}
%%%%%%%%%%%%%%%%%%%%%%%%%%%%%%%%%%%%%%%%%%%%%%%%%%%%%%%%%%%%%%%%
where, $|\bm{k}|=\sqrt{k_{1}^{2}+k_{2}^{2}}$. Then $\psi_{|\bf{k}|}(q)$ can be shown to satisfy the following one-dimensional Schr\"{o}dinger equation:
%%%%%%%%%%%%%%%%%%%%%%%%%%%%%%%%%%%%%%%%%%%%%%%%%%%%%%%%%%%%%%%%
\begin{align}\label{shear_WdW}
\left[-\frac{1}{2 M V_3}\left(\frac{\partial^2}{\partial q^{2}}\right)+V_{3}U_{\rm total}(q)\right]\psi_{|\bf{k}|}(q)=0~,
\end{align}  
%%%%%%%%%%%%%%%%%%%%%%%%%%%%%%%%%%%%%%%%%%%%%%%%%%%%%%%%%%%%%%%%
where, the total effective potential $U_{\rm total}(q)$ is given by,
%%%%%%%%%%%%%%%%%%%%%%%%%%%%%%%%%%%%%%%%%%%%%%%%%%%%%%%%%%%%%%%%
\begin{align}
U_{\rm total}(q)\equiv U_{\rm eff}(q)-\frac{\rho_{\theta}}{q^2}~,
\end{align}
%%%%%%%%%%%%%%%%%%%%%%%%%%%%%%%%%%%%%%%%%%%%%%%%%%%%%%%%%%%%%%%%
with the density $\rho_{\theta} \equiv (|\bm{k}|^{2}/2)$. We see that this is consistent with our classical analysis of the problem. Once again, we obtain the conditions \ref{bounce_cond_I} and \ref{bounce_cond_II}, respectively, for the bounce to occur in the linear and quadratic models. The typical form of the total effective potential $U_{\rm total}(q)$, when these conditions are met, is shown in \ref{U_T_graph}.

%%%%%%%%%%%%%%%%%%%%%%%%%%%%%%%%%%%%%%%%%%%%%%
%%%%%%%%%%%%%%%%%%%%%%%%%%%%%%%%%%%%%%%%%%%%%%
%%%%%%%%%%%%%%%%%%%%%%%%%%%%%%%%%%%%%%%%%%%%%%
%%%%%%%%%%%%%%%%%%%%%%%%%%%%%%%%%%%%%%%%%%%%%%
%%%%%%%%%%%%%%%%%%%%%%%%%%%%%%%%%%%%%%%%%%%%%%
%%%%%%%%%%%%%%%%%%%%%%%%%%%%%%%%%%%%%%%%%%%%%%
%%%%%%%%%%%%%%%%%%%%%%%%%%%%%%%%%%%%%%%%%%%%%%
%%%%%%%%%%%%%%%%%%%%%%%%%%%%%%%%%%%%%%%%%%%%%%

\begin{figure}[t]
	\centering
	\includegraphics[scale=.3]{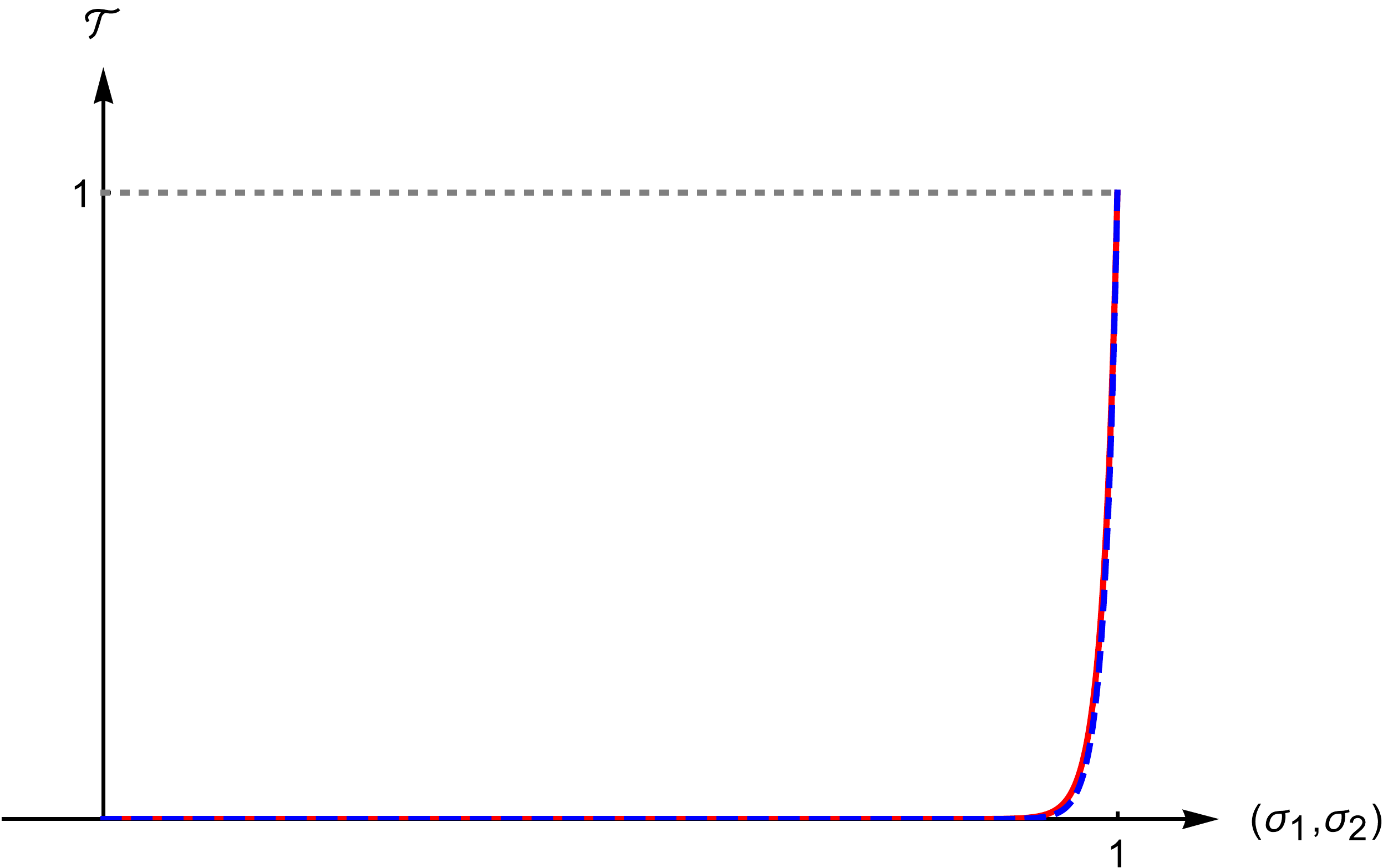}
	\caption{ The red graph shows $\mathcal{T}(\sigma_1)$ and the blue dashed graph shows $\mathcal{T}(\sigma_2)$, with $V_3\sqrt{2M\rho_0}$ fixed to a value of 50 in both the cases. Note, however, that the WKB tunnelling probability cannot be trusted for $\sigma_{1},\sigma_{2}\approx 1$, since, in this regime WKB approximation fails.}
	\label{T_sigma_graph}
\end{figure}

Even though the bouncing scenario can be realized classically for the parameter ranges presented in \ref{bounce_cond_I} and \ref{bounce_cond_II}, it is evident from \ref{U_T_graph} that there is also a region that corresponds to a classical crunch. Classically, however, a solution to Einstein's equations will correspond to either one of the two scenarios. Quantum mechanically, on the other hand, tunnelling is possible from one region to another. Similar quantum tunnelling scenarios in cosmology has been previously considered in the literature, for instance, in the context of a Brane-like universe \cite{davidson1999wavefunction}. We reserve the exact analysis of the current problem for a future publication, however, we present here the WKB probability $\mathcal{T}$ for tunnelling from the bouncing regime to the regime depicting classical crunch:  
%%%%%%%%%%%%%%%%%%%%%%%%%%%%%%%%%%%%%%%%%%%%%%%%%%%%%%%%%%%%%%%%
\begin{align}\label{tunelling_prob}
\mathcal{T}=\exp\left(-2\int_{q_{<}}^{q_{>}}V_3\sqrt{2MU_{\rm total}(q)}~dq\right)\equiv\exp\left[-2V_3\sqrt{2M\rho_0}~\Gamma\right]~,
\end{align}
%%%%%%%%%%%%%%%%%%%%%%%%%%%%%%%%%%%%%%%%%%%%%%%%%%%%%%%%%%%%%%%%
where, $q_{<}$ and $q_{>}$ are classical turning points such that, $0<q_{<}<q_{>}$ (see also \ref{U_T_graph}). In the light of \ref{bounce_cond_I} and \ref{bounce_cond_II}, it is convenient to define the following parameters: $\sigma_{1}\equiv\sqrt{(27\rho_{\theta}/4\rho_{0})}$ and $\sigma_{2}\equiv\sqrt{(4\rho_{\theta}/\rho_{0})}$, respectively, for the linear and quadratic models, such that $0<\sigma_1,\sigma_2<1$. It is now straightforward to evaluate the defining integral in the exponential of \ref{tunelling_prob} and hence determine $\Gamma(\sigma_{1})$ and $\Gamma(\sigma_{2})$ respectively. These can be expressed in terms of the Elliptic integrals with the following expressions:  
%%%%%%%%%%%%%%%%%%%%%%%%%%%%%%%%%%%%%%%%%%%%%%%%%%%%%%%%%%%%%%%%
\begin{align}
\Gamma(\sigma_1)&\approx \left(\alpha_{1}-\alpha_{2}\sigma_{1}\right)+\sigma_{1}^{2}\left(\beta_{1}+\beta_{2}\ln\sigma_1\right)
\\
\Gamma(\sigma_2)&=\frac{\pi}{4}(1-\sigma_2)
\end{align}
%%%%%%%%%%%%%%%%%%%%%%%%%%%%%%%%%%%%%%%%%%%%%%%%%%%%%%%%%%%%%%%%
where, the coefficients appearing in the expression for $\Gamma(\sigma_{1})$ has the following numerical expressions: $\alpha_{1}=0.666706$, $\alpha_{2}=0.607517$, $\beta_{1}=-0.059189$ and $\beta_{2}=0.028686$, respectively. For illustration purpose, for a fixed value of $V_{3}\sqrt{2M\rho_0}$, the typical behaviour of the transition probability $\mathcal{T}$ as function of $\sigma_{1}$ and $\sigma_{2}$, for the linear and the quadratic model are given in \ref{T_sigma_graph}.  As evident, when both $\sigma_{1}$ and $\sigma_{2}$ are near unity, i.e., when $\rho_{\theta}$ is comparable to $\rho_{0}$, the tunnelling probability increases. However, we caution that the WKB tunnelling amplitude cannot be trusted for $\sigma_{1},\sigma_{2}$ in the neighbourhood of 1, since, there the WKB approximation essentially fails. Moreover, by virtue of an effect analogous to the `reflection over-the-barrier' in standard quantum mechanics, the tunnelling probability $\mathcal{T}$ for $\sigma_{1},\sigma_{2}>1$ is \textit{less} that 1, despite the fact that in this parameter range classical bounce is not possible. This interesting aspect, which is a direct consequence of quantum cosmology, warrant further study and, hence, shall be explored in a future publication. 

%%%%%%%%%%%%%%%%%%%%%%%%%%%%%%%%%%%%%%%%%%%%%%
%%%%%%%%%%%%%%%%%%%%%%%%%%%%%%%%%%%%%%%%%%%%%%
%%%%%%%%%%%%%%%%%%%%%%%%%%%%%%%%%%%%%%%%%%%%%%
%%%%%%%%%%%%%%%%%%%%%%%%%%%%%%%%%%%%%%%%%%%%%%
%%%%%%%%%%%%%%%%%%%%%%%%%%%%%%%%%%%%%%%%%%%%%%
%%%%%%%%%%%%%%%%%%%%%%%%%%%%%%%%%%%%%%%%%%%%%%
%%%%%%%%%%%%%%%%%%%%%%%%%%%%%%%%%%%%%%%%%%%%%%
%%%%%%%%%%%%%%%%%%%%%%%%%%%%%%%%%%%%%%%%%%%%%%

%%%%%%%%%%%%%%%%%%%%%%%%%%%%%%%%%%%%%%%%%%%%%%%%%%%%%%%%%%%%%%%%%%%%%%%%%%%%%%%%%%%%%%%%%%%%%%%%%%%
%%%%%%%%%%%%%%%%%%%%%%%%%%%%%%%%%%%%%%%%%%%%%%%%%%%%%%%%%%%%%%%%%%%%%%%%%%%%%%%%%%%%%%%%%%%%%%%%%%%
%%%%%%%%%%%%%%%%%%%%%%%%%%%%%%%%%%%%%%%%%%%%%%%%%%%%%%%%%%%%%%%%%%%%%%%%%%%%%%%%%%%%%%%%%%%%%%%%%%%
\section{Introducing matter field}\label{intro_matter}

So far, we have discussed the quantum cosmology of a universe with the matter source being a perfect fluid with energy density $\rho_{\rm eff}(a)$, given by \ref{rho_I} and \ref{rho_II}, respectively, such that the action functional becomes at most quadratic in the dynamical variable $q$. However, there can be further fundamental fields living in the spacetime, and for simplicity, we choose a single conformally coupled scalar field. This will drive home the essential features arising out of addition of such matter fields in the quantum analysis of the bouncing model. Thus our starting point will be the Wheeler-DeWitt equation presented in \ref{WdW}, on which we will apply the WKB expansion and hence shall determine the behaviour of the matter fields in the classically allowed and in the classically forbidden regions explicitly. First of all, we will provide the key equations arising out of the WKB expansion of the Wheeler-DeWitt equation in the next section.

%%%%%%%%%%%%%%%%%%%%%%%%%%%%%%%%%%%%%%%%%%%%%%%%%%
%%%%%%%%%%%%%%%%%%%%%%%%%%%%%%%%%%%%%%%%%%%%%%%%%%
%%%%%%%%%%%%%%%%%%%%%%%%%%%%%%%%%%%%%%%%%%%%%%%%%%
%%%%%%%%%%%%%%%%%%%%%%%%%%%%%%%%%%%%%%%%%%%%%%%%%%
\subsection{Wheeler-DeWitt equation with matter and the WKB expansion}

As we have already remarked, we shall be treating the matter field $\phi$, which is a conformally coupled scalar field appearing in \ref{Full_Action_Final}, as a test field in a given background. The precise sense in which one can perform this analysis is through WKB expansion and it will be convenient for our purpose to follow the approach of \cite{Singh:1989ct,Padmanabhan:1990fn}. The starting point being the following ansatz for the Wheeler-DeWitt wave function $\Psi(q;\{\phi_{k}\})$ as a solution to \ref{WdW}, with the effective potentials of the form given in \ref{U_eff_I} or \ref{U_eff_II}: 
%%%%%%%%%%%%%%%%%%%%%%%%%%%%%%%%%%%%%%%%%%%%%%%%%%%%%%%%%%%%%%%%
\begin{align}
\Psi(q;\{\phi_{k}\})=\exp\left[\frac{iV_3}{\ell_{\rm p}^2}\sum_{j=0}^{\infty}\ell_{\rm p}^{2j}S_j(q,\{\phi_{k}\})\right]~.
\end{align}
%%%%%%%%%%%%%%%%%%%%%%%%%%%%%%%%%%%%%%%%%%%%%%%%%%%%%%%%%%%%%%%%
We now demand that $\Psi(q;\{\phi_{k}\})$ solves the Wheeler-DeWitt equation at all orders in $\ell_{\rm p}^2$. At the first three leading orders, namely $\ell_{\rm p}^{-4}$, $\ell_{\rm p}^{-2}$ and $\ell_{\rm p}^{0}$,  implies:
%%%%%%%%%%%%%%%%%%%%%%%%%%%%%%%%%%%%%%%%%%%%%%%%%%%%%%%%%%%%%%%%
\begin{align}
\partial_{\phi_{k}}S_0&=0~;
\\
\left(\partial_{q}S^{\rm (I)}_{0}\right)^{2}&+\frac{h_{n}^{2}}{\left(\frac{n}{6}-1\right)^{4}}\left(1-q\right)=0~;
\label{HJ_equation_01}
\\
\left(\partial_{q}S^{\rm (II)}_{0}\right)^{2}&+\left(\frac{16}{9}\right)\frac{h_{n}^{2}}{\left(\frac{n}{6}-1\right)^{4}}\left(1-q^{2}\right)=0~;
\label{HJ_equation_02}
\\
\label{T-dependent_Schrodinger}
-\frac{2i\left(\partial_{q}S_{0}\right)}{b^{2}}\partial_{q}\zeta(q,\{\phi_{k}\})&=V_3\int \frac{d^3p}{(2\pi)^3}\left(-\frac{1}{2\mu(q)}\partial_{\phi_{p}}^2+\frac{1}{2}\mu(q)\omega_p(q)^2\phi_{p}^2\right)\zeta(q,\{\phi_{k}\})~,
\end{align}
%%%%%%%%%%%%%%%%%%%%%%%%%%%%%%%%%%%%%%%%%%%%%%%%%%%%%%%%%%%%%%%%
where the function $\zeta(q,\{\phi_{k}\})$ is defined as,
%%%%%%%%%%%%%%%%%%%%%%%%%%%%%%%%%%%%%%%%%%%%%%%%%%%%%%%%%%%%%%%%
\begin{align}
\zeta(q,\{\phi_{k}\})\equiv\frac{\sqrt{S_0'(q)}}{\sqrt{V_3}}e^{iV_3S_1(q,\{\phi_{k}\})}~.
\end{align}
%%%%%%%%%%%%%%%%%%%%%%%%%%%%%%%%%%%%%%%%%%%%%%%%%%%%%%%%%%%%%%%%
The first equation $\partial_{\phi_{k}}S_{0}=0$, arising from setting the coefficient of $\ell_{\rm p}^{-4}$ to zero, merely tells us that the zeroth order contribution to the action, i.e., $S_{0}$ is independent of the matter degrees of freedom $\phi_{k}$. In addition, \ref{HJ_equation_01} and \ref{HJ_equation_02} are the Hamilton-Jacobi equations for the two scenarios of interest, as discussed in earlier sections. Finally, one can introduce the time coordinate $t$ in the present formalism via the following relation:
%%%%%%%%%%%%%%%%%%%%%%%%%%%%%%%%%%%%%%%%%%%%%%%%%%%%%%%%%%%%%%%%
\begin{align}\label{def_time}
-\frac{2\left(\partial_{q}S_{0}\right)}{b^{2}}=\frac{dq}{dt}~.
\end{align}
%%%%%%%%%%%%%%%%%%%%%%%%%%%%%%%%%%%%%%%%%%%%%%%%%%%%%%%%%%%%%%%%
With this definition, it is easy to see that \ref{HJ_equation_01} and \ref{HJ_equation_02} are equivalent to the classical constraint equation, given by \ref{classical_constraint}, with the matter field neglected. Moreover, the third equation, namely \ref{T-dependent_Schrodinger}, takes form of the following time-dependent Schr\"{o}dinger equation:
%%%%%%%%%%%%%%%%%%%%%%%%%%%%%%%%%%%%%%%%%%%%%%%%%%%%%%%%%%%%%%%%
\begin{align}
\label{T-dependent_Schrodinger_2}
i\partial_{t}\zeta&=V_3\int \frac{d^3p}{(2\pi)^3}\left(-\frac{1}{2\mu(q(t))}\partial_{\phi_{p}}^2+\frac{1}{2}\mu(q(t))\omega^2_p(q(t))\phi_{p}^2\right)\zeta
\end{align}
%%%%%%%%%%%%%%%%%%%%%%%%%%%%%%%%%%%%%%%%%%%%%%%%%%%%%%%%%%%%%%%%
where, $q(t)$ corresponds to the classical solutions of the purely gravitational part arising from \ref{HJ_equation_01} and \ref{HJ_equation_02}, respectively. Therefore, \ref{T-dependent_Schrodinger_2} mathematically describes the paradigm of quantum field theory, with $\phi_{k}$ being the Fourier modes of the field in a time dependent background spacetime, which has been discussed extensively in the literature (see, for example, \cite{birrell1984quantum,parker2009quantum,wald1994quantum,fulling1989aspects}). It is instructive to take the following Gaussian ansatz for the solution of \ref{T-dependent_Schrodinger_2},
%%%%%%%%%%%%%%%%%%%%%%%%%%%%%%%%%%%%%%%%%%%%%%%%%%%%%%%%%%%%%%%%
\begin{align}\label{zeta_gaussian}
\zeta(q(t);\{\phi_{k}\})\propto \exp\left[V_3\int \frac{d^3k}{(2\pi)^3}\left(i \mu(t)\frac{\dot{u}_k}{2u_k}\phi_{k}^2\right)\right]~.
\end{align}
%%%%%%%%%%%%%%%%%%%%%%%%%%%%%%%%%%%%%%%%%%%%%%%%%%%%%%%%%%%%%%%% 
For the above ansatz to solve \ref{T-dependent_Schrodinger_2}, the function $u_{k}(t)$ must satisfy the following differential equation:
%%%%%%%%%%%%%%%%%%%%%%%%%%%%%%%%%%%%%%%%%%%%%%%%%%%%%%%%%%%%%%%%
\begin{align}\label{KG1}
\ddot{u}_k+\frac{\dot{\mu}}{\mu}\dot{u}_k+\omega_k^2u_k=0~,
\end{align}
%%%%%%%%%%%%%%%%%%%%%%%%%%%%%%%%%%%%%%%%%%%%%%%%%%%%%%%%%%%%%%%%
where, $\mu(t)\equiv\mu(q(t))$, $\omega_k^2(t)\equiv \omega_k^2(q(t))$ and the dot denotes a derivative with respect to the time coordinate $t$ introduced in \ref{def_time}. Since, we want the solution for $\zeta(q(t);\{\phi_{k}\})$, presented in \ref{zeta_gaussian} to depict a stable situation, the real part of the argument of the exponential must be negative \cite{WADA1986729} and thus, in addition to \ref{KG1}, the solution $u_k$ must satisfy the following condition:
%%%%%%%%%%%%%%%%%%%%%%%%%%%%%%%%%%%%%%%%%%%%%%%%%%%%%%%%%%%%%%%%
\begin{align}\label{stability_cond}
\textrm{Re}\left[i \mu(t)\frac{\dot{u}_k}{u_k}\right]<0~.
\end{align}
%%%%%%%%%%%%%%%%%%%%%%%%%%%%%%%%%%%%%%%%%%%%%%%%%%%%%%%%%%%%%%%%
In summary, we find that the leading order solution in $\ell_{\rm p}$ of the Wheeler-DeWitt equation can be written as:
%%%%%%%%%%%%%%%%%%%%%%%%%%%%%%%%%%%%%%%%%%%%%%%%%%%%%%%%%%%%%%%%
\begin{align}\label{Sol_WDW_matter}
\Psi(q,\{\phi_{k}\})\approx\frac{\sqrt{V_3}}{\sqrt{\partial_{q}S_{0}}}e^{i\frac{V_{3}}{\ell_{\rm p}^{2}}S_{0}(q)}\zeta(q,\{\phi_{k}\})~,
\end{align}
%%%%%%%%%%%%%%%%%%%%%%%%%%%%%%%%%%%%%%%%%%%%%%%%%%%%%%%%%%%%%%%%
with $S_{0}(q)$ satisfying \ref{HJ_equation_01} or, \ref{HJ_equation_02} depending on the nature of the effective potential and $\zeta$ satisfies \ref{T-dependent_Schrodinger_2}. In the subsequent sections, we will analyze the nature of the solution $u_{k}$, which sits in the function $\zeta(q(t);\{\phi_{k}\}) $ to completely understand the behaviour of $\Psi(q,\{\phi_{k}\})$. Since the effective potentials $U_{\rm eff}(q)$ has classical turning points, pertaining to the bouncing models we are considering, it will be convenient to analyze the WKB solutions in the classically allowed and forbidden regions separately.

%%%%%%%%%%%%%%%%%%%%%%%%%%%%%%%%%%%%%%%%%%%%%%%%%%
%%%%%%%%%%%%%%%%%%%%%%%%%%%%%%%%%%%%%%%%%%%%%%%%%%
%%%%%%%%%%%%%%%%%%%%%%%%%%%%%%%%%%%%%%%%%%%%%%%%%%
%%%%%%%%%%%%%%%%%%%%%%%%%%%%%%%%%%%%%%%%%%%%%%%%%%
\subsection{Classically allowed region}

For the bouncing models under consideration, the classically allowed region corresponds to the following range of the dynamical variable: $1<q<\infty$. As evident from \ref{Ueff_plot}, the effective potentials appearing in the Wheeler-DeWitt equation, i.e., $U_{\rm eff}^{\rm (I)}(q)$ and $U_{\rm eff}^{\rm (II)}(q)$, respectively take negative values and therefore, the solutions to \ref{HJ_equation_01} and \ref{HJ_equation_02}, respectively, takes the form:
%%%%%%%%%%%%%%%%%%%%%%%%%%%%%%%%%%%%%%%%%%%%%%%%%%%%%%%%%%%%%%%%
\begin{align}\label{S_0_integral}
S^{\rm (I)}_{0,\pm}(q)=\pm\beta_{n}\ell^2_p\int_{1}^{q(>1)}\sqrt{-(1-q')}dq'~; \qquad	S^{\rm (II)}_{0,\pm}(q)=\pm\frac{4}{3}\beta_{n}\ell^2_p\int_{1}^{q(>1)}\sqrt{-(1-q'^2)}dq'~.
\end{align} 
%%%%%%%%%%%%%%%%%%%%%%%%%%%%%%%%%%%%%%%%%%%%%%%%%%%%%%%%%%%%%%%%
Here, we have used the definition of $\beta_{n}$, from \ref{def_beta_n}. From \ref{def_time}, it follows that for $(\partial_{q}S_{0})<0$, we have $(dq/dt)>0$, i.e., as the action decreases with $q$, the universe expands. Thus we notice that, $S^{\rm (I,II)}_{0,-}(q)$ corresponds to the expanding phase, while, $S^{\rm (I,II)}_{0,+}(q)$ corresponds to the contracting phase of the bouncing universe. Keeping this in mind, we integrate, \ref{S_0_integral} and then substituted the same in \ref{Sol_WDW_matter}, to obtain the general solution to the Wheeler-DeWitt equation at $\mathcal{O}(\ell_{\rm p}^{0})$ as:
%%%%%%%%%%%%%%%%%%%%%%%%%%%%%%%%%%%%%%%%%%%%%%%%%%%%%%%%%%%%%%%%
\begin{align}
\Psi^{\rm (I)}(q;\{\phi_{k}\})&\approx 
A^{\rm (I)}_{+}\frac{e^{\frac{2i}{3}\alpha_{n}(q-1)^{\frac{3}{2}}}}{\left[\alpha_{n}^{2}\left(q-1\right)\right]^{\frac{1}{4}}}\zeta^{\rm (I)}_+(q;\{\phi_{k}\})
+A^{\rm (I)}_{-}\frac{e^{-\frac{2i}{3}\alpha_{n}(q-1)^{\frac{3}{2}}}}{\left[\alpha_{n}^{2}\left(q-1\right)\right]^{\frac{1}{4}}}\zeta^{\rm (I)}_{-}(q;\{\phi_{k}\})~;
\label{WDW_m1}
\\
\Psi^{\rm (II)}(q;\{\phi_{k}\})&\approx 
A^{\rm (II)}_{+}\frac{e^{\frac{4i}{3}\alpha_{n}\xi(q)}}{\left[\alpha_{n}^{2}\left(q^{2}-1\right)\right]^{\frac{1}{4}}}\zeta^{\rm (II)}_{+}(q;\{\phi_{k}\})
+A^{\rm (II)}_{-}\frac{e^{-\frac{4i}{3}\alpha_{n}\xi(q)}}{\left[\alpha_{n}^{2}\left(q^{2}-1\right)\right]^{\frac{1}{4}}}\zeta^{\rm (II)}_{-}(q;\{\phi_{k}\})~.
\label{WDW_m2}
\end{align}   
%%%%%%%%%%%%%%%%%%%%%%%%%%%%%%%%%%%%%%%%%%%%%%%%%%%%%%%%%%%%%%%%
In the above solutions of the Wheeler-DeWitt equation with matter fields, $A^{\rm (I)}_{\pm}$ and $A^{\rm (II)}_{\pm}$ are arbitrary constants, $\alpha_{n}\equiv V_{3}\beta_{n}$, with $\beta_{n}$ defined in \ref{def_beta_n} and $\xi(q)$ is a function of the expansion parameter $q$, given by \ref{def_xi_q}. In order to fix the unknown constants, we may impose the boundary condition that in the absence of the conformally coupled scalar field, the Wheeler-DeWitt wave function must take the form presented in \ref{HH_wavefunction_I} and \ref{HH_wavefunction_II}, respectively for the two choices of the effective potential. This yields the following conditions on the unknown coefficients:  $A^{\rm (I)}_{+}=A^{\rm (I)}_{-}\equiv A^{\rm (I)}$ and $A^{\rm (II)}_{+}=A^{\rm (II)}_{-}\equiv A^{\rm (II)}$. As emphasized above, these conditions ensure that the wave functions reduce to that of the Hawking-Hartle prescription when matter fields are neglected.

What remains, is to find out the appropriate solutions $\zeta(t)$ to \ref{T-dependent_Schrodinger_2}.  As remarked earlier, we shall take $\zeta(t)$ to be of the Gaussian form presented in \ref{zeta_gaussian} and hence the task of finding a solution for $\zeta(t)$ boils down to finding a solution for $u_{k}(t)$ from \ref{KG1}. Let us denote the independent solutions to \ref{KG1} for the two effective potentials, as $u^{\rm (I)}_{k,\pm}$ and $u^{\rm (II)}_{k,\pm}$, respectively. These in turn will provide us the functions $\zeta^{\rm (I)}_{\pm}$ and $\zeta^{\rm (II)}_{\pm}$, used in \ref{WDW_m1} and \ref{WDW_m2}, respectively. Thus the time dependent harmonic oscillator equation, \ref{KG1} for the functions $u^{\rm (I)}_{k,\pm}$ and  $u^{\rm (II)}_{k,\pm}$ take the following form: 
%%%%%%%%%%%%%%%%%%%%%%%%%%%%%%%%%%%%%%%%%%%%%%%%%%%%%%%%%%%%%%%%
\begin{align}
\ddot{u}^{\rm (I)}_{k,\pm}+\frac{\dot{\mu}^{\rm (I)}}{\mu^{\rm (I)}}\dot{u}^{\rm (I)}_{k,\pm}+\left[\omega^{\rm (I)}_{k,\pm}\right]^2u^{\rm (I)}_{k,\pm}=0~,
\label{KG_I}
\\
\ddot{u}^{\rm (II)}_{k,\pm}+\frac{\dot{\mu}^{\rm (II)}}{\mu^{\rm (II)}}\dot{u}^{\rm (II)}_{k,\pm}+\left[\omega^{\rm (II)}_{k,\pm}\right]^2u^{\rm (II)}_{k,\pm}=0~,
\label{KG_II}
\end{align}
%%%%%%%%%%%%%%%%%%%%%%%%%%%%%%%%%%%%%%%%%%%%%%%%%%%%%%%%%%%%%%%%
where, the time dependent mass function $\mu(t)$ and the time dependent frequency function $\omega_{k}(t)$ has the following expressions, for the two choices of the effective potentials, appearing in the Wheeler-DeWitt equation:
%%%%%%%%%%%%%%%%%%%%%%%%%%%%%%%%%%%%%%%%%%%%%%%%%%%%%%%%%%%%%%%%
\begin{align}
\mu^{\rm (I)}(t)&=\left(1+h_{n}^{2}t^{2}\right)^{\frac{2 (n-3)}{(n-6)}}~;
\qquad
\omega^{\rm (I)}_{k,\pm}(t)^{2}=\left(1+h_{n}^{2}t^{2}\right)^{-\frac{4(n-3)}{(n-6)}}\left[k^{2}+m^{2}\left(1+h_{n}^{2}t^{2}\right)^{\frac{6}{6-n}}\right]~.
\\
\mu^{\rm (II)}(t)&=\left[\cosh\left(\frac{3}{2}h_{n}t\right)\right]^{\frac{2 (n-2)}{(n-6)}}~;
\qquad
\omega^{\rm (II)}_{k,\pm}(t)^{2}=\left[\cosh\left(\frac{3}{2}h_nt\right)\right]^{-\frac{4 (n-2)}{(n-6)}}\left\{k^2+m^2\left[\cosh\left(\frac{3}{2}h_nt\right)\right]^{\frac{8}{6-n}}\right\}~.
\end{align}
%%%%%%%%%%%%%%%%%%%%%%%%%%%%%%%%%%%%%%%%%%%%%%%%%%%%%%%%%%%%%%%%
The solutions to \ref{KG_I} and \ref{KG_II}, which may lead to a stable wave function for the Wheeler-DeWitt equation are obtained by imposing the additional condition as in \ref{stability_cond} on $u_{k}$. We shall now study various properties of the function $u_{k}$, leading to stable Wheeler-DeWitt wave functions.

For notational convenience, we shall momentarily ignore the subscripts `(I)' and `(II)' until absolutely necessary. We closely follow the strategy described in \cite{Padmanabhan:2017rsj}, where solution generating technique of a time-dependent harmonic oscillator has been studied in great detail.  The general solution to a time-dependent harmonic oscillator equation of the form \ref{KG1} can be written in terms of a complex function $v_{k}(t)$ as follows:
%%%%%%%%%%%%%%%%%%%%%%%%%%%%%%%%%%%%%%%%%%%%%%%%%%%%%%%%%%%%%%%%
\begin{align}
u_{k}(t)= v^*_k(t)+B_k\,v_k(t)~,
\end{align}
%%%%%%%%%%%%%%%%%%%%%%%%%%%%%%%%%%%%%%%%%%%%%%%%%%%%%%%%%%%%%%%%
where, $B_{k}$ is a constant and can be treated as the ratio between two Bogoliubov coefficients among the mode functions $u_{k}$ and $v_{k}$. Additionally, we consider the functions $v_k$ and $v_k^*$ as linearly independent solutions of \ref{KG1}, that satisfies the following Wronskian condition:
%%%%%%%%%%%%%%%%%%%%%%%%%%%%%%%%%%%%%%%%%%%%%%%%%%%%%%%%%%%%%%%%
\begin{align}\label{wronsk_cond}
i\mu(t)\left[v^*_k\dot{v}_k-v_k\dot{v}^*_k\right]=1~.
\end{align}
%%%%%%%%%%%%%%%%%%%%%%%%%%%%%%%%%%%%%%%%%%%%%%%%%%%%%%%%%%%%%%%%
For the solution $u_{k}$, presented above, we obtain: $\textrm{Re}[i\mu(t)(\dot{u}_{k}/u_{k})]=\{(|B_{k}|^{2}-1)/2|u_{k}|^{2}\}$. Thus imposing the stability condition as presented in \ref{stability_cond}, on the solution $u_{k}$, we obtain,
%%%%%%%%%%%%%%%%%%%%%%%%%%%%%%%%%%%%%%%%%%%%%%%%%%%%%%%%%%%%%%%%
\begin{align}
|B_{k}|^{2}<1~.
\end{align}
%%%%%%%%%%%%%%%%%%%%%%%%%%%%%%%%%%%%%%%%%%%%%%%%%%%%%%%%%%%%%%%%
It is worth mentioning that one can also express the functions $v_k(t)$ that satisfy \ref{wronsk_cond} in terms of two real functions $\chi_{k}(t)$ and $\Theta_{k}(t)$ as follows: $v_k(t)=\chi_{k}(t)e^{-i \Theta_k(t)}$. Here, the phase factor $\Theta_{k}(t)$ is related to the amplitude $\chi_{k}(t)$ through the following relation: $\dot{\Theta}_k(t)=\{1/2\mu(t)\chi^2_k(t)\}$ and the amplitude $\chi_{k}(t)$ satisfies the differential equation of a time-dependent harmonic oscillator with a source term, $\{1/4\mu(t)^{2}\chi^{3}_{k}(t)\}$. With this notation, general solutions $u^{\rm (I)}_{k,\pm}$ and  $u^{\rm (II)}_{k,\pm}$ can be written as,
%%%%%%%%%%%%%%%%%%%%%%%%%%%%%%%%%%%%%%%%%%%%%%%%%%%%%%%%%%%%%%%%
\begin{align}\label{gen_sol_matter}
u^{\rm (I,II)}_{k,\pm}(t)=\chi^{\rm (I,II)}_{k}(t)e^{i\Theta^{\rm (I,II)}_{k}(t)}+B^{\rm (I,II)}_{k,\pm}\,\chi^{\rm (I,II)}_{k}(t)e^{-i\Theta^{\rm (I,II)}_{k}(t)}~,
\end{align}
%%%%%%%%%%%%%%%%%%%%%%%%%%%%%%%%%%%%%%%%%%%%%%%%%%%%%%%%%%%%%%%%
and the stability condition further implies the following restriction on the function $B^{\rm (I,II)}_{k,\pm}$: $|B^{\rm (I,II)}_{k,\pm}|^{2}<1$. Since the function $q(t)$, depicting expansion of the universe must be continuous at the bounce, i.e., at $q=1$, it follows that: $\zeta^{\rm (I,II)}_{+}(q=1;\{\phi_{k}\})=\zeta^{\rm (I,II)}_{-}(q=1;\{\phi_{k}\})$ \cite{WADA1986729,Vachaspati:1988as}. This condition translates to
%%%%%%%%%%%%%%%%%%%%%%%%%%%%%%%%%%%%%%%%%%%%%%%%%%%%%%%%%%%%%%%%
\begin{align}
B^{\rm (I,II)}_{k,+}=B^{\rm (I,II)}_{k,-}=B^{\rm (I,II)}_{k}~,
\end{align}
%%%%%%%%%%%%%%%%%%%%%%%%%%%%%%%%%%%%%%%%%%%%%%%%%%%%%%%%%%%%%%%%
and all of these coefficients must have an absolute value less than unity. Thus a general solution to the Wheeler-DeWitt equation up to $\mathcal{O}(\ell_{\rm p}^{0})$, in the classical regime is given by \ref{WDW_m1} and \ref{WDW_m2}, respectively, with the $\zeta(t)$ given by \ref{zeta_gaussian} and \ref{gen_sol_matter}, keeping in mind the condition $|B_{k}|^{2}<1$, for stability. This completes our analysis of the solution to the Wheeler DeWitt equation in the classically allowed region. Next, we shall see how a similar analysis can be performed for the classically forbidden region as well. 

%%%%%%%%%%%%%%%%%%%%%%%%%%%%%%%%%%%%%%%%%%%%%%%%%%
%%%%%%%%%%%%%%%%%%%%%%%%%%%%%%%%%%%%%%%%%%%%%%%%%%
%%%%%%%%%%%%%%%%%%%%%%%%%%%%%%%%%%%%%%%%%%%%%%%%%%
%%%%%%%%%%%%%%%%%%%%%%%%%%%%%%%%%%%%%%%%%%%%%%%%%%
\subsection{Classically forbidden region}

For completeness we present here an analysis of the wave function with matter field in the classically forbidden region. For the bouncing models we are considering, the classically forbidden region corresponds to the range $0<q<1$. In this range, the effective potentials $U_{\rm eff}^{\rm (I)}(q)$ and $U_{\rm eff}^{\rm (II)}(q)$ take positive values, as shown in \ref{Ueff_plot}. Thus using the same parametrizations as in the case of classically allowed region, the solutions to \ref{HJ_equation_01} and \ref{HJ_equation_02} can, therefore, be written as,
%%%%%%%%%%%%%%%%%%%%%%%%%%%%%%%%%%%%%%%%%%%%%%%%%%%%%%%%%%%%%%%%
\begin{align}
i\bar{S}^{\rm (I)}_{0,\pm}(q)=\pm \beta_{n}\ell^2_{\rm p}\int_{1}^{q(<1)}dq'~\sqrt{1-q'}~;\qquad i\bar{S}^{\rm (II)}_{0,\pm}(q)=\pm\frac{4}{3} \beta_{n}\ell^2_{\rm p}\int_{1}^{q(<1)}dq'~\sqrt{1-q'^2}~.
\end{align} 
%%%%%%%%%%%%%%%%%%%%%%%%%%%%%%%%%%%%%%%%%%%%%%%%%%%%%%%%%%%%%%%%
Note that the action $\bar{S}_{0,\pm}^{\rm (I,II)}(q)$ is related to $S_{0,\pm}^{\rm (I,II)}(q)$, derived in the previous section, through analytic continuation of the dynamical variable $q(t)$. Similarly, the time coordinate $t$ must also be analytically continued, and hence following \ref{def_time}, it is convenient to define the Euclidean time coordinates $\tau_{\pm}$ in the following manner:
%%%%%%%%%%%%%%%%%%%%%%%%%%%%%%%%%%%%%%%%%%%%%%%%%%%%%%%%%%%%%%%%
\begin{align}
\frac{dq}{d\tau_{\pm}}=\mp i\frac{2\left(\partial_q\bar{S}_{0,\pm}\right)}{b^{2}}~.
\end{align}
%%%%%%%%%%%%%%%%%%%%%%%%%%%%%%%%%%%%%%%%%%%%%%%%%%%%%%%%%%%%%%%%
It then follows that the Euclidean time coordinates $\tau_{\pm}$ defined above corresponds to the Wick rotations of the original Lorentzian time coordinate $t$, such that: $t\rightarrow i\tau_{+}$ and $t\rightarrow -i\tau_{-}$. Performing the Wick rotation and analytic continuation to the complex $q$ plane, we can now write down the general solution to the Wheeler-DeWitt equation to $\mathcal{O}(\ell_p^{0})$, in the classically forbidden region (i.e., $0<q<1$) as,
%%%%%%%%%%%%%%%%%%%%%%%%%%%%%%%%%%%%%%%%%%%%%%%%%%%%%%%%%%%%%%%%
\begin{align}
\Psi^{\rm (I)}(q;\{\phi_{k}\})&\approx \bar{A}^{\rm (I)}_{+}\frac{e^{-\frac{2}{3}\alpha_{n}(1-q)^{\frac{3}{2}}}}{\left[\alpha_{n}^{2}(1-q)\right]^{\frac{1}{4}}}\bar{\zeta}^{\rm (I)}_+(q;\{\phi_{k}\})+\bar{A}^{\rm (I)}_{-}\frac{e^{\frac{2}{3}\alpha_{n}(1-q)^{\frac{3}{2}}}}{\left[\alpha_{n}^{2}(1-q)\right]^{\frac{1}{4}}}\bar{\zeta}^{\rm (I)}_{-}(q;\{\phi_{k}\})~,
\\
\Psi^{\rm (II)}(q;\{\phi_{k}\})&\approx \bar{A}^{\rm (II)}_{+}\frac{e^{-\frac{4}{3}\alpha_{n}\bar{\xi}(q)}}{\left[\alpha_{n}^{2}(1-q^2)\right]^{\frac{1}{4}}}\bar{\zeta}^{\rm (II)}_{+}(q;\{\phi_{k}\})+\bar{A}^{\rm (II)}_{-}\frac{e^{\frac{4}{3}\alpha_{n}\bar{\xi}(q)}}{\left[\alpha_{n}^{2}(1-q^2)\right]^{\frac{1}{4}}}\bar{\zeta}^{\rm (II)}_{-}(q;\{\phi_{k}\})~,
\end{align}   
%%%%%%%%%%%%%%%%%%%%%%%%%%%%%%%%%%%%%%%%%%%%%%%%%%%%%%%%%%%%%%%%
where, the function $\bar{\xi}(q)$ is defined as: $\bar{\xi}(q)=(1/2)[q \sqrt{1-q^2}-\cos^{-1}(q)]$. Further, the matter wave functions $\bar{\zeta}^{\rm (I,II)}_{\pm}$ in the classically forbidden region in terms of the Euclidean time coordinate becomes,
%%%%%%%%%%%%%%%%%%%%%%%%%%%%%%%%%%%%%%%%%%%%%%%%%%%%%%%%%%%%%%%%
\begin{align}\label{zeta_gaussian_forbidden}
\bar{\zeta}^{\rm (I,II)}_{\pm}(\tau_{\pm})\propto \exp\left[\int \frac{d^3k}{(2\pi)^3}\left(\pm \mu^{\rm (I,II)}\frac{\bar{u}'^{\rm (I,II)}_{k,\pm}}{2\bar{u}^{\rm (I,II)}_{k,\pm}}\phi_{k}^2\right)\right]~,
\end{align}
%%%%%%%%%%%%%%%%%%%%%%%%%%%%%%%%%%%%%%%%%%%%%%%%%%%%%%%%%%%%%%%%
where the prime denotes derivative with respect to the appropriate Euclidean time coordinates $\tau_{\pm}$ and, the functions $\bar{u}^{\rm (I)}_{k,\pm}$ and $\bar{u}^{\rm (II)}_{k,\pm}$ are solutions of the Euclidean versions of \ref{KG_I} and \ref{KG_II}, respectively. For large values of $\alpha_{n}$, we expect the wave functions $\Psi^{\rm (I,II)}$ to exponentially decay to zero as $q\rightarrow 0$. Hence, we demand that the unknown coefficients should be such that, $\bar{A}_{-}^{\rm (I)}=\bar{A}_{-}^{\rm (II)}=0$. This implies that the only Wick rotation relevant to the matter sector of the no-boundary wave function is $t\rightarrow i\tau_{+}$. Therefore, for notational convenience we shall henceforth avoid the subscript `$+$' and express $\Psi^{\rm (I,II)}$ in the classically forbidden region as:
%%%%%%%%%%%%%%%%%%%%%%%%%%%%%%%%%%%%%%%%%%%%%%%%%%%%%%%%%%%%%%%%
\begin{align}
\Psi^{\rm (I)}(q;\{\phi_{k}\})&\approx \bar{A}^{(I)}\frac{e^{-\frac{2}{3}\alpha_{n}(1-q)^{\frac{3}{2}}}}{\left[\alpha_{n}^{2}(1-q)\right]^{\frac{1}{4}}}\bar{\zeta}^{\rm (I)}(q;\{\phi_{k}\})~;
\\
\Psi^{(II)}(q;\{\phi_{k}\})&\approx \bar{A}^{\rm (II)}\frac{e^{-\frac{4}{3}\alpha_{n}\bar{\xi}(q)}}{\left[\alpha_{n}^{2}(1-q^2)\right]^{\frac{1}{4}}}\bar{\zeta}^{\rm (II)}(q;\{\phi_{k}\})~;	
\end{align} 
%%%%%%%%%%%%%%%%%%%%%%%%%%%%%%%%%%%%%%%%%%%%%%%%%%%%%%%%%%%%%%%%
where, $\bar{\zeta}^{\rm (I,II)}(\tau)$ is given by \ref{zeta_gaussian_forbidden}, keeping in mind that only the `positive' branch of the solution is allowed. Momentarily dropping the superscripts (I) and (II) for notational convenience, following our analysis for the classically allowed region, we can express the general solution $\bar{u}_{k}$ appearing in \ref{zeta_gaussian_forbidden} in terms of two linearly independent solutions $\bar{v}_{k,1}$ and $\bar{v}_{k,2}$ of the time-dependent harmonic oscillator as:
%%%%%%%%%%%%%%%%%%%%%%%%%%%%%%%%%%%%%%%%%%%%%%%%%%%%%%%%%%%%%%%%
\begin{align}
\bar{u}_{k}=\bar{v}_{k,1}+\bar{B}_{k}\bar{v}_{k,2}~.
\end{align}
%%%%%%%%%%%%%%%%%%%%%%%%%%%%%%%%%%%%%%%%%%%%%%%%%%%%%%%%%%%%%%%%
Here, $\bar{B}_{k}$ is the ratio of the two Bogoliubov coefficients appearing from the rotation between the basis vectors $\bar{u}_{k}$ and $\bar{v}_{k}$. Following the analysis of the previous section, we may express the basis solutions $\bar{v}_{k,1}$ and $\bar{v}_{k,2}$ in terms of two real functions obtained by the Euclidean continuation of the solutions in the previous section, yielding, 
%%%%%%%%%%%%%%%%%%%%%%%%%%%%%%%%%%%%%%%%%%%%%%%%%%%%%%%%%%%%%%%%
\begin{align}
\bar{v}_{k,1}(\tau)=\bar{\chi}_k(\tau)e^{-\bar{\Theta}_k(\tau)}~;\qquad \bar{v}_{k,2}(\tau)=\bar{\chi}_k(\tau)e^{\bar{\Theta}_k(\tau)}~,
\end{align}
%%%%%%%%%%%%%%%%%%%%%%%%%%%%%%%%%%%%%%%%%%%%%%%%%%%%%%%%%%%%%%%%
where, $\bar{\chi}_k(\tau)\equiv \chi_k(i\tau)$ and $\bar{\Theta}'_k(\tau)\equiv\{1/2\mu(i\tau)\bar{\chi}^2_k(\tau)\}$. In terms of these two real functions along with the complex function $\bar{B}_{k}$, the stability condition presented in \ref{stability_cond}, for the classically forbidden region takes the following form:
%%%%%%%%%%%%%%%%%%%%%%%%%%%%%%%%%%%%%%%%%%%%%%%%%%%%%%%%%%%%%%%%
\begin{align}\label{stability_forbidden}
\textrm{Re}\left[\mu(i\tau)\frac{\bar{u}'_k}{\bar{u}_k}\right]=\mu (i\tau) \left(\frac{\bar{\chi}_k '(\tau)}{\bar{\chi}_k(\tau)}+\frac{\bar{\Theta}_k'(\tau) \left(e^{4 \bar{\Theta}_k (\tau)}|\bar{B}_k|^2-1\right)}{\left(e^{2 \bar{\Theta}_k (\tau)}+\textrm{Re}[\bar{B}_k]\right)^2+\textrm{Im}[\bar{B}_k]^2}\right)<0
\end{align}
%%%%%%%%%%%%%%%%%%%%%%%%%%%%%%%%%%%%%%%%%%%%%%%%%%%%%%%%%%%%%%%%
For the real function $\bar{\chi}_k(t)$, we may choose an initial condition, such that for some value of the Euclidean time $\tau_{1}$, $\{\bar{\chi}_k '(\tau_1)/\bar{\chi}_k(\tau_1)\}=0$ and therefore, the stability condition presented in \ref{stability_forbidden}, at the Euclidean time $\tau_{1}$, translates to:
%%%%%%%%%%%%%%%%%%%%%%%%%%%%%%%%%%%%%%%%%%%%%%%%%%%%%%%%%%%%%%%%
\begin{align}\label{restrict_B}
|\bar{B}^{\rm (I,II)}_k|^2<e^{-4\bar{\Theta}^{\rm (I,II)}_k(\tau_1)}
\end{align}
%%%%%%%%%%%%%%%%%%%%%%%%%%%%%%%%%%%%%%%%%%%%%%%%%%%%%%%%%%%%%%%%
where we have retained the superscripts for clarity. Additionally, the continuity at the classical turning point imposes the condition \cite{WADA1986729,Vachaspati:1988as}: $\bar{B}^{\rm (I,II)}_k=B^{\rm (I,II)}_{k}$. This implies that we cannot specify the matter-sector wave function uniquely, but only up to the constants $\bar{B}^{\rm (I, II)}_{k}$. A special choice of $\bar{B}^{\rm (I,II)}_{k}$ that corresponds to the condition that the solution $\bar{u}_k(\tau)$ vanish at $q=0$ is given by
%%%%%%%%%%%%%%%%%%%%%%%%%%%%%%%%%%%%%%%%%%%%%%%%%%%%%%%%%%%%%%%%
\begin{align}
\bar{B}^{\rm (I,II)}_{k}=-e^{-2\bar{\Theta}^{\rm (I,II)}_k(\tau_0)}
\end{align}
%%%%%%%%%%%%%%%%%%%%%%%%%%%%%%%%%%%%%%%%%%%%%%%%%%%%%%%%%%%%%%%%
where, $\tau_0$ is defined via $q(\tau_0)=0$. This choice may be considered as the generalization of the Euclidean vacuum introduced by Hartle and Hawking in the context of de Sitter spacetime. An interesting special case occurs when $\bar{\Theta}_k(\tau_0)\rightarrow\infty$ for all values of $k$. If this condition is satisfied, we can choose $\tau_1=\tau_0$ and, therefore, \ref{restrict_B} implies $\bar{B}^{(I,II)}_k=0$, which in turn implies that a unique wave function exists.

%%%%%%%%%%%%%%%%%%%%%%%%%%%%%%%%%%%%%%%%%%%%%%%%%%%%%%%%%%%%%%%%%%%%%%%%%%%%%%%%%%%%%%%%%%%%%%%%%%%
%%%%%%%%%%%%%%%%%%%%%%%%%%%%%%%%%%%%%%%%%%%%%%%%%%%%%%%%%%%%%%%%%%%%%%%%%%%%%%%%%%%%%%%%%%%%%%%%%%%
%%%%%%%%%%%%%%%%%%%%%%%%%%%%%%%%%%%%%%%%%%%%%%%%%%%%%%%%%%%%%%%%%%%%%%%%%%%%%%%%%%%%%%%%%%%%%%%%%%%
\section{Discussion}\label{discussion}

Bouncing cosmologies, generally, aim to solve the fundamental issues of the SBC without resorting to the inflation mechanism. Moreover, certain shortcomings of the inflationary paradigm, such as the `TransPlanckian' and the singularity problem, can be bypassed in a bouncing scenario. Consequently, cosmological models with a bounce, as against a singularity, are gaining interest as viable alternatives to the much-celebrated inflationary models. In light of this, we attempted to analyze the quantum aspects of bouncing cosmologies, which will be of relevance, especially, to the study of quantum seeds of structure formation. 

Since there are several, conceptually disparate, routes to realize a bounce, it is practically difficult to adopt a single framework to study the set of \textit{all} bouncing models in its entirety. Therefore, we employed a phenomenological approach, that accommodates the essential aspects of a wide class of bouncing models, while at the same time, render analytical calculations tractable. There are two key aspects to this approach; one concerns the kinematics, while the other concerns the dynamics of the set of bouncing universes of our current interest. A convenient parametrization of the FRLW spacetime in terms of the variable $q(t)$, as given in \ref{def_minisuperspace}, together with the condition $p=4-3b$, is the kinematic ingredient of our scheme. Two classes of effective perfect-fluids that source the background FRLW spacetime, which are described by their respective energy densities $\rho_{\rm eff}^{\rm (I)}$ and $\rho_{\rm eff}^{\rm (II)}$, constitute the second, dynamical aspect. The aforementioned, specific parametrization of the FRLW metric was so chosen, such that it leaves the gravitational part of the action a quadratic functional of $\dot{q}(t)$. On the other hand, the specific form of the energy densities $\rho_{\rm eff}^{\rm (I)}$ and $\rho_{\rm eff}^{\rm (II)}$, as presented in \ref{rho_I} and \ref{rho_II}, were so chosen such that: (1) the classical solutions of the corresponding equations of motion describe bouncing scenarios and (2) for an appropriate choice of the parameter $b$, the perfect-fluid part of the action is a linear/quadratic functional of $q(t)$, for densities $\rho_{\rm eff}^{\rm (I)}$ and $\rho_{\rm eff}^{\rm (II)}$, respectively. Hence, we refer to the class of bouncing universes described by $\rho_{\rm eff}^{\rm (I)}$  as the `linear model' and that by $\rho_{\rm eff}^{\rm (II)}$  as the `quadratic model'.  The kinematic and dynamic aspects of our approach, together, enable us to study two wide classes of bouncing cosmologies with considerable analytic comfort.

We explored the quantum aspects of our model using the framework of minisuperspaces. To this end, we have followed two different approaches: (1) using the appropriate minisuperspace Wheeler-de Witt equations and (2) using the minisuperspace path integral. Thanks to the specific choice of the dynamical variable $q(t)$ and the density of the effective fluid, we can find the solutions of the gravitational Wheeler-de Witt equation, for both classes of bouncing models we introduced, analytically. These solutions encode the initial conditions of the universe and hence, is of importance to early universe cosmology. In the context of de Sitter spacetime, a particular solution to the corresponding Wheeler-De Witt equation, namely, the no-boundary wave function, is widely utilized in cosmology for its several appealing features. On account of this, it is worthwhile to investigate the properties of the natural generalization of the no-boundary wave function for bouncing cosmologies. We presented the explicit expressions for the gravitational part of the bouncing-model counterpart of the no-boundary wave function, which are denoted by $\Psi^{\rm (I)}(q)$ and $\Psi^{\rm (II)}(q)$, respectively, for the linear and quadratic models. 

The no-boundary wave function was originally envisioned as arising out of a Euclidean path integral over compact and regular metrics. Recently, it has become clear that, when one attempts a more rigorous calculation, such a Euclidean path integral is ill-defined and divergent. Moreover, even though a well defined, convergent Lorentzian path integral, with the initial condition corresponding to a zero-sized universe, can give rise to the no-boundary wave function, it leads to unstable perturbation. These issues can, however, be circumvented by imposing an initial condition corresponding to a well defined Euclidean momentum associated with the scale factor. Motivated by this, we investigated the extension of this new path integral approach, to the case of bouncing cosmologies. We found that the dominant contribution to the bouncing-model counterpart of the no-boundary wave function comes from a geometry that corresponds to a spacetime that evolves from zero size to a finite size. One is, thus, reminded of the Hawking-Hartle saddle geometry that appears in the context of Euclidean path integral approach to de Sitter cosmology. But, in view of the fact that one is fixing the initial momentum, as against setting the initial size to zero, in the new approach, the off-shell geometries can have \textit{any} initial size. Consequently, one must not interpret the analogue of the no-boundary wave functions thus obtained, as the amplitudes for creation of the universe from `nothing'. A more sensible interpretation would be, that the wave functions are amplitudes for the transition from an initial state of well defined Euclidean momentum to a state of well-defined size. 

Finally, we introduced a real scalar field conformally coupled to the background FRLW spacetime. The quantum theory of the gravity-scalar system can be studied using the corresponding Wheeler-de Witt equation. Since exact solutions could not be found in this case, we progressed by assuming that the scalar fields are perturbations at the quantum level. To enforce this assumption mathematically, we started off by taking the solution to be exponential of a power series in $\ell_{\rm p}^{2}$, with the lowest power being $\ell_{\rm p}^{-2}$. Demanding that this wave function solves the corresponding Wheeler-de Witt equation at all orders of $\ell_{\rm p}^2$, implies that the coefficients of $\ell_{\rm p}^2$ in the power series that define the wave function satisfy a set of differential equations. At the first two leading orders, the corresponding differential equation satisfied by the coefficients can be shown to be equivalent to the Hamilton-Jacobi equation of the gravitational part. In the next order, we obtain a differential equation that essentially describes the paradigm of quantum field theory in curved spacetime. We then proceeded to find the bouncing-model analogue of the wave function that corresponds to the no-boundary proposal for the gravity-scalar system. Interestingly, we found that the wave function, in general, is not unique. Only when $\bar{\Theta}_k(\tau)$, a certain function of the Euclidean time $\tau$, diverges at the point of singularity, do we get a unique wave function. This has the consequence that the initial conditions of the universe have a certain level of arbitrariness that cannot, in general, be fixed without invoking further principles.   

%%%%%%%%%%%%%%%%%%%%%%%%%%%%%%%%%%%%%%%%%%%%%%%%%%%%%%%%%%%%%%%%%%%%%%%%%%%%%%%%%%%%%%%%%%%%%%%%%%%
%%%%%%%%%%%%%%%%%%%%%%%%%%%%%%%%%%%%%%%%%%%%%%%%%%%%%%%%%%%%%%%%%%%%%%%%%%%%%%%%%%%%%%%%%%%%%%%%%%%
%%%%%%%%%%%%%%%%%%%%%%%%%%%%%%%%%%%%%%%%%%%%%%%%%%%%%%%%%%%%%%%%%%%%%%%%%%%%%%%%%%%%%%%%%%%%%%%%%%%
\section*{Acknowledgements}

The authors gratefully acknowledge several helpful comments from L. Sriramkumar, which has improved the content of this work. Research of S.C. is funded by the INSPIRE Faculty fellowship from the DST, Government of India (Reg. No. DST/INSPIRE/04/2018/000893) and by the Start-Up Research Grant from SERB, DST, Government of India (Reg. No. SRG/2020/000409). Research of V.M. is funded by the INSPIRE fellowship from the DST, Government of India (Reg. No. DST/INSPIRE/03/2019/001887). 
%%%%%%%%%%%%%%%%%%%%%%%%%%%%%%%%%%%%%%%%%%%%%%%%%%%%%%%%%%%%%%%%%%%%%%%%%%%%%%%%%%%%%%%%%%%%%%%%%%%
%%%%%%%%%%%%%%%%%%%%%%%%%%%%%%%%%%%%%%%%%%%%%%%%%%%%%%%%%%%%%%%%%%%%%%%%%%%%%%%%%%%%%%%%%%%%%%%%%%%
%%%%%%%%%%%%%%%%%%%%%%%%%%%%%%%%%%%%%%%%%%%%%%%%%%%%%%%%%%%%%%%%%%%%%%%%%%%%%%%%%%%%%%%%%%%%%%%%%%%
\bibliography{Quantum_Bouncing_Cosmology_Journal_revised}

\providecommand{\href}[2]{#2}\begingroup\raggedright\begin{thebibliography}{10}

\bibitem{Guth:1980zm}
A.~H. Guth, ``{The Inflationary Universe: A Possible Solution to the Horizon
  and Flatness Problems},''
  \href{http://dx.doi.org/10.1103/PhysRevD.23.347}{{\em Phys. Rev. D}
  {\bfseries 23} (1981) 347--356}.

\bibitem{Sato:1980yn}
K.~Sato, ``{First Order Phase Transition of a Vacuum and Expansion of the
  Universe},'' {\em Mon. Not. Roy. Astron. Soc.} {\bfseries 195} (1981)
  467--479.

\bibitem{Linde:1981mu}
A.~D. Linde, ``{A New Inflationary Universe Scenario: A Possible Solution of
  the Horizon, Flatness, Homogeneity, Isotropy and Primordial Monopole
  Problems},'' \href{http://dx.doi.org/10.1016/0370-2693(82)91219-9}{{\em Phys.
  Lett. B} {\bfseries 108} (1982) 389--393}.

\bibitem{Albrecht:1982wi}
A.~Albrecht and P.~J. Steinhardt, ``{Cosmology for Grand Unified Theories with
  Radiatively Induced Symmetry Breaking},''
  \href{http://dx.doi.org/10.1103/PhysRevLett.48.1220}{{\em Phys. Rev. Lett.}
  {\bfseries 48} (1982) 1220--1223}.

\bibitem{Starobinsky:1980te}
A.~A. Starobinsky, ``{A New Type of Isotropic Cosmological Models Without
  Singularity},'' \href{http://dx.doi.org/10.1016/0370-2693(80)90670-X}{{\em
  Phys. Lett. B} {\bfseries 91} (1980) 99--102}.

\bibitem{Hawking:1969sw}
S.~W. Hawking and R.~Penrose, ``{The Singularities of gravitational collapse
  and cosmology},'' \href{http://dx.doi.org/10.1098/rspa.1970.0021}{{\em Proc.
  Roy. Soc. Lond. A} {\bfseries 314} (1970) 529--548}.

\bibitem{Borde:1993xh}
A.~Borde and A.~Vilenkin, ``{Eternal inflation and the initial singularity},''
  \href{http://dx.doi.org/10.1103/PhysRevLett.72.3305}{{\em Phys. Rev. Lett.}
  {\bfseries 72} (1994) 3305--3309},
  \href{http://arxiv.org/abs/gr-qc/9312022}{{\ttfamily arXiv:gr-qc/9312022}}.

\bibitem{Martin:2000xs}
J.~Martin and R.~H. Brandenberger, ``{The TransPlanckian problem of
  inflationary cosmology},''
  \href{http://dx.doi.org/10.1103/PhysRevD.63.123501}{{\em Phys. Rev. D}
  {\bfseries 63} (2001) 123501},
  \href{http://arxiv.org/abs/hep-th/0005209}{{\ttfamily arXiv:hep-th/0005209}}.

\bibitem{Brandenberger_2013}
R.~H. Brandenberger and J.~Martin, ``Trans-planckian issues for inflationary
  cosmology,'' \href{http://dx.doi.org/10.1088/0264-9381/30/11/113001}{{\em
  Classical and Quantum Gravity} {\bfseries 30} no.~11, (Apr, 2013) 113001}.
  \url{http://dx.doi.org/10.1088/0264-9381/30/11/113001}.

\bibitem{Finelli:2001sr}
F.~Finelli and R.~Brandenberger, ``{On the generation of a scale invariant
  spectrum of adiabatic fluctuations in cosmological models with a contracting
  phase},'' \href{http://dx.doi.org/10.1103/PhysRevD.65.103522}{{\em Phys. Rev.
  D} {\bfseries 65} (2002) 103522},
  \href{http://arxiv.org/abs/hep-th/0112249}{{\ttfamily arXiv:hep-th/0112249}}.

\bibitem{Brandenberger:2012zb}
R.~H. Brandenberger, ``{The Matter Bounce Alternative to Inflationary
  Cosmology},'' \href{http://arxiv.org/abs/1206.4196}{{\ttfamily
  arXiv:1206.4196 [astro-ph.CO]}}.

\bibitem{Gasperini:1992em}
M.~Gasperini and G.~Veneziano, ``{Pre - big bang in string cosmology},''
  \href{http://dx.doi.org/10.1016/0927-6505(93)90017-8}{{\em Astropart. Phys.}
  {\bfseries 1} (1993) 317--339},
  \href{http://arxiv.org/abs/hep-th/9211021}{{\ttfamily arXiv:hep-th/9211021}}.

\bibitem{Brandenberger:2016vhg}
R.~Brandenberger and P.~Peter, ``{Bouncing Cosmologies: Progress and
  Problems},'' \href{http://dx.doi.org/10.1007/s10701-016-0057-0}{{\em Found.
  Phys.} {\bfseries 47} no.~6, (2017) 797--850},
  \href{http://arxiv.org/abs/1603.05834}{{\ttfamily arXiv:1603.05834
  [hep-th]}}.

\bibitem{Battefeld:2014uga}
D.~Battefeld and P.~Peter, ``{A Critical Review of Classical Bouncing
  Cosmologies},'' \href{http://dx.doi.org/10.1016/j.physrep.2014.12.004}{{\em
  Phys. Rept.} {\bfseries 571} (2015) 1--66},
  \href{http://arxiv.org/abs/1406.2790}{{\ttfamily arXiv:1406.2790
  [astro-ph.CO]}}.

\bibitem{Brandenberger:2009yt}
R.~Brandenberger, ``{Matter Bounce in Horava-Lifshitz Cosmology},''
  \href{http://dx.doi.org/10.1103/PhysRevD.80.043516}{{\em Phys. Rev. D}
  {\bfseries 80} (2009) 043516},
  \href{http://arxiv.org/abs/0904.2835}{{\ttfamily arXiv:0904.2835 [hep-th]}}.

\bibitem{Bamba:2013fha}
K.~Bamba, A.~N. Makarenko, A.~N. Myagky, S.~Nojiri, and S.~D. Odintsov,
  ``{Bounce cosmology from $F(R)$ gravity and $F(R)$ bigravity},''
  \href{http://dx.doi.org/10.1088/1475-7516/2014/01/008}{{\em JCAP} {\bfseries
  01} (2014) 008}, \href{http://arxiv.org/abs/1309.3748}{{\ttfamily
  arXiv:1309.3748 [hep-th]}}.

\bibitem{Desai:2015haa}
S.~Desai and N.~J. Pop\l{}awski, ``{Non-parametric reconstruction of an
  inflaton potential from
  Einstein\textendash{}Cartan\textendash{}Sciama\textendash{}Kibble gravity
  with particle production},''
  \href{http://dx.doi.org/10.1016/j.physletb.2016.02.014}{{\em Phys. Lett. B}
  {\bfseries 755} (2016) 183--189},
  \href{http://arxiv.org/abs/1510.08834}{{\ttfamily arXiv:1510.08834
  [astro-ph.CO]}}.

\bibitem{ArkaniHamed:2003uy}
N.~Arkani-Hamed, H.-C. Cheng, M.~A. Luty, and S.~Mukohyama, ``{Ghost
  condensation and a consistent infrared modification of gravity},''
  \href{http://dx.doi.org/10.1088/1126-6708/2004/05/074}{{\em JHEP} {\bfseries
  05} (2004) 074}, \href{http://arxiv.org/abs/hep-th/0312099}{{\ttfamily
  arXiv:hep-th/0312099}}.

\bibitem{Cai:2008qw}
Y.-F. Cai, T.-t. Qiu, R.~Brandenberger, and X.-m. Zhang, ``{A Nonsingular
  Cosmology with a Scale-Invariant Spectrum of Cosmological Perturbations from
  Lee-Wick Theory},'' \href{http://dx.doi.org/10.1103/PhysRevD.80.023511}{{\em
  Phys. Rev. D} {\bfseries 80} (2009) 023511},
  \href{http://arxiv.org/abs/0810.4677}{{\ttfamily arXiv:0810.4677 [hep-th]}}.

\bibitem{Raveendran:2018why}
R.~N. Raveendran and L.~Sriramkumar, ``{Viable scalar spectral tilt and
  tensor-to-scalar ratio in near-matter bounces},''
  \href{http://dx.doi.org/10.1103/PhysRevD.100.083523}{{\em Phys. Rev. D}
  {\bfseries 100} no.~8, (2019) 083523},
  \href{http://arxiv.org/abs/1812.06803}{{\ttfamily arXiv:1812.06803
  [astro-ph.CO]}}.

\bibitem{Raveendran:2017vfx}
R.~N. Raveendran, D.~Chowdhury, and L.~Sriramkumar, ``{Viable tensor-to-scalar
  ratio in a symmetric matter bounce},''
  \href{http://dx.doi.org/10.1088/1475-7516/2018/01/030}{{\em JCAP} {\bfseries
  01} (2018) 030}, \href{http://arxiv.org/abs/1703.10061}{{\ttfamily
  arXiv:1703.10061 [gr-qc]}}.

\bibitem{Bonanno:2017gji}
A.~Bonanno, G.~Gionti, S.~J., and A.~Platania, ``{Bouncing and emergent
  cosmologies from Arnowitt\textendash{}Deser\textendash{}Misner RG flows},''
  \href{http://dx.doi.org/10.1088/1361-6382/aaa535}{{\em Class. Quant. Grav.}
  {\bfseries 35} no.~6, (2018) 065004},
  \href{http://arxiv.org/abs/1710.06317}{{\ttfamily arXiv:1710.06317 [gr-qc]}}.

\bibitem{Bamba:2014zoa}
K.~Bamba, A.~N. Makarenko, A.~N. Myagky, and S.~D. Odintsov, ``{Bounce universe
  from string-inspired Gauss-Bonnet gravity},''
  \href{http://dx.doi.org/10.1088/1475-7516/2015/04/001}{{\em JCAP} {\bfseries
  04} (2015) 001}, \href{http://arxiv.org/abs/1411.3852}{{\ttfamily
  arXiv:1411.3852 [hep-th]}}.

\bibitem{Basile:2021amb}
I.~Basile and A.~Platania, ``{Cosmological $\alpha'$-corrections from the
  functional renormalization group},''
  \href{http://arxiv.org/abs/2101.02226}{{\ttfamily arXiv:2101.02226
  [hep-th]}}.

\bibitem{Brandenberger:1988aj}
R.~H. Brandenberger and C.~Vafa, ``{Superstrings in the Early Universe},''
  \href{http://dx.doi.org/10.1016/0550-3213(89)90037-0}{{\em Nucl. Phys. B}
  {\bfseries 316} (1989) 391--410}.

\bibitem{Haro:2015oqa}
J.~Haro, A.~N. Makarenko, A.~N. Myagky, S.~D. Odintsov, and V.~K. Oikonomou,
  ``{Bouncing loop quantum cosmology in Gauss-Bonnet gravity},''
  \href{http://dx.doi.org/10.1103/PhysRevD.92.124026}{{\em Phys. Rev. D}
  {\bfseries 92} no.~12, (2015) 124026},
  \href{http://arxiv.org/abs/1506.08273}{{\ttfamily arXiv:1506.08273 [gr-qc]}}.

\bibitem{Ashtekar:2008ay}
A.~Ashtekar, ``{Singularity Resolution in Loop Quantum Cosmology: A Brief
  Overview},'' \href{http://dx.doi.org/10.1088/1742-6596/189/1/012003}{{\em J.
  Phys. Conf. Ser.} {\bfseries 189} (2009) 012003},
  \href{http://arxiv.org/abs/0812.4703}{{\ttfamily arXiv:0812.4703 [gr-qc]}}.

\bibitem{WilsonEwing:2012pu}
E.~Wilson-Ewing, ``{The Matter Bounce Scenario in Loop Quantum Cosmology},''
  \href{http://dx.doi.org/10.1088/1475-7516/2013/03/026}{{\em JCAP} {\bfseries
  03} (2013) 026}, \href{http://arxiv.org/abs/1211.6269}{{\ttfamily
  arXiv:1211.6269 [gr-qc]}}.

\bibitem{Cai:2014zga}
Y.-F. Cai and E.~Wilson-Ewing, ``{Non-singular bounce scenarios in loop quantum
  cosmology and the effective field description},''
  \href{http://dx.doi.org/10.1088/1475-7516/2014/03/026}{{\em JCAP} {\bfseries
  03} (2014) 026}, \href{http://arxiv.org/abs/1402.3009}{{\ttfamily
  arXiv:1402.3009 [gr-qc]}}.

\bibitem{Feldbrugge:2017kzv}
J.~Feldbrugge, J.-L. Lehners, and N.~Turok, ``{Lorentzian Quantum Cosmology},''
  \href{http://dx.doi.org/10.1103/PhysRevD.95.103508}{{\em Phys. Rev. D}
  {\bfseries 95} no.~10, (2017) 103508},
  \href{http://arxiv.org/abs/1703.02076}{{\ttfamily arXiv:1703.02076
  [hep-th]}}.

\bibitem{DiTucci:2019dji}
A.~Di~Tucci and J.-L. Lehners, ``{No-Boundary Proposal as a Path Integral with
  Robin Boundary Conditions},''
  \href{http://dx.doi.org/10.1103/PhysRevLett.122.201302}{{\em Phys. Rev.
  Lett.} {\bfseries 122} no.~20, (2019) 201302},
  \href{http://arxiv.org/abs/1903.06757}{{\ttfamily arXiv:1903.06757
  [hep-th]}}.

\bibitem{Lehners2019}
A.~Di~Tucci, J.-L. Lehners, and L.~Sberna, ``No-boundary prescriptions in
  lorentzian quantum cosmology,''
  \href{http://dx.doi.org/10.1103/PhysRevD.100.123543}{{\em Phys. Rev. D}
  {\bfseries 100} (Dec, 2019) 123543}.
  \url{https://link.aps.org/doi/10.1103/PhysRevD.100.123543}.

\bibitem{Louko:1988zb}
J.~Louko, ``{Semiclassical Path Measure and Factor Ordering in Quantum
  Cosmology},'' \href{http://dx.doi.org/10.1016/0003-4916(88)90170-4}{{\em
  Annals Phys.} {\bfseries 181} (1988) 318--373}.

\bibitem{Louko:1989up}
J.~Louko and P.~J. Ruback, ``{SPATIALLY FLAT QUANTUM COSMOLOGY},''
  \href{http://dx.doi.org/10.1088/0264-9381/8/1/013}{{\em Class. Quant. Grav.}
  {\bfseries 8} (1991) 91--122}.

\bibitem{Halliwell1988}
J.~J. Halliwell, ``Derivation of the wheeler-dewitt equation from a path
  integral for minisuperspace models,''
  \href{http://dx.doi.org/10.1103/PhysRevD.38.2468}{{\em Phys. Rev. D}
  {\bfseries 38} (Oct, 1988) 2468--2481}.
  \url{https://link.aps.org/doi/10.1103/PhysRevD.38.2468}.

\bibitem{Rajeev:2019ivq}
K.~Rajeev, ``{Complex time route to quantum backreaction},''
  \href{http://dx.doi.org/10.1140/epjc/s10052-019-7480-2}{{\em Eur. Phys. J. C}
  {\bfseries 79} no.~11, (2019) 959},
  \href{http://arxiv.org/abs/2001.02543}{{\ttfamily arXiv:2001.02543 [gr-qc]}}.

\bibitem{hawking_ellis_1973}
S.~W. Hawking and G.~F.~R. Ellis,
  \href{http://dx.doi.org/10.1017/CBO9780511524646}{{\em The Large Scale
  Structure of Space-Time}}.
\newblock Cambridge Monographs on Mathematical Physics. Cambridge University
  Press, 1973.

\bibitem{Cai:2007qw}
Y.-F. Cai, T.~Qiu, Y.-S. Piao, M.~Li, and X.~Zhang, ``{Bouncing universe with
  quintom matter},''
  \href{http://dx.doi.org/10.1088/1126-6708/2007/10/071}{{\em JHEP} {\bfseries
  10} (2007) 071}, \href{http://arxiv.org/abs/0704.1090}{{\ttfamily
  arXiv:0704.1090 [gr-qc]}}.

\bibitem{Hartle:1983ai}
J.~B. Hartle and S.~W. Hawking, ``Wave function of the universe,''
  \href{http://dx.doi.org/10.1103/PhysRevD.28.2960}{{\em Phys. Rev. D}
  {\bfseries 28} (Dec, 1983) 2960--2975}.
  \url{https://link.aps.org/doi/10.1103/PhysRevD.28.2960}.

\bibitem{WADA1986729}
S.~Wada, ``Quantum cosmological perturbations in pure gravity,''
  \href{http://dx.doi.org/https://doi.org/10.1016/0550-3213(86)90073-8}{{\em
  Nuclear Physics B} {\bfseries 276} no.~3, (1986) 729 -- 743}.
  \url{http://www.sciencedirect.com/science/article/pii/0550321386900738}.

\bibitem{Vilenkin:1987kf}
A.~Vilenkin, ``{Quantum Cosmology and the Initial State of the Universe},''
  \href{http://dx.doi.org/10.1103/PhysRevD.37.888}{{\em Phys. Rev. D}
  {\bfseries 37} (1988) 888}.

\bibitem{PhysRev.160.1113}
B.~S. DeWitt, ``Quantum theory of gravity. i. the canonical theory,''
  \href{http://dx.doi.org/10.1103/PhysRev.160.1113}{{\em Phys. Rev.} {\bfseries
  160} (Aug, 1967) 1113--1148}.
  \url{https://link.aps.org/doi/10.1103/PhysRev.160.1113}.

\bibitem{Damour:2019iyi}
T.~Damour and A.~Vilenkin, ``{Quantum instability of an oscillating
  universe},'' \href{http://dx.doi.org/10.1103/PhysRevD.100.083525}{{\em Phys.
  Rev. D} {\bfseries 100} no.~8, (2019) 083525},
  \href{http://arxiv.org/abs/1907.04029}{{\ttfamily arXiv:1907.04029 [gr-qc]}}.

\bibitem{olver2010nist}
F.~W. Olver, D.~W. Lozier, R.~F. Boisvert, and C.~W. Clark, {\em NIST handbook
  of mathematical functions hardback and CD-ROM}.
\newblock Cambridge university press, 2010.

\bibitem{DiazDorronsoro:2017hti}
J.~Diaz~Dorronsoro, J.~J. Halliwell, J.~B. Hartle, T.~Hertog, and O.~Janssen,
  ``{Real no-boundary wave function in Lorentzian quantum cosmology},''
  \href{http://dx.doi.org/10.1103/PhysRevD.96.043505}{{\em Phys. Rev. D}
  {\bfseries 96} no.~4, (2017) 043505},
  \href{http://arxiv.org/abs/1705.05340}{{\ttfamily arXiv:1705.05340 [gr-qc]}}.

\bibitem{DiazDorronsoro:2018wro}
J.~Diaz~Dorronsoro, J.~J. Halliwell, J.~B. Hartle, T.~Hertog, O.~Janssen, and
  Y.~Vreys, ``{Damped perturbations in the no-boundary state},''
  \href{http://dx.doi.org/10.1103/PhysRevLett.121.081302}{{\em Phys. Rev.
  Lett.} {\bfseries 121} no.~8, (2018) 081302},
  \href{http://arxiv.org/abs/1804.01102}{{\ttfamily arXiv:1804.01102 [gr-qc]}}.

\bibitem{Feldbrugge:2017fcc}
J.~Feldbrugge, J.-L. Lehners, and N.~Turok, ``{No smooth beginning for
  spacetime},'' \href{http://dx.doi.org/10.1103/PhysRevLett.119.171301}{{\em
  Phys. Rev. Lett.} {\bfseries 119} no.~17, (2017) 171301},
  \href{http://arxiv.org/abs/1705.00192}{{\ttfamily arXiv:1705.00192
  [hep-th]}}.

\bibitem{Feldbrugge:2017mbc}
J.~Feldbrugge, J.-L. Lehners, and N.~Turok, ``{No rescue for the no boundary
  proposal: Pointers to the future of quantum cosmology},''
  \href{http://dx.doi.org/10.1103/PhysRevD.97.023509}{{\em Phys. Rev. D}
  {\bfseries 97} no.~2, (2018) 023509},
  \href{http://arxiv.org/abs/1708.05104}{{\ttfamily arXiv:1708.05104
  [hep-th]}}.

\bibitem{Feldbrugge:2018gin}
J.~Feldbrugge, J.-L. Lehners, and N.~Turok, ``{Inconsistencies of the New
  No-Boundary Proposal},''
  \href{http://dx.doi.org/10.3390/universe4100100}{{\em Universe} {\bfseries 4}
  no.~10, (2018) 100}, \href{http://arxiv.org/abs/1805.01609}{{\ttfamily
  arXiv:1805.01609 [hep-th]}}.

\bibitem{Bojowald:2018gdt}
M.~Bojowald and S.~Brahma, ``{Loops rescue the no-boundary proposal},''
  \href{http://dx.doi.org/10.1103/PhysRevLett.121.201301}{{\em Phys. Rev.
  Lett.} {\bfseries 121} no.~20, (2018) 201301},
  \href{http://arxiv.org/abs/1810.09871}{{\ttfamily arXiv:1810.09871 [gr-qc]}}.

\bibitem{Bojowald:2020kob}
M.~Bojowald and S.~Brahma, ``{Loop quantum gravity, signature change, and the
  no-boundary proposal},''
  \href{http://dx.doi.org/10.1103/PhysRevD.102.106023}{{\em Phys. Rev. D}
  {\bfseries 102} no.~10, (2020) 106023},
  \href{http://arxiv.org/abs/2011.02884}{{\ttfamily arXiv:2011.02884 [gr-qc]}}.

\bibitem{Cai_2013}
Y.-F. Cai, R.~Brandenberger, and P.~Peter, ``Anisotropy in a non-singular
  bounce,'' \href{http://dx.doi.org/10.1088/0264-9381/30/7/075019}{{\em
  Classical and Quantum Gravity} {\bfseries 30} no.~7, (Mar, 2013) 075019}.
  \url{http://dx.doi.org/10.1088/0264-9381/30/7/075019}.

\bibitem{davidson1999wavefunction}
A.~Davidson, D.~Karasik, and Y.~Lederer, ``Wavefunction of a brane-like
  universe,'' {\em Classical and Quantum Gravity} {\bfseries 16} no.~4, (1999)
  1349.

\bibitem{Singh:1989ct}
T.~Singh and T.~Padmanabhan, ``{NOTES ON SEMICLASSICAL GRAVITY},''
  \href{http://dx.doi.org/10.1016/0003-4916(89)90180-2}{{\em Annals Phys.}
  {\bfseries 196} (1989) 296--344}.

\bibitem{Padmanabhan:1990fn}
T.~Padmanabhan and T.~Singh, ``{On the semiclassical limit of the
  Wheeler-DeWitt equation},''
  \href{http://dx.doi.org/10.1088/0264-9381/7/3/015}{{\em Class. Quant. Grav.}
  {\bfseries 7} (1990) 411--426}.

\bibitem{birrell1984quantum}
N.~Birrell, N.~Birrell, and P.~Davies, {\em Quantum Fields in Curved Space}.
\newblock Cambridge Monographs on Mathematical Physics. Cambridge University
  Press, 1984.

\bibitem{parker2009quantum}
L.~Parker and D.~Toms, {\em Quantum field theory in curved spacetime: quantized
  fields and gravity}.
\newblock Cambridge university press, 2009.

\bibitem{wald1994quantum}
R.~M. Wald, {\em Quantum field theory in curved spacetime and black hole
  thermodynamics}.
\newblock University of Chicago press, 1994.

\bibitem{fulling1989aspects}
S.~A. Fulling {\em et~al.}, {\em Aspects of quantum field theory in curved
  spacetime}.
\newblock No.~17. Cambridge university press, 1989.

\bibitem{Padmanabhan:2017rsj}
T.~Padmanabhan, ``{Demystifying the constancy of the Ermakov\textendash{}Lewis
  invariant for a time-dependent oscillator},''
  \href{http://dx.doi.org/10.1142/S0217732318300057}{{\em Mod. Phys. Lett. A}
  {\bfseries 33} no.~07n08, (2018) 1830005},
  \href{http://arxiv.org/abs/1712.07328}{{\ttfamily arXiv:1712.07328
  [physics.class-ph]}}.

\bibitem{Vachaspati:1988as}
T.~Vachaspati and A.~Vilenkin, ``{On the Uniqueness of the Tunneling Wave
  Function of the Universe},''
  \href{http://dx.doi.org/10.1103/PhysRevD.37.898}{{\em Phys. Rev. D}
  {\bfseries 37} (1988) 898}.

\end{thebibliography}\endgroup

\bibliographystyle{utphys1}
%%%%%%%%%%%%%%%%%%%%%%%%%%%%%%%%%%%%%%%%%%%%%%%%%%%%%%%%%%%%%%%%%%%%%%%%%%%%%%%%%%%%%%%%%%%%%%%%%%%
%%%%%%%%%%%%%%%%%%%%%%%%%%%%%%%%%%%%%%%%%%%%%%%%%%%%%%%%%%%%%%%%%%%%%%%%%%%%%%%%%%%%%%%%%%%%%%%%%%%
%%%%%%%%%%%%%%%%%%%%%%%%%%%%%%%%%%%%%%%%%%%%%%%%%%%%%%%%%%%%%%%%%%%%%%%%%%%%%%%%%%%%%%%%%%%%%%%%%%%

\end{document}